\documentclass[aip,reprint,amsmath,amssymb]{revtex4-2}

\usepackage{hyperref}
\usepackage[dvipsnames]{xcolor}
\usepackage{tikz}

\usepackage{graphicx}

\usepackage{caption}
\usepackage[labelformat=simple]{subcaption}

\usepackage{ragged2e}
\DeclareCaptionJustification{justified}{\justifying}

\DeclareMathOperator{\trace}{Tr} 
\newcommand\bra[1] {\langle {#1} |} 
\newcommand\ket[1] {| {#1} \rangle} 
\newcommand\scp[2] {\langle{#1}|{#2}\rangle}

\newcommand{\argmin}{\mathop{\mathrm{argmin}}}

\newcommand{\rhoref}{\rho_{\mathrm{ref}}}

\newcommand\myvec[1] {\vec{#1}}

\newcommand\unitcell {\mathcal{C}}
\newcommand\dBvKzone {\mathcal{C}_{\mathrm{BvK}}'}
\newcommand\cBvKzone {\mathcal{C}_{\mathrm{BvK}}}
\newcommand\RealLattice {\mathcal{L}}
\newcommand\ReciprocalLattice {\mathcal{RL}}
\newcommand\BvKLattice {\mathcal{L}_{\mathrm{BvK}}}
\newcommand\BvKRecLattice {\mathcal{RL}_{\mathrm{BvK}}}
\newcommand\cBZ {\mathcal{BZ}}
\newcommand\dBZ {\mathcal{BZ}'}
\newcommand\WaveFunSpace[1] {\mathcal{W}_{#1}}
\newcommand\MixedStateSpace[1] {\mathcal{D}_{#1}}

\newcommand\CoulQform {E_{\mathrm{H}}}
\newcommand\Xbvk {X_{\mathrm{BvK}}}

\newcommand\pairing[2] {\langle {#1}, {#2} \rangle}
\newcommand\Jbvk {J^{\text{BvK}}}
\newcommand\Jdual {J}

\newcommand\Nbvk {N_{\mathrm{BvK}}}
\newcommand\invNbvk {\eta}

\begin{document}

\title{Regularised density-potential inversion for periodic systems: application to exact exchange in one dimension}

\author{Oliver M. Bohle}
\affiliation{Hylleraas Centre for Quantum Molecular Sciences, Department of Chemistry, University of Oslo, 0315 Oslo, Norway}

\author{Maryam Lotfigolian}
\affiliation{Department of Computer Science, Oslo Metropolitan University, 0130 Oslo, Norway}

\author{Andre Laestadius}
\affiliation{Department of Computer Science, Oslo Metropolitan University, 0130 Oslo, Norway}
\affiliation{Hylleraas Centre for Quantum Molecular Sciences, Department of Chemistry, University of Oslo, 0315 Oslo, Norway}

\author{Erik I. Tellgren}
\email[Electronic address:\;]{erik.tellgren@kjemi.uio.no}
\affiliation{Hylleraas Centre for Quantum Molecular Sciences, Department of Chemistry, University of Oslo, 0315 Oslo, Norway}

\begin{abstract}
    A detailed convex analysis-based formulation of density-functional theory for periodic systems in arbitrary dimensions is presented. The electron-electron interaction is taken to be of Yukawa type, harmonising with underlying function spaces for densities and wave functions. Moreau--Yosida regularisation of the underlying non-interacting density functionals is then considered, allowing us to recast the Hohenberg--Kohn mapping in a form that is insensitive to perturbations (non-expansiveness) and lends itself to numerical implementation. The general theory is exemplified with a numerical Hartree--Fock implementation for one-dimensional systems. We discuss in particular the challenge of self-consistent field optimisation in calculations related to the regularised noninteracting Hohenberg--Kohn map. The implementation is used to demonstrate that it is practically feasible to recover local Kohn--Sham potentials reproducing the effects of exact exchange within this scheme, which provides a proof-of-principle for recovering the exchange-correlation potential at more accurate levels of theory. Error analysis is performed for 
    the regularised inverse Kohn--Sham algorithm 
    by quantifying, both theoretically and numerically, how perturbations of the input ground-state density propagate through the regularised density-to-potential map. 
\end{abstract}

\maketitle

\section{Introduction}
\label{sec:intro}

Density-functional theory (DFT) reformulates the many-electron ground-state problem in terms of the electron density, circumventing the complicated many-body wavefunction. In practice, the key method is Kohn--Sham density-functional theory (KS-DFT), which replaces the interacting system by a non-interacting one subject to an effective, local potential~\cite{KS1965}. The speed and accuracy of KS-DFT has cemented it as a pillar of quantum chemistry, computational materials science and solid-state physics~\cite{Burke2012,Verma_2020,Teale2022}. The Hohenberg--Kohn theorems provided the first general justification for DFT~\cite{Hohenberg1964}. Although the subsequent Levy--Lieb constrained search formulation\cite{PERCUS_IJQC13_89,levy1979} and Lieb's convex analysis formulation~\cite{Lieb1983} further strengthened the foundations of DFT, the Hohenberg--Kohn mapping from ground-state densities to potentials, also known as density-potential inversion, remains of great interest~\cite{PenzPartI2023}. This is particularly true when it is employed to find the effective, local (Kohn--Sham) potential for a ground-state density of the interacting system. This is referred to as inverse Kohn--Sham (iKS). Besides its theoretical significance, the ability to use the density-potential inversion in this way provides valuable data, facilitating the construction of computationally tractable approximate density functionals.

However, the Hohenberg--Kohn mapping is difficult to approximate in a controlled manner~\cite{GarrigueCMP386_1803}. Mathematically, the application to reconstruct local potentials for interacting ground-state densities depend on the assumption of noninteracting $v$-representability. No tractable method exists to verify this assumption. It is also sensitive to perturbations in that a $v$-representable and non-$v$-representable densities can be arbitrarily close to each other. Nonetheless, a number of numerical methods have been explored~\cite{Aryasetiawan1988,Knorr-Godby-1992,Gorling1992,ZP1993,Wang1993,ZP1993,ZMP1994,Knorr_1994,vanleeuwen1994exchange,Yang_Wu_2002, Wu-Yang-2003,Peirs2003,Kadantsev2004,Bulat2007,Gaiduk2013,Wagner2014,JensenWasserman2017,Zhang2018,Ou2018,kumar2019universal,Kanungo2019,Kumar2020,Garrick2020,Callow2020,Shi2021,Erhard2022,Gould2023}. We highlight in particular the variational principle studied by Foulkes and Haydock~\cite{FOULKES_PRB39_12520,Kadantsev2004}, the work by Wang and Parr that self-consistently iterated an expression for the local potential in terms of the Kohn--Sham orbitals~\cite{Wang1993}, the penalty term approach by Zhao, Morrison, and Parr (ZMP) to enforce the density constraint~\cite{ZMP1994}, and the work by Wu and Yang that exploited Lieb's variational formulation to systematically approximate the local potential~\cite{Wu-Yang-2003}. Kumar, \textit{et al.}, attempted to subsume several methods within a unified scheme~\cite{kumar2019universal}. Other works have focused on improving the accuracy of the asymptotic behavior of the local potential or Kohn--Sham properties like shell structure~\cite{vanleeuwen1994exchange,Peirs2003} and on smoothing out spurious oscillations due to finite basis set effects~\cite{Bulat2007,Gaiduk2013,Callow2020}. Jensen and Wasserman surveyed iKS methods, emphasising algorithm design and error analysis, and benchmarked the performance on model systems~\cite{JensenWasserman2017}. Despite the attention the problem has received, many open problems and challenges remain~\cite{Shi2021,Erhard2022,Shi2022,Crisostomo2023,Wrighton2023,Gould2023}.

Some of these methods can in principle be employed in a rather broad sense of density-potential inversion and even be used to compute what is essentially the exact functional except for a basis set discretisation~\cite{TEALE_JCP130_104111,Wagner2014}. They can also be used to compute, for example, the adiabatic connection for a given density at an accuracy equivalent to wave function methods such as MP2, the coupled-cluster method or full CI level and approximate density functionals such as range-separated hybrid functionals~\cite{TEALE_JCP132_164115,TEALE_JCP133_164112}. Applications to a relatively simple model of the interacting system, such as limiting to only exact exchange, while significant in their own right, can thus also serve as a proof of concept for later applications to more accurate treatments.

An important development was the recasting of DFT~\cite{Kvaal2014} within the mathematical theory of Moreau--Yosida (MY) regularisation~\cite{Bauschke_2017,Barbu-Precepanu,Barbu_2010}. This not only circumvents the non-differentiability of the universal functional~\cite{Lieb1983,Lammert2007,Kvaal2014} and imposes practically useful regularity within the exact setting~\cite{KSpaper2018,MY-CDFTpaper2019,PRLerrata,Penz_2023,Penz2026}, but also enables a rigorous framework for penalty methods such as the ZMP~\cite{Kvaal2014,Penz_2023,Herbst2025} and the van Leeuwen and Baerends approach~\cite{vanLeeuwen_1994}. MY regularisation additionally circumvents the need to assume $v$-representability, since it extends the domain to non-$v$-representable and non-$N$-representable densities. In the present work, we use MY regularisation to study density-potential inversion in periodic systems. As most mathematical work has focused on either finite systems or simple periodic boundary conditions, it is of interest to provide a detailed formulation for the case of Born--von K\'arm\'an periodic boundary conditions allowing for conventional band structures. We focus on the case where the electron-electron repulsion operator is directly related to the chosen function space for wave functions and densities. Specifically, we take the electron-electron repulsion to be of Yukawa type~\cite{ROWLINSON_PA156_15} and the density space to only include densities with finite Yukawa interaction. The relevant theory is developed in detail and a numerical implementation corresponding to the special case of Hartree--Fock theory---or exact exchange at the DFT level---in one dimension is presented. Using the numerical implementation, we investigate the role of including both the Hartree and external energy for the noninteracting functional to be regularised. We moreover explore the rate of convergence toward the local exchange potential in the limit of the regularisation parameter to zero. This is also combined with an error analysis where we include perturbations to the input ground-state density. Thus, the mathematical analysis is demonstrated and connected to the numerical results.

The rest of the article is structured as follows. In Sec.~\ref{sec:theory} we present a suitable mathematical formulation for a convex analysis formulation of DFT and MY regularisation in the periodic setting. Sec.~\ref{secNumImpl} describes a specialisation of the general theory to the case of Hartree--Fock theory with a focus on its numerical implementation. In Sec.~\ref{secNumRes}, we report numerical results and error analysis for one-dimensional systems. Finally, Sec.~\ref{sec:concl} summarises the theoretical and numerical results.

\section{Theory}
\label{sec:theory}

Here we give a self-contained presentation of exact density-functional
theory, Hartree--Fock theory and Moreau--Yosida (MY) regularisation in a
periodic setting. As the electron-electron interaction, we consider
the Yukawa potential, which is chosen since it is well adapted to the MY
formalism. The setting is very general and allows for spatial domains
in $d$ dimensions, with $p \leq d$ periodic directions. This subsumes,
e.g., Luttinger liquids ($p=d=1$), polymers ($p=1,d=3$), confined
two-dimensional systems ($p=d=2$), two-dimensional sheets or surfaces embedded in three-dimensions ($p=2,d=3$),
and three-dimensional crystalline materials ($p=d=3$). Our periodic formalism
is mostly standard, with the exception of an unconventional choice of
the basic density variable in the density-functional formulation. This
choice arises as the natural way to accomodate both Lieb's 
formulation of DFT based on convex analysis~\cite{Lieb1983} and the standard notion of band structures
within the first Brillouin zone.

\subsection{Periodic boundary conditions}\label{subsec:PBC}

We consider the spatial domain $\mathbb{R}^d$, $d\geq 1$, and impose Born--von K\'arm\'an periodic boundary conditions in $p \leq d$ dimensions specified by lattice vectors $\myvec{a}_1,\ldots,\myvec{a}_p$ and additional basis vectors $\myvec{a}_{p+1},\ldots,\myvec{a}_d$ in the nonperiodic directions. The periodic and nonperiodic basis vectors are orthogonal, i.e.
\begin{equation}
    \myvec{a}_i\cdot\myvec{a}_j = 0, \quad 1\leq i \leq p, \quad p < j \leq d.
\end{equation}
The reciprocal lattice vectors $\myvec{b}_1,\ldots,\myvec{b}_d$ are dual vectors, defined by
\begin{equation}
    \myvec{b}_i\cdot\myvec{a}_j = 2\pi \delta_{ij}.
\end{equation}
Due to the orthogonality of the periodic and nonperiodic directions, we sometimes write a position vector as $\myvec{r} = \myvec{r}' + \myvec{r}''$ and think of the periodic part $\myvec{r}' = \sum_{i=1}^p \lambda_i \myvec{a}_i$ as an element of $\mathbb{R}^p$ and the nonperiodic part $\myvec{r}'' = \sum_{i=p+1}^d \lambda_i \myvec{a}_i$ as an element of $\mathbb{R}^{d-p}$. Mutatis mutandis for vectors in reciprocal space that enjoy the same orthogonality properties.

A primitive unit cell is any translation of
\begin{equation}
  \unitcell = \Bigg\{\sum_{i=1}^d \lambda_i \myvec{a}_i \Big| \lambda_1,\ldots,\lambda_p \in (0,1], \ \lambda_{p+1},\ldots,\lambda_d \in \mathbb{R} \Bigg\}
\end{equation}
by a lattice vector and the real-space lattice is the integer multiples of the periodic lattice vectors,
\begin{equation}
    \RealLattice = \Big\{ \sum_{i=1}^p m_i \myvec{a}_i \Big| m_i \in \mathbb{Z} \Big\}.
\end{equation}
The reciprocal lattice is analogously
\begin{equation}
    \ReciprocalLattice = \Big\{ \sum_{i=1}^p m_i \myvec{b}_i \Big| m_i \in \mathbb{Z} \Big\}
\end{equation}
and the (continuous) first Brillouin zone is the set of all points in reciprocal space that are closer to the origin than to any other lattice site
\begin{equation}
    \cBZ = \Big\{ \myvec{q} \in \mathbb{R}^d \, \Big| \, |\myvec{q}| \leq \min_{\myvec{G}\in\ReciprocalLattice} |\myvec{q}+\myvec{G}| \Big\}.
\end{equation}

\begin{figure}
  \begin{center}
    \includegraphics[width=0.95\columnwidth]{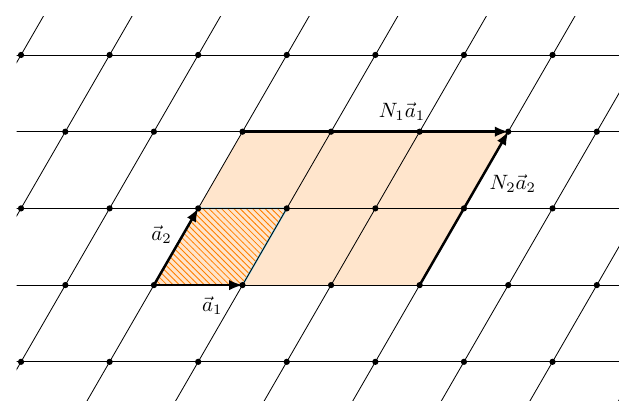} \\
    \includegraphics[width=0.95\columnwidth]{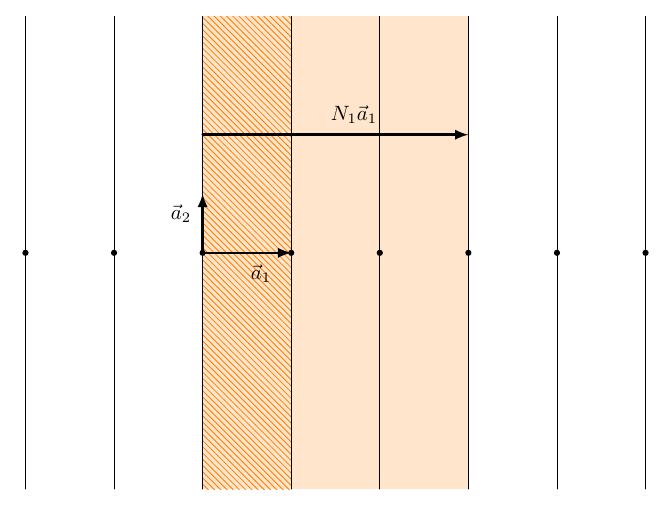}
  \end{center}
  
  \caption{\label{figLattice} Two examples of real-space lattices. Black dots mark the points in $\RealLattice$. Top: $p=d=2$ dimensions, the hatched area is the unit cell and the filled orange area is the $3\times 2$ Born--von K\'arm\'an zone. Bottom: $p=1$, $d=2$ dimensions, with a Born--von K\'arm\'an zone consisting of three unit cells.}
\end{figure}

A Born--von K\'arm\'an zone is chosen as a $N_1\times \ldots \times N_p$ supercell, and represented by the discrete set
\begin{equation}
  \dBvKzone = \Bigg\{\sum_{i=1}^p m_i \myvec{a}_i \Big| 0 \leq m_i < N_i \Bigg\} \subset \RealLattice
\end{equation}
and the continuous
\begin{equation}
  \cBvKzone = \Bigg\{\myvec{r}+\myvec{G} \Big| \myvec{r}\in\unitcell, \ \myvec{G}\in\dBvKzone \Bigg\} \subset \mathbb{R}^d.
\end{equation}
We also define the Born--von K\'arm\'an lattice
\begin{equation}
    \BvKLattice = \Big\{ \sum_{i=1}^p m_i N_i \myvec{a}_i \Big| m_i \in \mathbb{Z} \Big\} \subset \RealLattice.
\end{equation}
and its dual in reciprocal space 
\begin{equation}
    \BvKRecLattice = \Big\{ \sum_{i=1}^p m_i \frac{\myvec{b}_i}{N_i} \Big| m_i \in \mathbb{Z} \Big\} \supset \ReciprocalLattice.
\end{equation}
The discretised first Brillouin zone is given by
\begin{equation}
    \dBZ = \cBZ \cap \BvKRecLattice.
\end{equation}
In addition, it is possible that two points $\myvec{k},\myvec{k}'$ lie on opposite sides of the boundary of $\cBZ$ so that they differ by a reciprocal lattice vector $\myvec{k}-\myvec{k}' = \myvec{b}_i$. In such cases the redundacy should be pruned away by excluding one of $\myvec{k},\myvec{k}'$ from $\dBZ$; alternatively, the elments of $\dBZ$ could be understood as equivalence classes of vectors that are equal up to multiples of the reciprocal lattice vectors. For later use, we introduce
\begin{align}
  \Nbvk & = N_1 N_2 \cdots N_p = \frac{|\cBvKzone|}{|\unitcell|} = |\dBZ|, \\
   \invNbvk & = \frac{1}{\Nbvk} .
\end{align}

As the function space for single-electron wave functions we choose the Sobolev space
\begin{equation}
    \WaveFunSpace{1} = \mathcal H^1(\cBvKzone) \otimes \mathbb{C}^2,
\end{equation}
where the contribution from $\mathbb{C}^2$ represents the spin components, and the inner product is given by
\begin{equation}
    \label{eqSpinOrbInnerProd}
    \scp{\phi}{\psi} = \int_{\cBvKzone} \phi(\myvec{r})^{\dagger} \psi(\myvec{r}) d\myvec{r}.
\end{equation}

Turning to many-electron states, a state corresponding to $n$ electrons per unit cell is formally represented by a state with $M = n N_1\cdots N_p$ electrons defined on the Born--von K\'arm\'an zone. Hence, the relevant function space is
\begin{equation}
  \WaveFunSpace{M} = \bigwedge_{i=1}^M \WaveFunSpace{1}.
\end{equation}
For any $\Psi \in \WaveFunSpace{M}$, the density is given by
\begin{equation}
  \rho_{\Psi}(\myvec{r}_1) = M \int_{{\cBvKzone}^{M-1}} |\Psi(\myvec{x}_1,\ldots,\myvec{x}_M)|^2 d\myvec{r}_2\cdots d\myvec{r}_M
\end{equation}
where $\myvec{x}_i = (\myvec{r}_i,\sigma_i)$ is joint spatial-spin coordinate and summation over all spin degrees of freedom is implied. The factor $M$ ensures that the density corresponds to $M$ electrons per Born--von K\'arm\'an zone and, equivalently, $n$ electrons per unit cell. Note that wave functions are only required to be periodic in the Born--von K\'arm\'an zone, $\cBvKzone$, and not necessarily in the unit cell, $\unitcell$. In order for $\WaveFunSpace{M}$ to be a vector space, closed under superposition, the space contains wave functions that are not merely nonperiodic due to a Bloch phase factor, but also giving rise to a nonperiodic densities:
\begin{equation}
 \rho_{\Psi}(\myvec{r}+N_i \myvec{a}_i) = \rho_{\Psi}(\myvec{r}) \neq \rho_{\Psi}(\myvec{r}+\myvec{a}_i), \qquad 1\leq i \leq p.
\end{equation}
In our density-functional context, it is natural to define also the translation symmetrised density
\begin{equation}
  \bar{\rho}_{\Psi}(\myvec{r}) = \frac{1}{\Nbvk} \sum_{\myvec{g}\in\dBvKzone} \rho_{\Psi}(\myvec{r}+\myvec{g}),
\end{equation}
which always satisfies
\begin{equation}
  \bar{\rho}_{\Psi}(\myvec{r}) = \bar{\rho}_{\Psi}(\myvec{r}+\myvec{a}_i), \qquad 1\leq i \leq p.
\end{equation}
A direct consequence of the above is that
\begin{align}
  \label{eqRhoLOne}
  \rho_{\Psi} & \in L^1(\cBvKzone), \\
  \bar{\rho}_{\Psi} & \in L^1(\unitcell).    \label{eqRhoBarLOne}
\end{align}
Being integrable functions, the densities have Fourier representations $\hat{\rho}_{\Psi}$ and $\hat{\bar{\rho}}_{\Psi}$. The latter is defined by
\begin{equation}
  \bar{\rho}_{\Psi}(\myvec{r}) = \sum_{\myvec{G}' \in \ReciprocalLattice} \int_{\mathbb{R}^{d-p}} d\myvec{G}'' \, e^{i(\myvec{G}'+\myvec{G}'')\cdot\myvec{r}} \hat{\bar{\rho}}_{\Psi}(\myvec{G}'+\myvec{G}'')
\end{equation}
and
\begin{equation}
  \hat{\bar{\rho}}_{\Psi}(\myvec{G}) = \frac{1}{(2\pi)^{d-p} |\unitcell|} \int_{\unitcell} d\myvec{r} \, e^{-i\myvec{G}\cdot\myvec{r}} \bar{\rho}_{\Psi}(\myvec{r}).
  \label{eqDensBvKSobolev}
\end{equation}
Here $|\unitcell|$ is the volume of the projection of the unit cell on $\mathbb{R}^p$.

For reasons that will be apparent below, the densities will be considered to be elements of the Sobolev spaces
\begin{align}
    \rho_{\Psi} & \in \Xbvk = H^{-1}(\cBvKzone), \\
    \bar{\rho}_{\Psi} & \in X = H^{-1}(\unitcell).
\end{align}
We take the norms of these spaces to depend on a parameter $\gamma > 0$ that will be further specified below. The spaces may be equipped with the norms
\begin{widetext}
\begin{equation}
    \begin{split}
        \|\rho\|_{\Xbvk} = \sqrt{ (2\pi)^{d-p} |\unitcell| \sum_{\myvec{q}'\in \BvKRecLattice} \int_{\mathbb{R}^{d-p}} d\myvec{q}'' \frac{|\hat{\rho}(\myvec{q}'+\myvec{q}'')|^2}{\gamma^2 + |\myvec{q}'+\myvec{q}''|^2} },
    \end{split}
\end{equation}
\begin{equation}
   \label{eqXNorm}
  \|\bar{\rho}\|_X = \sqrt{  (2\pi)^{d-p} |\unitcell|  \sum_{\myvec{G}'\in \ReciprocalLattice} \int_{\mathbb{R}^{d-p}} d\myvec{G}'' \frac{|\hat{\bar{\rho}}(\myvec{G}'+\myvec{G}'')|^2}{\gamma^2 + |\myvec{G}'+\myvec{G}''|^2} }.
\end{equation}
\end{widetext}
These norms are chosen so that $\|\bar{\rho}\|_X = \|\bar{\rho}\|_{\Xbvk}$, where we do not distinguish between a function restricted to a single unit cell (as would be sufficient to evaluate the left-hand side) and its periodic extension to all space or the Born--von K\'arm\'an zone (as would be needed to evaluate the right-hand side).

\subsection{The Hamiltonian}
\label{subsec:Hamiltonian}

On $\WaveFunSpace{M}$, we define the Hamiltonian
\begin{equation}
   \begin{split}
     H^{\mathrm{BvK}}_{\lambda}(v) & = -\frac{1}{2} \sum_{i=1}^M \nabla_i^2 + \sum_{i=1}^M v(\myvec{r}_i) + \lambda \sum_{i<j} w_{\mathrm{BvK}}(\myvec{r}_i-\myvec{r}_j)
  \end{split}
\end{equation}
where the potential $v$ has the periodicity of the crystal lattice and belongs to a Sobolev space $X^* = H^1(\unitcell)$ that is the dual of the density space $X = H^{-1}(\unitcell)$. The scale factor $\lambda\in \mathbb{R}$ in front of the electron-electron interaction is there to ensure a unified treatment of both fully interacting ($\lambda=1$) and the noninteracting Kohn--Sham ($\lambda=0$) systems. The above Hamiltonian gives the energy per Born--von K\'arm\'an zone. However, we are primarily interested in the energy per unit cell in the limit $N_1,N_2,\ldots, N_p \to \infty$ of an infinite Born--von K\'arm\'an zone. Therefore, we introduce
\begin{equation}
   \begin{split}
     H_{\lambda}(v) & = \frac{1}{N_1N_2\cdots N_p} H^{\mathrm{BvK}}_{\lambda}(v) \\
     & = -\frac{1}{2} \sum_{i=1}^n \nabla_i^2 + \sum_{i=1}^n v(\myvec{r}_i) + \lambda \sum_{i=1}^n \sum_{j=i+1}^M w_{\mathrm{BvK}}(\myvec{r}_i-\myvec{r}_j)
     \\
     & = T + V + \lambda W,
  \end{split}
\end{equation}
where the restricted summation ranges are possible due to the permutation symmetry of $\WaveFunSpace{M}$ and the last line is the standard decomposition into kinetic energy, external potential, and electron-electron interaction. The electron-electron interaction $w_{\mathrm{BvK}}$ has the periodicity of the Born--von K\'arm\'an lattice and thus has a Fourier representation. Moreover, it can be usefully thought of as the sum of the pairwise interactions between particle $i$ and all the periodic images of particle $j$. Hence,
\begin{equation}
  \label{eqwBvKIntro}
  \begin{split}
    w_{\mathrm{BvK}}(\myvec{r}_{12}) & = \sum_{\myvec{q}\in\BvKRecLattice} \hat{w}_{\mathrm{BvK}}(\myvec{q}) \, e^{i\myvec{q}\cdot\myvec{r}_{12}}
    \\
    & =  \sum_{\myvec{M} \in \BvKLattice} w_{\text{pair}}(\myvec{r}_{12}+\myvec{M})
  \end{split}
\end{equation}
where $w_{\text{pair}}$ is non-periodic.  The Fourier series coefficients of $w_{\mathrm{BvK}}$,
\begin{equation}
 \begin{split}
  \hat{w}_{\mathrm{BvK}}(\myvec{q}) & = \frac{1}{(2\pi)^{d-p} |\cBvKzone|} \int_{\cBvKzone} e^{-i\myvec{q}\cdot\myvec{r}_{12}} w_{\mathrm{BvK}}(\myvec{r}_{12}) d\myvec{r}_{12}
     \\
     & = \frac{1}{(2\pi)^{d-p} |\cBvKzone|} \int_{\mathbb{R}^d} e^{-i\myvec{q}\cdot\myvec{r}_{12}} w_{\text{pair}}(\myvec{r}_{12}) d\myvec{r}_{12},
 \end{split}
 \label{eqWpairToFourier}
\end{equation}
are seen to be proportional to the Fourier transform of $w_{\text{pair}}$.

\begin{figure}
  \begin{center}
    \includegraphics[width=0.97\columnwidth]{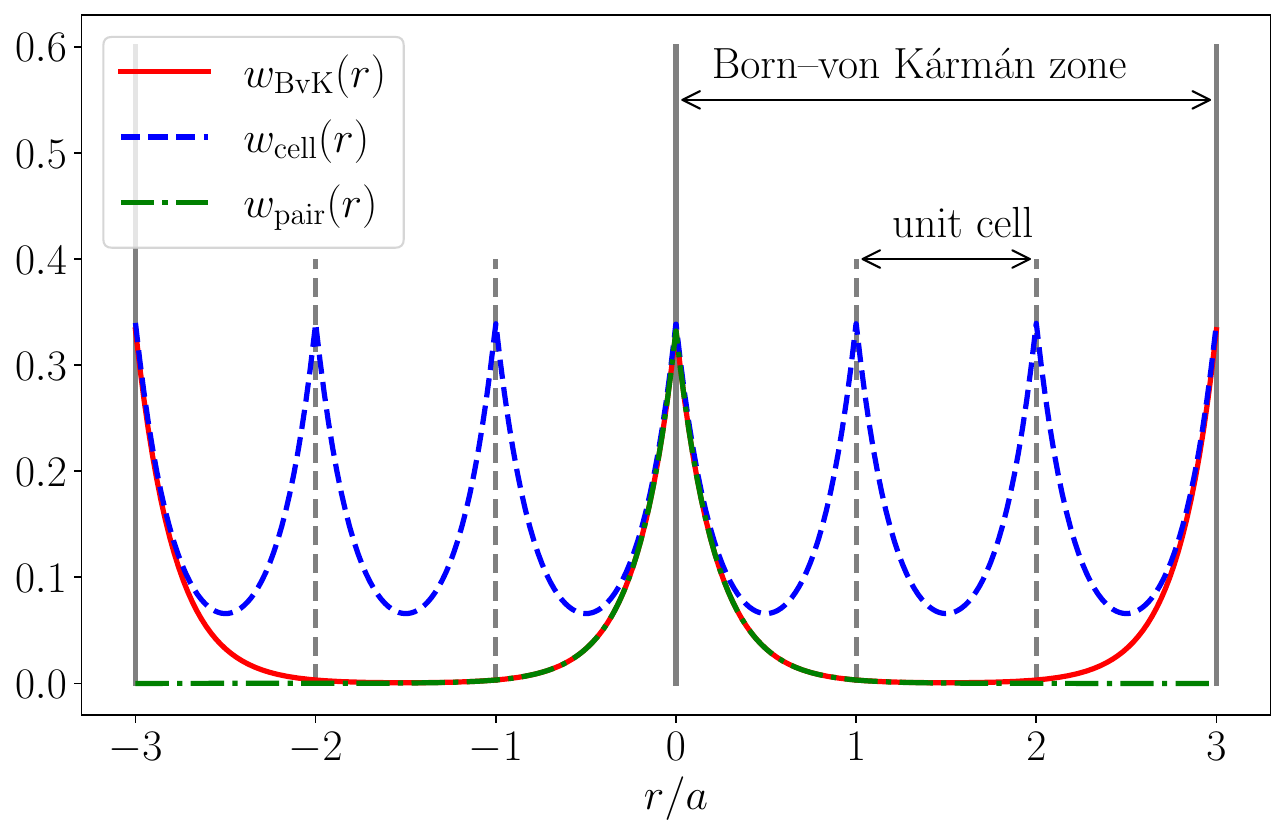}
  \end{center}
  \caption{\label{figYukawa} The Yukawa interaction illustrated for a one-dimensional ($p=d=1$) system with arbitrary lattice constant $a$ and Yukawa parameter $\gamma = 1.5 a$.}
\end{figure}

In the present work we focus on the Yukawa potential~\cite{ROWLINSON_PA156_15}, defined by its Fourier representation
\begin{equation}
  \label{eqYukawaDef}
  \hat{w}_{\mathrm{BvK}}(\myvec{q}) = \frac{S_d}{(2\pi)^{d-p} |\cBvKzone|} \cdot \frac{1}{\gamma^2 + q^2}, \quad \myvec{q} \in \BvKRecLattice,
\end{equation}
where $S_d$ is the surface area of a unit sphere in $d$ dimensions. It is also useful to introduce the restriction to $\ReciprocalLattice$ with a different pre-factor,
\begin{equation}
  \label{eqYukawaCellDef}
  \hat{w}_{\mathrm{cell}}(\myvec{G}) = \frac{S_d}{(2\pi)^{d-p} |\unitcell|} \cdot \frac{1}{\gamma^2 + G^2}, \quad \myvec{G} \in \ReciprocalLattice.
\end{equation}
In real space, the pair potential and the periodic potentials are related via Eq.~\eqref{eqwBvKIntro} and
\begin{align}
  w_{\mathrm{cell}}(\myvec{r}) & = \sum_{\myvec{m}\in\RealLattice} w_{\mathrm{pair}}(\myvec{r}+\myvec{m}) = \sum_{\myvec{m} \in \dBvKzone} w_{\mathrm{BvK}}(\myvec{r}+\myvec{m}).
\end{align}
These relations are illustrated in Fig.~\ref{figYukawa}.

In dimensions $d \leq 3$, the Yukawa pair potential is given by
\begin{align}
  \label{eqWPAIRone}
  w_{\text{pair}}(\vec{r}_{12}) & = \frac{e^{-\gamma |\myvec{r}_{12}|}}{2\gamma} \qquad (d=1), \\
  w_{\text{pair}}(\vec{r}_{12}) & = \sqrt{\frac{2}{\pi}} K_0(\gamma |\myvec{r}_{12}|), \qquad (d=2), \\
  w_{\text{pair}}(\vec{r}_{12}) & =  \frac{e^{-\gamma |\myvec{r}_{12}|}}{|\myvec{r}_{12}|} \qquad (d= 3),
  \label{eqWPAIR}
\end{align}
where $K_0$ is the modified Bessel function of order 0.

\begin{figure}[h]
    \includegraphics[width=0.45\textwidth]{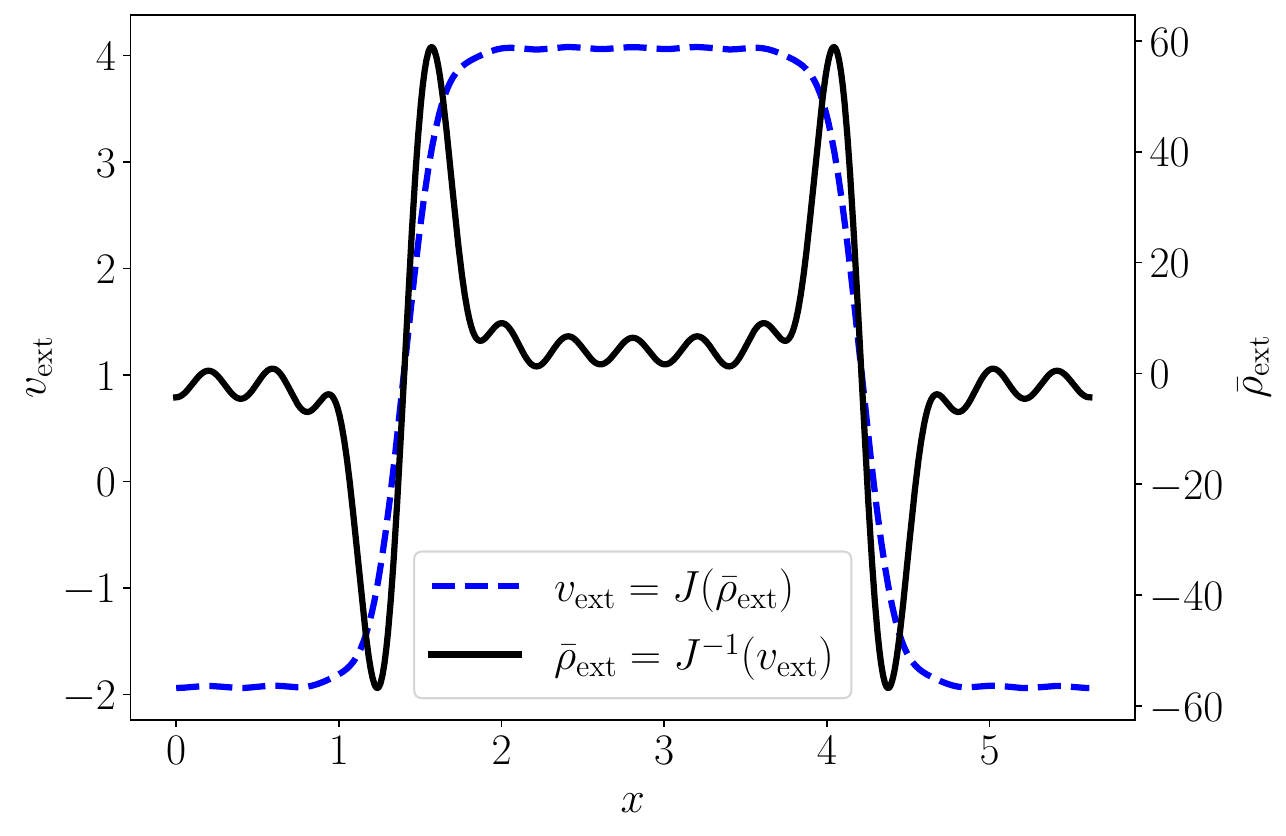}
    \caption{\label{fig:vext_and_rho_ext_twin_N6} One-dimensional illustration of the effect of the duality map on potentials and densities. The horizontal axis is the position within a unit cell.}
\end{figure}

For densities $\rho \in \Xbvk$, $\bar{\rho} \in X$, and potentials $V \in \Xbvk^* = H^1(\cBvKzone)$, $v \in X^* = H^1(\unitcell)$ in the respective dual spaces, we define the \emph{dual pairings}
\begin{widetext}
\begin{equation}
    \pairing{V}{\rho}_{\mathrm{BvK}} = \invNbvk \int_{\cBvKzone} V(\myvec{r}) \rho(\myvec{r}) d\myvec{r} = (2\pi)^{d-p} |\unitcell| \sum_{\myvec{q}'\in\BvKRecLattice} \int_{\mathbb{R}^{d-p}} d\myvec{q}'' \, \hat{V}(\myvec{q}'+\myvec{q}'')^* \, \hat{\rho}(\myvec{q}'+\myvec{q}''),
\end{equation}
\begin{equation}
    \label{eqDualPairing}
    \pairing{v}{\bar{\rho}} = \int_{\unitcell} v(\myvec{r}) \bar{\rho}(\myvec{r}) d\myvec{r} = (2\pi)^{d-p} |\unitcell| \sum_{\myvec{G}'\in\ReciprocalLattice} \int_{\mathbb{R}^{d-p}} d\myvec{G}'' \, \hat{v}(\myvec{G}'+\myvec{G}'')^* \, \hat{\bar{\rho}}(\myvec{G}'+\myvec{G}''),
\end{equation}
and the norms
\begin{equation}
    \|V\|_{\Xbvk} = \sup_{\rho \in \Xbvk} \frac{|\pairing{V}{\rho}|}{\|\rho\|_X} = \sqrt{(2\pi)^{d-p} |\unitcell| \sum_{\myvec{q}'\in\BvKRecLattice} \int_{\mathbb{R}^{d-p}} d\myvec{q}'' \, (\gamma^2 + |\myvec{q}'+\myvec{q}''|^2) |\hat{V}(\myvec{q}'+\myvec{q}'')|^2},
\end{equation}
\begin{equation}
   \label{eqXDualNorm}
    \|v\|_{X^*} = \sup_{\bar{\rho} \in X} \frac{|\pairing{v}{\bar{\rho}}|}{\|\bar{\rho}\|_X} = \sqrt{(2\pi)^{d-p} |\unitcell| \sum_{\myvec{G}'\in\ReciprocalLattice} \int_{\mathbb{R}^{d-p}} d\myvec{G}'' \, (\gamma^2 + |\myvec{G}'+\myvec{G}''|^2) |\hat{v}(\myvec{G}'+\myvec{G}'')|^2}.
\end{equation}
\end{widetext}
The above choices also give rise to the duality maps $\Jbvk : \Xbvk \to \Xbvk^*$ and $\Jdual : X \to X^*$,
\begin{align}
  \label{eqDualMapDef}
  \Jbvk(\rho) & = \{ V \in \Xbvk^* \, | \, \pairing{V}{\rho} = \| V \|_{\Xbvk^*}^2= \|\rho\|_{\Xbvk}^2 \}, \\
  \Jdual(\bar{\rho}) & = \{ v \in X^* \, | \, \pairing{v}{\bar{\rho}} = \| v \|_{X^*}^2= \|\bar{\rho}\|_X^2 \}.
\end{align}
Although the duality maps are set-valued, their values are always a singleton sets with our specific choice of function spaces. We note that, the unique elements $U \in \Jbvk(\bar{\rho})$ and $u \in \Jdual(\bar{\rho})$ are given by the explicit formulas
\begin{align}
    U(\myvec{r}) & = \invNbvk \int_{\cBvKzone} d\myvec{r}'  w_{\mathrm{BvK}}(\myvec{r}-\myvec{r}') \bar{\rho}(\myvec{r}'), \\
    \hat{U}(\myvec{q}) & = \frac{\hat{\bar{\rho}}(\myvec{q})}{\gamma^2 + q^2}, \quad \myvec{q} \in \BvKRecLattice, \\
    u(\myvec{r}) & = \int_{\unitcell} d\myvec{r}'  w_{\mathrm{cell}}(\myvec{r}-\myvec{r}') \bar{\rho}(\myvec{r}'), \\
    \hat{u}(\myvec{G}) & = \frac{\hat{\bar{\rho}}(\myvec{G})}{\gamma^2 + G^2}, \quad \myvec{G} \in \ReciprocalLattice.
\end{align}
Hence, the duality maps give the (Yukawa-)Hartree potentials due to the densities. The effect of the duality map is illustrated for $p=d=1$ in Fig~\ref{fig:vext_and_rho_ext_twin_N6}. The duality map enables us to interpret the classical interaction energy between two charge distributions $\rho,\sigma$ in terms of the dual pairing,
\begin{equation}
  \begin{split}
    \CoulQform(\rho,\sigma) & = \frac{\invNbvk^2}{2} \int_{\cBvKzone^2} \rho(\myvec{r}_1) w_{\mathrm{BvK}}(\myvec{r}_{12}) \sigma(\myvec{r}_2) d\myvec{r}_1 d\myvec{r}_2 \\
    & =\frac{\invNbvk^2}{2} \pairing{\Jbvk(\rho)}{\sigma}_{\mathrm{BvK}},
  \end{split}
\end{equation}
where the abuse of notation on the last line does not lead to any misunderstandings.

Returning to the electron-electron repulsion operator, a finite Yukawa interaction energy is ensured by the bound
\begin{equation}
  \begin{split}
    \bra{\Psi} W \ket{\Psi} & = \frac{N(N-1)}{2} \sum_{\myvec{K}_1,\ldots,\myvec{K}_N\in\BvKRecLattice} \frac{|\hat{\Psi}(\myvec{K}_1,\ldots,\myvec{K}_N)|^2}{\gamma^2+|\myvec{K}_1-\myvec{K}_2|^2}
    \\
    & \leq \frac{N(N-1)}{2} \cdot \frac{\|\Psi\|_2^2}{\gamma^2}.
  \end{split}
\end{equation}
Additionally, at least for Slater determinants $\Phi$, we have the tighter bound
\begin{equation}
  \begin{split}
    \bra{\Phi} W \ket{\Phi} \leq  \CoulQform(\rho_{\Phi}, \rho_{\Phi}) & = \frac{1}{2} \pairing{\Jbvk(\rho_{\Phi})}{\rho_{\Phi}}_{\mathrm{BvK}} \\
    & = \frac{1}{2} \|\rho_{\Phi}\|_{\Xbvk}^2,
  \end{split}
\end{equation}
where the right-hand side is the Hartree energy, i.e. the classical electrostatic self-interaction.
Notably, convexity of norms yields
\begin{equation}
  \|\bar{\rho}_{\Phi}\|_{X} =  \|\bar{\rho}_{\Phi}\|_{\Xbvk} \leq  \|\rho_{\Phi}\|_{\Xbvk}
\end{equation}
and thus we find that the translation symmetrisation always lowers the Hartree energy, $\CoulQform(\bar{\rho}_{\Phi},\bar{\rho}_{\Phi}) \leq \CoulQform(\rho_{\Phi},\rho_{\Phi})$.

\subsubsection{Bloch's theorem}

Spontaneous translation-symmetry breaking, where the ground-state density $\rho$ does not have the periodicity of the lattice $\RealLattice$ would present a challenge for the present formalism with its focus on the translation-symmetrised $\bar{\rho}$. However, the possibility of such symmetry breaking can be ruled out, at least at the level of the exact theory.

To see this, let $T_{\myvec{g}}$, with $\myvec{g} \in \RealLattice$, denote the translation operator defined by rigid translation of all electron coordinates,
\begin{equation}
 \begin{split}
  T_{\myvec{g}} & \Psi(\myvec{r}_1,\sigma_1, \myvec{r}_2,\sigma_2,\ldots, \myvec{r}_M,\sigma_M) 
  \\
  & = \Psi(\myvec{r}_1+\myvec{g},\sigma_1, \myvec{r}_2+\myvec{g},\sigma_2,\ldots, \myvec{r}_M+\myvec{g},\sigma_M).
  \end{split}
\end{equation}
All such operators form a commutative group, and since we allow only periodic potentials $v$, every $T_{\myvec{g}}$ commutes with the Hamiltonian $H_{\lambda}(v)$. Consequently, one may always choose the exact eigenstates of $H_{\lambda}(v)$ to be translation symmetric,
\begin{equation}
  T_{\myvec{g}} \Psi = e^{i\myvec{k}\cdot\myvec{g}} \Psi,
\end{equation}
where it follows that $\myvec{k} \in \BvKRecLattice$ from the Born--von K\'arm\'an boundary conditions, since $T_{N_i\myvec{a}_i}$ must be the identity operator. Moreover, every non-translation symmetric eigenstate of $H_{\lambda}(v)$ remains an eigenstate, sharing the same eigenvalue, after it is translation symmetrised.

\subsection{Formal Kohn--Sham density-functional theory}
\label{subsec:DFT}

Let $\MixedStateSpace{M}$ denote the mixed states, or density operators, that can be formed from the wave functions in $\WaveFunSpace{M}$. Formally, $\MixedStateSpace{M}$ is the set of trace-class operators on $\WaveFunSpace{M}$ which are normalised so that $\trace{\Gamma} = 1$ for all $\Gamma \in \MixedStateSpace{M}$. For any fixed, non-empty subset $A \subseteq \MixedStateSpace{M}$ of the density operators, we may use the variational principle,
\begin{equation}
  \label{eqRayRitzVarPrinc}
  E_{\lambda;A}(v) = \inf_{\Gamma \in A} \trace{ \Gamma H_{\lambda}(v) }
\end{equation}
to define a functional that is an upper bound on the true ground-state energy $E_{\lambda;\MixedStateSpace{M}}(v)$. Though the formalism here is very general, we highlight a few special cases: (a) Hartree--Fock theory, $\lambda=1$, $A$ is the set of projectors $\ket{\Phi}\bra{\Phi}$ formed from Slater determinants $\Phi$, (b) Kohn--Sham theory, $\lambda=0$, $A$ is the set of of projectors formed from Slater determinants, (c) exact density functional theory, $\lambda=1$, $A$ is either the entire $\MixedStateSpace{M}$ or the subset of pure states. Introducing the constrained-search functional,
\begin{equation}
  F^{\text{cs}}_{\lambda;A}(\bar{\rho})= \inf_{\substack{\Gamma \in A \\ \bar{\rho}_{\Gamma} = \bar{\rho}}} \trace{\Gamma H_{\lambda}(0)}
\end{equation}
we have the Hohenberg--Kohn variational principle
\begin{align}
  \label{eqHKVarPrinc}
  E_{\lambda;A}(v) & = \inf_{\bar{\rho} \in X} \left( \pairing{v}{\bar{\rho}} + F^{\text{cs}}_{\lambda;A}(\bar{\rho}) \right).
\end{align}
We remark, firstly, that the minimisation domain $X$ may appear excessively large since many ``densities'' in this space are not $N$-representable, i.e. do not arise from any state in $A$ or even in $\MixedStateSpace{M}$. However, recalling that the infimum of an empty set is $+\infty$, this extension of the domain to non-representable densities is harmless. Secondly, the functional $E_{\lambda;A}(v)$ is generally concave in $v$, irrespective of the choice of $A$. Thirdly, when $A$ is a convex set, the constrained-search functional $F^{\text{cs}}_{\lambda;A}(\bar{\rho})$ is convex in $\bar{\rho}$.

The density space $X$ and its dual, the potential space $X^*$, defined above are a Banach spaces, which have useful properties related to convergence and topology. The associated norms are given in Eqs.~\eqref{eqXNorm} and \eqref{eqXDualNorm} and the dual pairing in Eq.~\eqref{eqDualPairing}. We recall that convergence (in norm) of a sequence $\bar{\rho}_n \to \bar{\rho}$ in $X$ is then the statement $\|\bar{\rho}_n - \bar{\rho}\|_X \to 0$, and similarly for convergence in $X^*$.  Weak convergence
$\bar{\rho}_n\rightharpoonup \bar{\rho}$ in $X$ means that for all $v \in X^*$ we have 
$\pairing{v}{\bar{\rho}_n} \to \pairing{v}{\bar{\rho}}$. Convergence in norm implies weak convergence.

A functional $G$ on $X$ is \emph{lower semicontinuous} if whenever $\bar{\rho}_n\to \bar{\rho}$ in $X$ (i.e., $\| \bar{\rho}_n - \bar{\rho} \|_X \to 0$)
\begin{equation}
    \liminf_{n\to\infty} G(\bar{\rho}_n) \geq G(\bar{\rho}). 
\end{equation}
For a convex functional, lower semicontinuity is equivalent to weak lower semicontinuity (i.e., the above definition but with $\bar{\rho}_n\rightharpoonup \bar{\rho}$ in $X$). Similarly, upper semicontinuity is defined by reversing the above inequality.

The functional $E_{\lambda;A}(v)$ is generally upper semicontinuous and concave, as this is a general property of functions defined as a pointwise infimum like in Eq.~\eqref{eqRayRitzVarPrinc}. 
That $X$ is a Banach space combined with the concavity and upper semicontinuity of $E_{\lambda;A}$ is sufficient to guarantee the existence of a Lieb functional $F_{\lambda;A}$ such that
\begin{align}
  \label{eqHKVarPrincII} 
  E_{\lambda;A}(v) & = \inf_{\bar{\rho} \in X} \left( \pairing{v}{\bar{\rho}} + F_{\lambda;A}(\bar{\rho}) \right),
  \\
  F_{\lambda;A}(\bar{\rho})& = \sup_{v \in X^*} \left( E_{\lambda;A}(v) - \pairing{v}{\bar{\rho}}  \right).
  \label{eqLiebVarPrinc}
\end{align}
The Lieb functional is by construction convex and lower semicontinuous and satisfies
\begin{equation}
  F_{\lambda;A}(\bar{\rho}) \leq F^{\text{cs}}_{\lambda;A}(\bar{\rho}).
\end{equation}
If $A$ is such that $F^{\text{cs}}_{\lambda;A}(\bar{\rho})$ is convex and lower semicontinuous too, the above relation reduces to an equality.

Up to this point, the formalism corresponds to orbital-free density-functional theory. Let us now consider the interacting system ($\lambda=1$) described by states in $A$ in parallel with a fictitious noninteracting system ($\lambda=0$) described by states in some subset $A' \subseteq \MixedStateSpace{M}$. Typically $A'$ is either the set of projectors $\ket{\Phi}\bra{\Phi}$ made up from Slater determinants $\Phi$ or the convex hull of these projectors. The former choice corresponds to Kohn--Sham theory and the latter choice to ensemble Kohn--Sham theory, where fractional occupation numbers are allowed. Many other choices are possible too. The main consideration is that it is desirable that representability of densities by states in $A$ is equivalent to representability by states in $A'$:
\begin{equation}
  \label{eqConditionAAprime}
  \{ \bar{\rho}_{\Gamma} | \Gamma \in A\} = \{ \bar{\rho}_{\Gamma'} | \Gamma' \in A'\}.
\end{equation}
This condition is necessary and sufficient for $F^{\text{cs}}_{1;A}(\bar{\rho})$ and $F^{\text{cs}}_{0;A'}(\bar{\rho})$ to be finite for the same densities.

The density functional for the $\lambda=0$ system is the noninteracting kinetic energy, typically denoted
\begin{equation}
  T_{\mathrm{s}}(\bar{\rho}) = F_{0;A'}(\bar{\rho})
\end{equation}
Introducing the Hartree-exchange-correlation functional,
\begin{equation}
  F^{\text{cs}}_{\text{Hxc}}(\bar{\rho}) = F^{\text{cs}}_{1;A}(\bar{\rho}) - F^{\text{cs}}_{0;A'}(\bar{\rho}) = F^{\text{cs}}_{1;A}(\bar{\rho}) - T_{\mathrm{s}}(\bar{\rho}),
\end{equation}
where we suppress the dependence on $A$ and $A'$ from the notation, we trivially obtain the Kohn--Sham decomposition
\begin{equation}
  F^{\text{cs}}_{1;A}(\bar{\rho})  = T_{\mathrm{s}}(\bar{\rho}) + F_{\text{Hxc}}(\bar{\rho}).
\end{equation}
which can be further decomposed by splitting out the Hartree term, $F_{\text{Hxc}}(\bar{\rho}) = \CoulQform(\bar{\rho},\bar{\rho})  + F_{\text{xc}}(\bar{\rho})$.

The forward problem in orbital-free density-functional theory is to determine the minimiser $\bar{\rho}_{\mathrm{gs}}$, if it exists, in the Hohenberg--Kohn variation principle (Eqs.~\eqref{eqHKVarPrinc} or \eqref{eqHKVarPrincII}) for the interacting system,
\begin{equation}
  \begin{split}
  E_{1;A}(v_{\mathrm{ext}}) & = F^{\text{cs}}_{1;A}(\bar{\rho}_{\mathrm{gs}}) + \pairing{v_{\mathrm{ext}}}{\bar{\rho}_{\mathrm{gs}}}
  \\
  & = F_{1;A}(\bar{\rho}_{\mathrm{gs}}) + \pairing{v_{\mathrm{ext}}}{\bar{\rho}_{\mathrm{gs}}}
  \end{split}
\end{equation}
Assuming Eq.~\eqref{eqConditionAAprime} holds, the variation principle can also be rewritten as
\begin{equation}
  \begin{split}
    E_{1;A}(v_{\mathrm{ext}}) & = \inf_{\bar{\rho} \in X} \left( T_{\mathrm{s}}(\bar{\rho}) + F_{\text{Hxc}}(\bar{\rho}) + \pairing{v_{\mathrm{ext}}}{\bar{\rho}} \right)
    \\
    & = \inf_{\bar{\rho} \in X} \left( \inf_{\substack{\Gamma \in A' \\ \bar{\rho}_{\Gamma} = \bar{\rho}}} \trace{\Gamma T} + F_{\text{Hxc}}(\bar{\rho}) + \pairing{v_{\mathrm{ext}}}{\bar{\rho}} \right)
    \\
    & = \inf_{\Gamma \in A'} \left( \trace{\Gamma T} + F_{\text{Hxc}}(\bar{\rho}_{\Gamma}) + \pairing{v_{\mathrm{ext}}}{\bar{\rho}_{\Gamma}} \right).
  \end{split}
\end{equation}
The minimiser $\Gamma_{\mathrm{s}}$, if it exists, of the last expression is the Kohn--Sham state. Computing the Kohn--Sham state $\Gamma_{\mathrm{s}}$, or at least its density $\bar{\rho}_{\Gamma_{\mathrm{s}}} = \bar{\rho}_{\mathrm{gs}}$, for a given $v$ is the forward problem in Kohn--Sham theory. The \emph{inverse problem} is to, given a density $\bar{\rho}_{\mathrm{gs}}$, determine for which potential $v_{\mathrm{s}}$, if any, this density is noninteracting 
ground-state density. That is, for which $v_{\mathrm{s}}$ the given density is the minimiser in the Hohenberg--Kohn variation principle for the noninteracting system,
\begin{equation}
  \begin{split}
  E_{0;A'}(v_{\mathrm{s}}) & = T_{\mathrm{s}}(\bar{\rho}_{\mathrm{gs}}) + \pairing{v_{\mathrm{s}}}{\bar{\rho}_{\mathrm{gs}}}
  \\
  & = F_{0;A'}(\bar{\rho}_{\mathrm{gs}}) + \pairing{v_{\mathrm{s}}}{\bar{\rho}_{\mathrm{gs}}}.
 \end{split}
\end{equation}
The inverse problem does not arise in typical density-functional calculations, but it is important in the development of density-functional approximations and as a computationally feasible way of studying the Hohenberg--Kohn mapping numerically.
From a formal viewpoint, the desired potential $v_{\mathrm{s}}$ may be found as the maximiser in Lieb's variation principle for the noninteracting system,
\begin{equation}
  \begin{split}
    F_{0;A'}(\bar{\rho}_{\mathrm{gs}}) & = \sup_{v'\in X} \left( E_{0;A'}(v') - \pairing{v'}{\bar{\rho}_{\mathrm{gs}}}\right)
    \\
    & = E_{0;A'}(v_{\mathrm{s}}) - \pairing{v_{\mathrm{s}}}{\bar{\rho}_{\mathrm{gs}}}.
  \end{split}
\end{equation}
From the formulations as convex optimisation problems, a characterisation in terms of the sub- and supergradients is possible:
\begin{align}
  \label{eqSubgradHK}
   -v_{\mathrm{s}} & \in \underline{\partial} F_{0;A'}(\rho_{\mathrm{gs}}), \\
  \rho_{\mathrm{gs}} & \in \overline{\partial} E_{0;A'}(v_{\mathrm{s}}).
\end{align}
However, the inverse problem is ill-posed~\cite{GarrigueCMP386_1803} due to lack of coercivity and is very challenging in numerical implementations.
A pragmatic alternative is to include a ZMP-like penalty term in a modified variation principle for the Kohn--Sham system,
\begin{equation}
  \label{eqEnergyZMP}
  \widetilde{E}_{0;A'} = \inf_{\Gamma \in A'} \left( G(\Gamma) + \frac{\|\bar{\rho}_{\Gamma} - \bar{\rho}_{\mathrm{gs}}\|_X^2}{2\varepsilon} \right),
\end{equation}
where $G(\Gamma)$ is some simple energy term that would typically include at least the kinetic energy $\trace{ \Gamma H_{0}}$.
In addition, it might also contain the potential energy $\trace{v_\mathrm{ext} \Gamma}$ from the potential for which 
$\bar{\rho}_{\mathrm{gs}}$ is the interacting ground-state density as well as additional terms, e.g., the Hartree energy, the simple Fermi--Almadi correction of the self-energy, or similar.
For small enough penalty parameter $\varepsilon > 0$ (or large $1/\varepsilon$ in the language of ZMP iKS), and irrespective of chosen $h(\Gamma)$, one may intuitively expect this to force the Kohn--Sham state to approximately reproduce the given density, $\bar{\rho}_{\Gamma_{\mathrm{s}}} \approx \bar{\rho}_{\mathrm{gs}}$. As described further below, this approach enables the Kohn--Sham potential $v_{\mathrm{s}}$ to be determined in the limit $\varepsilon\to 0^+$ that can be made mathematically rigorous~\cite{Penz_2023}. We will turn to this state of affair in the next section.

\subsection{Moreau--Yosida regularisation in DFT}\label{subsec:MY}

\subsubsection{Moreau--Yosida density-potential inversion}

Moreau--Yosida regularisation of the universal density functional has been shown to restore functional differentiability, give quantitative results for Hohenberg--Kohn mappings~\cite{Penz2026}, provide a rigorous foundation for ZMP-type penalty term methods~\cite{Kvaal2014,KSpaper2018,Penz_2023,Herbst2025} as well as to enable rigorous convergence guarantees for self-consistent field iterations~\cite{KSpaper2018,KS_PRL_2019,PRLerrata}. It can be readily employed for any density space that is a reflexive and uniformly convex Banach space~\cite{KSpaper2018,MY-CDFTpaper2019}. In our present setting, the MY regularised universal functional is given by
\begin{equation}
  \label{eqMYdef}
  \begin{split}
   F^{\varepsilon}_{\lambda;A} (\bar{\rho}_0) & = \min_{\bar{\rho} \in X}\left( F_{\lambda;A}(\bar{\rho})+ \frac{1}{2\varepsilon} \|\bar{\rho}-\bar{\rho}_0\|_{X}^2 \right) 
   \\
   & = \min_{\bar{\rho} \in X}G^{\varepsilon}_{\lambda;A}(\bar{\rho},\bar{\rho}_0),
 \end{split}
\end{equation}
with the obvious definition of $G^{\varepsilon}_{\lambda;A}(\bar{\rho},\bar{\rho}_0)$ as the expression in parenthesis in the first equality. The minimiser
\begin{equation}
  \bar{\rho}^{\varepsilon} = \underset{\varepsilon F_{\lambda;A}}{\mathrm{prox}} (\bar{\rho}_0)
\end{equation}
is called the \emph{proximal point} or \emph{proximal density} to the input density $\bar{\rho}_0$.  Unlike the unregularised $F_{\lambda;A}$, the regularised functional $F^{\varepsilon}_{\lambda;A}$ is finite everywhere on $X$, including for non-$N$-representable densities. We recall that the unregularised functional does not have a conventional (Gateaux or Frech\'et) functional derivative, but its convexity admits a subdifferential
\begin{equation}
    \begin{split}
        \underline{\partial} F_{\lambda;A}(\bar{\rho}) = \{v \in X^* | & F_{\lambda;A}(\bar{\rho}') \geq F_{\lambda;A}(\bar{\rho}) \\
        & + \pairing{v}{\bar{\rho}'-\bar{\rho}}, \forall \bar{\rho}' \in X \}.
    \end{split}
\end{equation}
Similarly, $G^{\varepsilon}_{\lambda;A}(\bar{\rho},\bar{\rho}_0)$ regarded as a functional of $\bar{\rho}$ for a fixed $\bar{\rho}_0$ is convex and admits a subgradient. At the global minimum of any convex functional, its subgradient contains $0 \in X$. Hence,
\begin{equation}\label{eq:statMY}
  0 \in \underline{\partial} G^{\varepsilon}_{\lambda;A}(\bar{\rho}^{\varepsilon},\bar{\rho}_0) = \underline{\partial} F_{\lambda;A}(\bar{\rho}^{\varepsilon})+ \frac{1}{\varepsilon}\Jdual(\bar{\rho}^{\varepsilon}-\bar{\rho}_0),
\end{equation}
where $+$ here denotes the Minkowski sum. Hence, for $-v^{\varepsilon} \in \underline{\partial} F_{\lambda;A}(\bar{\rho}^{\varepsilon})$, we obtain
\begin{equation}
  v^{\varepsilon} = \frac{1}{\varepsilon} \Jdual(\bar{\rho}^{\varepsilon}-\bar{\rho}_0), \quad 
  \bar{\rho}^{\varepsilon}  = \underset{\varepsilon F_{\lambda;A}}{\mathrm{prox}} (\bar{\rho}_0).
\end{equation}
This is an approximation to the potential in the Hamiltonian $H_{\lambda}(v)$ that has $\bar{\rho}_0$ as its ground-state density. An important property of MY regularisation is that the proximal density converges to the input density $\bar{\rho}_0$, whenever the latter is $N$-representable~\cite{Penz_2023},
\begin{equation}
   \lim_{\varepsilon \to 0^+} \|\bar{\rho}^{\varepsilon} - \bar{\rho}_0\|_X = 0.
\end{equation}
The potential $-v^{\varepsilon}$ converges to the element of $\underline{\partial} F_{\lambda;A}(\bar{\rho}_0)$ with minimal norm, whenever this set is non-empty~\cite{Penz_2023}. When the input density is not $N$-representable, the proximal density converges to the, in some sense, closest $N$-representable density.

\subsubsection{Moreau--Yosida inverse Kohn--Sham}

We will in the remaining part of this section discuss the choice $\lambda=0$ that will be important for our numerical density-potential inversion scheme. 
Let $v_\mathrm{ext}$ be given and assume thay $\bar{\rho}_\mathrm{gs}$ is a ground-state density of the Hamiltonian $H_1(v_\mathrm{ext})$.  
By definition, a Kohn--Sham potential $v_{\mathrm{s}}$ fulfills
\begin{equation}
    \bar{\rho}_\mathrm{gs} = \argmin_{\bar{\rho} \in X}\left( T_{\mathrm{s}}(\bar{\rho})+ \langle v_{\mathrm{s}},\bar{\rho} \rangle   \right)  .
\end{equation}
The stationarity condition then gives 
\begin{equation}
    \label{eqSubgradHKb}
    \partial T_{\mathrm{s}}(\bar{\rho}_\mathrm{gs}) + v_{\mathrm{s}} \ni 0.
\end{equation}
By the fact that $\bar{\rho}^\varepsilon =  \underset{\varepsilon F_{0;A}}{\mathrm{prox}}(\bar{\rho}_\mathrm{gs})  \to \bar{\rho}_\mathrm{gs}$ we can link the effective KS potential to the Moreau--Yosida regularisation of $T_{\mathrm{s}} =F_{0;A}$ (compare the above equation with Eq.~\eqref{eq:statMY} with $\lambda=0$ and $\bar\rho_0=\bar\rho_\mathrm{gs}$)
and obtain~\cite{Penz_2023}
\begin{equation}
  \label{eqVsAsLimit}
v_{\mathrm{s}} = \lim_{\varepsilon\to 0^+} \frac{1}{\varepsilon} \Jdual(\bar{\rho}^{\varepsilon}-\bar{\rho}_\mathrm{gs}).
\end{equation}
Furthermore, returning to the ZMP-like energy in Eq.~\eqref{eqEnergyZMP}, we specialise the model energy to $G(\Gamma) = \trace{H_0(0) \, \Gamma} + \alpha \pairing{v_{\mathrm{ext}}}{\bar{\rho}_{\Gamma}} + \tfrac{\xi}{2} v_{\mathrm{H}}(\bar{\rho}_{\Gamma})$. Switching to the density as a variable yields
\begin{equation}
    \widetilde{E}_{0;A'} = \inf_{\bar{\rho}} \left( F_{\alpha,\xi}^\mathrm{mod}(\bar{\rho}) + \frac{\|\bar{\rho} - \bar{\rho}_{\mathrm{gs}}\|_X^2}{2\epsilon} \right)
\end{equation}
where we introduced the model functional
\begin{equation}
    \label{eq:mod_func_alp_xi}
    F^{\mathrm{mod}}_{\alpha,\xi}(\bar{\rho}) = T_{\mathrm{s}}(\bar{\rho}) + \alpha \pairing{v_{\mathrm{ext}}}{\bar{\rho}} + \xi\,E_{\mathrm{H}}(\bar{\rho},\bar{\rho}),
\end{equation}
with parameters $\alpha$ and $\xi$.
The idea is now to regularise this functional in the pursuit of an effective potential being able to 
reproduce a ground-state density of an interacting system in a field generated by $v_\mathrm{ext}$. 
The stationarity condition of the infimal MY convolution then gives by above
\begin{equation}
  0 \in \underline{\partial} T_\mathrm{s}(\bar{\rho}^{\varepsilon})+ \alpha v_\mathrm{ext} + \xi v_\mathrm{H}(\bar{\rho}^{\varepsilon}) + \frac{1}{\varepsilon} \Jdual(\bar{\rho}^{\varepsilon} - \bar{\rho}_{\mathrm{gs}}).
\end{equation}
Again, comparing with Eq.~\eqref{eqSubgradHKb}, we obtain~\cite{Penz_2023}
\begin{equation} \label{eq:vseps}
\begin{aligned}
v_\mathrm{s} &=  \alpha v_\mathrm{ext} + \xi v_\mathrm{H}(\bar{\rho}^{\varepsilon}) +  \lim_{\varepsilon\to^+} 
    \frac{1}{\varepsilon} \Jdual(\bar{\rho}^{\varepsilon}-\bar{\rho}_\mathrm{gs}), \\
    \bar{\rho}^{\varepsilon} &= \underset{\varepsilon F_{\alpha,\xi}^\mathrm{mod} }{\mathrm{prox}} (\bar{\rho}_{\mathrm{gs}}) .
\end{aligned}
\end{equation}
Although not emphasised in the notation, the proximal density $\bar{\rho}^{\varepsilon}$ in the above equation depends on $\alpha$ and $\xi$. Hence, for large regularisation parameters $\varepsilon$, these parameters could either introduce a helpful bias towards $\bar{\rho}_{\mathrm{gs}}$ or an unhelpful bias away from it. However, the same limit will be recovered as $\varepsilon \to 0^+$ irrespective of parameter values.

\section{Numerical implementation}
\label{secNumImpl}

As a practical implementation of the above theory, we describe the interacting system at the spin-restricted Hartree--Fock level. The Hartree--Fock model is suitable for several reasons. Firstly, it provides a non-trivial and important reference model as the corresponding Kohn--Sham description must model the effect of nonlocal, exact exchange using a local potential. Secondly, any post-Hartree--Fock method that is both variational and, importantly for periodic systems, size extensive requires much more implementation work and the computational cost is much higher. Thirdly, the ZMP penalty term gives rise to a Hartree term which is trivial to incorporate at the Hartree--Fock level whereas its nonlinear nature requires larger modifications to standard implementations of other wave function methods.

For simplicity the implementation is specific to the one-dimensional ($p=d=1$) case, though we continue to write equations valid for any dimension. The one-dimensional setting is sufficient to illustrate our theoretical formalism above and lets us avoid complications due to pseudopotentials that are needed to reduce computational cost for realistic systems in three dimensions.

In terms of the very general theory presented in Sec.~\ref{subsec:DFT}, our implementation corresponds to the concrete choice of letting the sets of interacting $A$ and noninteracting states $A'$ be the Slater determinants that can be obtained in a finite, plane-wave basis.

\subsection{Periodic Hartree--Fock model}\label{subsec:PBCHF}

\newcommand\EnergyHF{\mathcal{E}_{\text{HF}}}
\newcommand\EnergyHFmod{\widetilde{\mathcal{E}}_{\text{HF}}}

The one-electron spatial basis functions are chosen as plane waves,
\begin{equation}
  \chi_{\myvec{G}}(\myvec{r})= \frac{e^{i\myvec{G}\cdot\myvec{r}}}{\sqrt{|\cBvKzone|}}, \quad \myvec{G} \in \ReciprocalLattice, |\myvec{G}| \leq G_{\mathrm{cut}},
\end{equation}
which are normalised in the inner product given in Eq.~\eqref{eqSpinOrbInnerProd} (with obvious modification for spatial orbitals without any spin part). The parameter $G_{\mathrm{cut}}$ is a momentum cut off that controls the number of basis functions. The canonical Hartree--Fock orbitals are expanded as
\begin{align} 
  \label{eqCrystOrbExp}
  \phi_{l\myvec{k}}(\myvec{r}) = e^{i\myvec{k}\cdot\myvec{r}} \sum_{\substack{\myvec{G} \in \ReciprocalLattice \\ |\myvec{G}| \leq G_{\mathrm{cut}}}} C_{\myvec{G},l}(\myvec{k}) \, \chi_{\myvec{G}}(\myvec{r}), \qquad \myvec{k} \in \dBZ,
\end{align}
where $l$ is a band index. These orbitals are translation-symmetric in the Bloch sense, i.e.\ $\phi_{l\myvec{k}}(\myvec{r}+\myvec{a}_j) =  e^{i\myvec{k}\cdot\myvec{a}_j} \phi_{l\myvec{k}}(\myvec{r})$, with the crystal momentum $\myvec{k}$ functioning as a symmetry index. As is standard, we do not allow Hartree--Fock orbitals that are superpositions of different translation symmetries. This amounts to a restriction of the superposition principle at the level of one-electron states, in addition to the restriction at the $M$-electron level due to the restriction to Slater determinants.

Occupation numbers $n_{l}(\myvec{k}) \in \{0,2\}$ such that $\sum_{l}\sum_{\myvec{k}\in\dBZ} n_{l}(\myvec{k}) = M = nN_1N_2 \cdots N_p$ is the total occupation in the Born--von K\'arm\'an zone ($n$ electrons per unit cell) define a Slater determinant $\Phi \in \WaveFunSpace{M}$. Our implementation is based on the one-particle reduced density operator,
\begin{align}
  D^{\mathrm{BvK}} & = \sum_{l} \sum_{\myvec{k}\in\dBZ} n_l(\myvec{k}) \, \ket{\phi_{l\myvec{k}}} \, \bra{\phi_{l\myvec{k}}},
                            \\
  D & = \invNbvk D^{\mathrm{BvK}},
                            \\
  D_{\myvec{G}\myvec{k},\myvec{G}' \myvec{q}} & = \bra{\chi_{\myvec{G}\myvec{k}}} D \ket{\chi_{\myvec{G}'\myvec{q}}} = \delta_{\myvec{k},\myvec{q}} D_{\myvec{G},\myvec{G}'}(\myvec{k}),
\end{align}
where the Kronecker delta arises due to the above mentioned imposition of translation symmetry, leading to a block diagonal structure of $D_{\myvec{G}\myvec{k},\myvec{G}' \myvec{q}}$. The blocks are given by
\begin{equation}
   \label{eq:Dmat_elem}
  D_{\myvec{G},\myvec{G}'}(\myvec{k}) = \invNbvk \sum_{l} \sum_{\myvec{k}\in\dBZ} n_l(\myvec{k}) \, C_{\myvec{G},l\myvec{k}} \, C_{\myvec{G}',l\myvec{k}}^*.
\end{equation}
Formulated in terms of the reduced density matrix, the Hartree--Fock energy is given by
\begin{equation}
    \EnergyHF(D) = \trace(h D) + \lambda \mathcal{E}_{\mathrm{H}}(D) - \lambda \mathcal{E}_{\mathrm{x}} (D)
\end{equation}
where $h = -\tfrac{1}{2} \nabla^2 + v(\myvec{r})$ is the core Hamiltonian, the trace $\trace(h D)$ gives the energy contribution per unit cell, and both the Hartree energy and the the exchange energy are obtained as simple, quadratic functionals of $D$,
\begin{widetext}
\begin{equation}
  \begin{split}
    \mathcal{E}_{\mathrm{H}}(D)  = \CoulQform(\bar{\rho}_D,\bar{\rho}_D) = \frac{1}{2} \sum_{\myvec{G}_1,\myvec{G}_2,\myvec{G}_3,\myvec{G}_4} \sum_{\myvec{k},\myvec{q}\in\dBZ} D_{\myvec{G}_2\myvec{G}_1}(\myvec{k}) \,  (\myvec{G}_1\myvec{k},\myvec{G}_2\myvec{k}|\myvec{G}_3\myvec{q},\myvec{G}_4\myvec{q})\, D_{\myvec{G}_4\myvec{G}_3}(\myvec{q}),
   \end{split}
\end{equation}
\begin{equation}
  \mathcal{E}_{\mathrm{x}}(D) = \frac{1}{4} \sum_{\myvec{G}_1,\myvec{G}_2,\myvec{G}_3,\myvec{G}_4} \sum_{\myvec{k},\myvec{q}\in\dBZ} D_{\myvec{G}_4\myvec{G}_1}(\myvec{k}) \,  (\myvec{G}_1\myvec{k},\myvec{G}_2\myvec{q}|\myvec{G}_3\myvec{q},\myvec{G}_4\myvec{k})\, D_{\myvec{G}_2\myvec{G}_3}(\myvec{q})
\end{equation}
\end{widetext}
with the two-electron matrix element, in Mulliken notation, given by
\begin{equation}
  \begin{split}
    (\myvec{G}_1\myvec{k},\myvec{G}_2\myvec{q}|\myvec{G}_3\myvec{q},\myvec{G}_4\myvec{k}) & = \bra{\myvec{G}_1\myvec{k},\myvec{G}_3\myvec{q}} W \ket{\myvec{G}_2\myvec{q},\myvec{G}_4\myvec{k}}
    \\
    & = \frac{\delta_{\myvec{G}_1-\myvec{G}_2,\myvec{G}_4-\myvec{G}_3}}{\gamma^2 + |\myvec{G}_1-\myvec{G}_2|^2}.
  \end{split}
\end{equation}

To perform density-potential inversion, we also introduce the modified energy
\begin{equation}
  \label{eqEHFmod}
  \begin{split}
      \EnergyHFmod(D) & = \trace((T+\alpha v) D) + \lambda_{\mathrm{H}} \mathcal{E}_{\mathrm{H}}(D) - \lambda_{\mathrm{x}} \mathcal{E}_{\mathrm{x}}(D) 
      \\ 
      & \ \ \ \ \ \ + \frac{\|\bar{\rho}_D- \bar{\rho}_{\mathrm{ref}}\|_X^2}{2\varepsilon}, 
   \end{split}
\end{equation}
where $\bar{\rho}_{\mathrm{ref}}$ is the density to be reproduced and the parameters $\alpha$, $\lambda_{\mathrm{H}}$ and $\lambda_{\mathrm{x}}$ unify the various calculations we want to perform.
Exploiting the relation between the norm and the Hartree energy, this energy can also be written
\begin{equation}
    \label{eqModifiedHFenergy}
  \begin{split}
    \EnergyHFmod(D) &= \trace{(T+\alpha v - \frac{1}{\varepsilon} v_{\mathrm{H,ref}}) D} + \left( \lambda_{\mathrm{H}} + \frac{1}{\varepsilon} \right) \mathcal{E}_{\mathrm{H}}(D) \\
    & \ \ \ \ - \lambda_{\mathrm{x}} \mathcal{E}_{\mathrm{x}}(D) + \frac{\|\bar{\rho}_{\mathrm{ref}}\|_X^2}{2\varepsilon} \\
    & = \trace{h' D} + \left( \lambda_{\mathrm{H}} + \frac{1}{\varepsilon} \right) \mathcal{E}_{\mathrm{H}}(D) - \lambda_{\mathrm{x}} \mathcal{E}_{\mathrm{x}}(D)
    \\
    & \qquad + \frac{\|\bar{\rho}_{\mathrm{ref}}\|_X^2}{2\varepsilon},
  \end{split}
\end{equation}
where $v_{\mathrm{H,ref}} = \Jdual(\bar{\rho}_{\mathrm{ref}})$ is the Hartree potential due to the reference density and $h' = -\tfrac{1}{2} \nabla^2 + \alpha v - \tfrac{1}{\varepsilon} v_{\mathrm{H,ref}}$. Allowing $1/\varepsilon = 0$ as a possible value, this energy is flexible enough to capture standard Hartree--Fock calculations, noninteracting Kohn--Sham calculations, and the various density-potential inversion approaches, depending on the parameter values $1/\varepsilon,\alpha,\lambda_{\mathrm{H}},\lambda_{\mathrm{x}}$. In particular, minimisation of this energy with respect to $D$ allows us to determine the proximal density of a given $\bar{\rho}_{\mathrm{ref}}$, which is crucial for our inversion procedure, and also yields the modified ground-state-energy in Eq.~\eqref{eqEnergyZMP} above corresponding to the choice of model functional given by Eq.~\eqref{eq:mod_func_alp_xi}. In our forward calculations, we set $\alpha=1$, $v=v_{\mathrm{ext}}$, $\lambda_{\mathrm{H}} = \lambda_{\mathrm{x}} = 1$, and $1/\varepsilon=0$. In our inverse Kohn--Sham calculations, we set $\lambda_{\mathrm{x}} = 0$, $\lambda_{\mathrm{H}} = \xi$ 
and $v= v_{\mathrm{ext}}$, while exploring different values of $\alpha$, $\xi$, and $1/\varepsilon$.

\subsection{Self-consistent field optimisation}
\label{subsec:eng_min}

The Fock matrix is given by
\begin{equation}
  \begin{split}
    \mathcal{F}_{\myvec{G}'\myvec{G}}(\myvec{k}) & = \frac{\partial \EnergyHFmod(D)}{\partial D_{\myvec{G}\myvec{G}'}(\myvec{k})}
    \\
    & = \eta h'_{\myvec{G}'\myvec{G}} + \left( \lambda_{\mathrm{H}} + \frac{1}{2\varepsilon} \right) \mathcal{J}_{\myvec{G}'\myvec{G}}(\myvec{k})- \lambda_{\mathrm{x}} \mathcal{K}_{\myvec{G}'\myvec{G}}(\myvec{k})
   \end{split}
\end{equation}
with
\begin{equation}
  \begin{split}
    \mathcal{J}_{\myvec{G}'\myvec{G}}(\myvec{k}) = \invNbvk \sum_{\myvec{G}_3,\myvec{G}_4} \sum_{\myvec{q}\in\dBZ} (\myvec{G}'\myvec{k},\myvec{G}\myvec{k}|\myvec{G}_3\myvec{q},\myvec{G}_4\myvec{q})\, D_{\myvec{G}_4\myvec{G}_3}(\myvec{q}),
   \end{split}
\end{equation}
\begin{equation}
  \mathcal{K}_{\myvec{G}'\myvec{G}}(\myvec{k}) = \frac{\invNbvk}{2} \sum_{\myvec{G}_2,\myvec{G}_3} \sum_{\myvec{q}\in\dBZ} (\myvec{G}'\myvec{k},\myvec{G}_2\myvec{q}|\myvec{G}_3\myvec{q},\myvec{G}\myvec{k})\, D_{\myvec{G}_2\myvec{G}_3}(\myvec{q}).
\end{equation}
The self-consistency condition, i.e.\ the Roothaan--Hall equations,
\begin{equation}
  \mathcal{F}(\myvec{k}) \, \mathbf{C}(\myvec{k}) = \mathbf{C}(\myvec{k}) \, \boldsymbol{\varepsilon}(\myvec{k})
\end{equation}
determine the coefficients $\mathbf{C}(\myvec{k})$ in Eq.~\eqref{eqCrystOrbExp} and the band energies $\boldsymbol{\varepsilon}(\myvec{k})$. For each point $\myvec{k} \in \dBZ$, we enforce a local aufbau principle, i.e.\ the $n/2$ lowest crystal orbitals are occupied. In all reported runs, this turned out to be equivalent to a global aufbau principle.

We implemented two methods for self-consistent field (SCF) optimisation. The first was a standard Direct Inversion in Iterative Subspaces (DIIS) method~\cite{PULAY_CPL73_393,PULAY_JCC3_556}, where we tried both $\myvec{k}$-point specific error vectors and $\myvec{k}$-point averaged error vectors.

The second SCF method is a simplified second-order method based on orbital rotations and line search. Letting $P = D^{(j)}$ denote the density operator in SCF iteration $j$, the next density operator is written in the exponential parametrisation~\cite{HELGAKER00,HELGAKER_CPL327_397} as
\begin{equation}
  D(X) = e^{-X} P e^{X} = P + [P,X] + \frac{1}{2} [[P,X],X] + \ldots,
\end{equation}
where $X = -X^{\dagger}$ is an anti-hermitian operator. To second order in $X$ we obtain
\begin{equation}
  \begin{split}
    \EnergyHFmod(D(X))& \approx \EnergyHFmod(P) + \trace{\mathcal{F} \, ([P,X] + \tfrac{1}{2} [[P,X],X])}
    \\
    & \ \ \ \ + \left( \lambda_{\mathrm{H}} + \frac{1}{\varepsilon} \right) \mathcal{E}_{\mathrm{H}}([P,X]) - \lambda_{\mathrm{x}} \mathcal{E}_{\mathrm{x}}([P,X]),
  \end{split}
\end{equation}
where the Fock matrix $\mathcal{F}$ is computed from $P$. Next we consider a line search along the gradient
\begin{equation}
    Y := \frac{\partial \EnergyHFmod(D(X))}{\partial X} \Big|_{X=0} = [P, \mathcal{F}],
\end{equation}
i.e.\ we set $X = -tY$. To second order, the energy becomes
\begin{equation}
  \EnergyHFmod(D(-tY))  \approx \EnergyHFmod(P) - Bt + At^2,
\end{equation}
with
\begin{align}
  B & = \trace{\mathcal{F} \, [P,Y]},
  \\
  A & = \tfrac{1}{2} \trace{\mathcal{F} \, [[P,X],X]} +  \left( \lambda_{\mathrm{H}} + \frac{1}{\varepsilon} \right) \mathcal{E}_{\mathrm{H}}([P,Y]) \nonumber\\
  & \qquad - \lambda_{\mathrm{x}} \mathcal{E}_{\mathrm{x}}([P,Y]).
\end{align}
When $A>0$, the optimal step length is $t = \tfrac{B}{2A}$, and otherwise we take a small gradient step. Hence, the next density matrix becomes
\begin{equation}
    D^{(j+1)} = D^{(j)} - t [D^{(j)},Y] + \frac{t^2}{2} [[D^{(j)},Y],Y].
\end{equation}

\section{Numerical results}
\label{secNumRes}

The above one-dimensional Hartree--Fock scheme was implemented in new open source code Sable~\cite{SableCode}. In this section we employ this implementation to present illustrative numerical results using the MY-based inversion method.

In the present section, we write $\bar{\rho}_{\mathrm{ref}}$ instead of $\bar{\rho}_{\mathrm{gs}}$ for the input density to the inversion procedure. We consider non-$N$-representable densities in Sec.~\ref{secNonNRep}. Elsewhere, all reference densities are Hartree--Fock ground-state densities, i.e. $\bar{\rho}_{\mathrm{ref}} = \bar{\rho}_{\mathrm{gs}}^\mathrm{HF}$ corresponding to $v_{\mathrm{ext}}$ and obtained in a forward calculation as described above.

In the inversion procedure, we use the model functional 
Eq.~\eqref{eq:mod_func_alp_xi} and always with matching $v_{\mathrm{ext}}$ for the ground-state density being inverted (with exception in Sec.~\ref{secNonNRep} for the non-$N$-representable case). 
At the end of the procedure, we obtain in accordance with Eq.~\eqref{eq:vseps} the Kohn--Sham potential
\begin{equation}
  v_{\mathrm{s}}^{\varepsilon} = \alpha v_{\mathrm{ext}} + \xi v_\mathrm{H}(\bar{\rho}^{\varepsilon}) + u^{\varepsilon},
\end{equation}
where
\begin{equation}
    \label{eqUpotFromInv}
    u^{\varepsilon} = \frac{1}{\varepsilon} \Jdual(\bar{\rho}^{\varepsilon} - \bar{\rho}_{\mathrm{ref}}), \quad
    \bar{\rho}^{\varepsilon} = \underset{\varepsilon F_{\alpha,\xi}^\mathrm{mod} }{\mathrm{prox}} (\bar{\rho}_{\mathrm{ref}}).
\end{equation}
Combining this with the standard Kohn--Sham decomposition (making use that $v_\mathrm{ext}$ is the potential of the interacting density being inverted), which in the present lacks the correlation term, i.e.\ $v_{\mathrm{s}}= v_{\mathrm{ext}} + v_{\mathrm{H}} + v_{\mathrm{x}}$, we furthermore obtain the exchange potential as
\begin{equation}
  v_{\mathrm{x}}^{\varepsilon} = (\alpha-1) v_{\mathrm{ext}} + (\xi-1) v_\mathrm{H}(\bar{\rho}^{\varepsilon}) + u^{\varepsilon}.
\end{equation}

\begin{figure}
  \begin{center}
    \includegraphics[width=0.99\columnwidth]{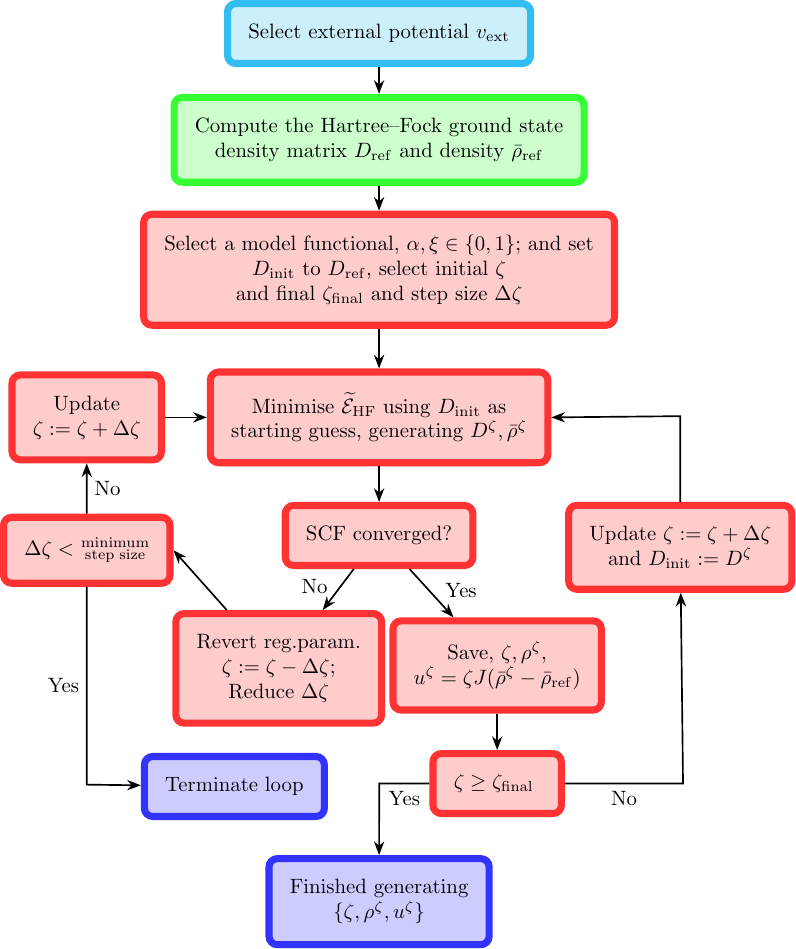}
  \end{center}
    \caption{\label{figFlowchart} Summary of the forward Hartree--Fock (top two boxes) and the Moreau--Yosida-based inversion procedure used in this work. Note that $\zeta = 1/\varepsilon$ is the inverse regularisation parameter and the model functional parameters are defined in Eq.~\eqref{eq:mod_func_alp_xi}. The output is the proximal point $\rho^{\zeta}$ and the contribution $u^{\zeta}$ to the KS potential.}
\end{figure}

A general pattern in our calculations is that the DIIS method struggles to reach SCF convergence when the factor $\lambda_{H} + \tfrac{1}{\varepsilon}$ in front of the Hartree term in Eq.~\eqref{eqModifiedHFenergy} is large. For $\varepsilon \lesssim 10^{-3}$, SCF convergence was not feasible with this method. We therefore relied on the simple orbital rotation-based method described in Sec.~\ref{subsec:eng_min}. Combined with the use of the converged density matrix from a similar, but larger value of $\varepsilon$, as the initial guess, this method was able to reliably reach SCF convergence. We thus constructed a sequence $\varepsilon_1,\varepsilon_2,\ldots$ of values of the regularisation parameter. The sequence is constructed dynamically using the inverse values $\zeta_i = 1/\varepsilon_i$ and a simple update $\zeta_{i+1} = \zeta_i + \Delta\zeta$. In the event of failure to reach SCF convergence for a given regularisation parameter, we reduced the update $\Delta\zeta$. Since the contribution to factor in front of the Hartree term is $1/\varepsilon$ the main limitation is that the same inverse update $\Delta\zeta = 1/\varepsilon_{i+1}-1/\varepsilon_i$ results in diminishing returns for the update $\varepsilon_{i+1}-\varepsilon_i$. Nonetheless, by this method we could perform calculations down to regularisation parameters of $\varepsilon \approx 10^{-8}-10^{-7}$. The typical procedure is summarised in Fig.~\ref{figFlowchart}.

\subsection{Computational details}

Unless otherwise noted, we set $\gamma = 1$, $a = 5.613$, $\Nbvk = 41$, $G_{\mathrm{cut}}/|b_1| = 61$.

\subsection{Effect of parameters on convergence}

  \subsubsection{Model dependence}
  \label{sec:AnsatzDep}
  
\begin{figure}[h]
    \includegraphics[width=0.45\textwidth]{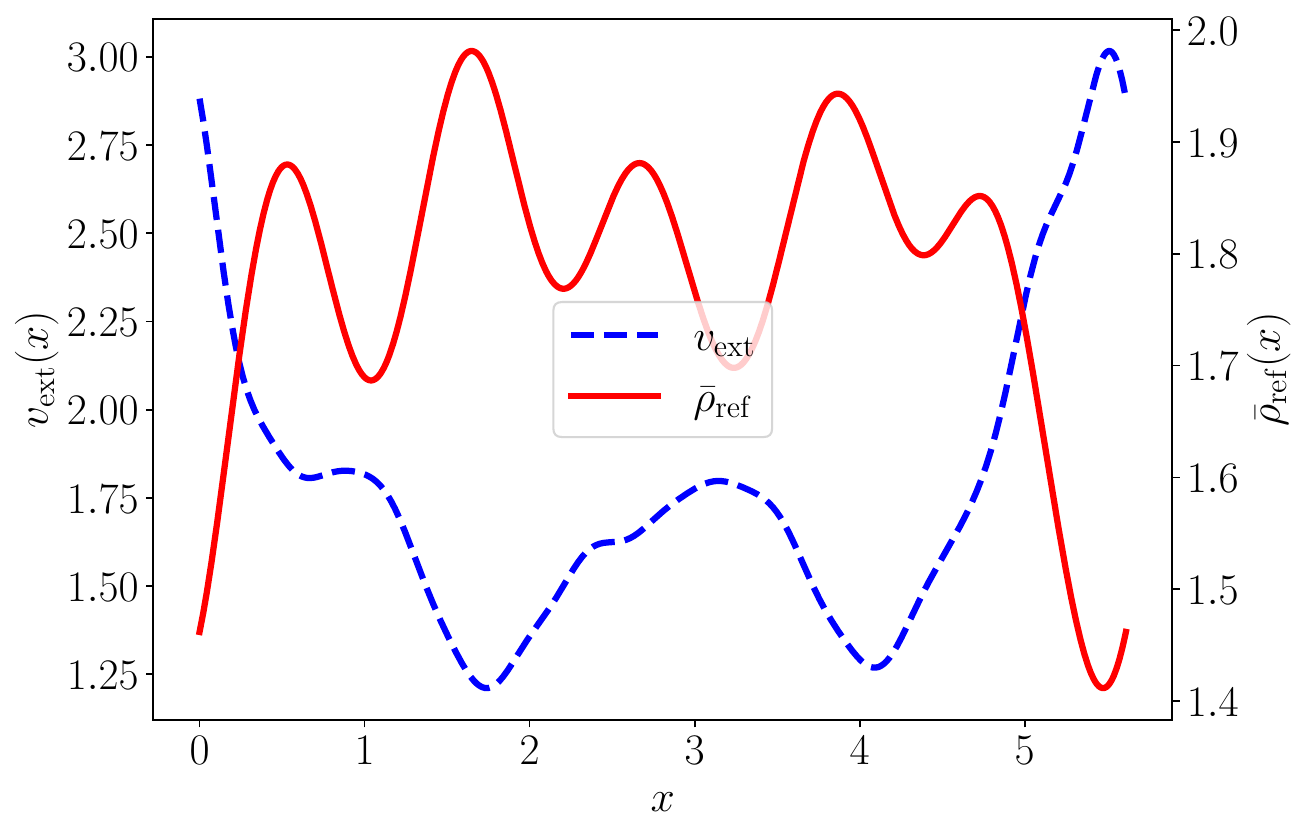}
    \caption{An external potential, $v_{\mathrm{ext}}$, and the corresponding Hartree--Fock ground-state density, $\rhoref$ for a $n=10$ electron system. Note the twin vertical axes.}
    \label{fig:vext_and_rho_twin_N10}
\end{figure}

\begin{figure}[h]
    \begin{subfigure}[b]{0.9\linewidth}
            \centering
            \caption{}
            \includegraphics[width=0.98\linewidth]{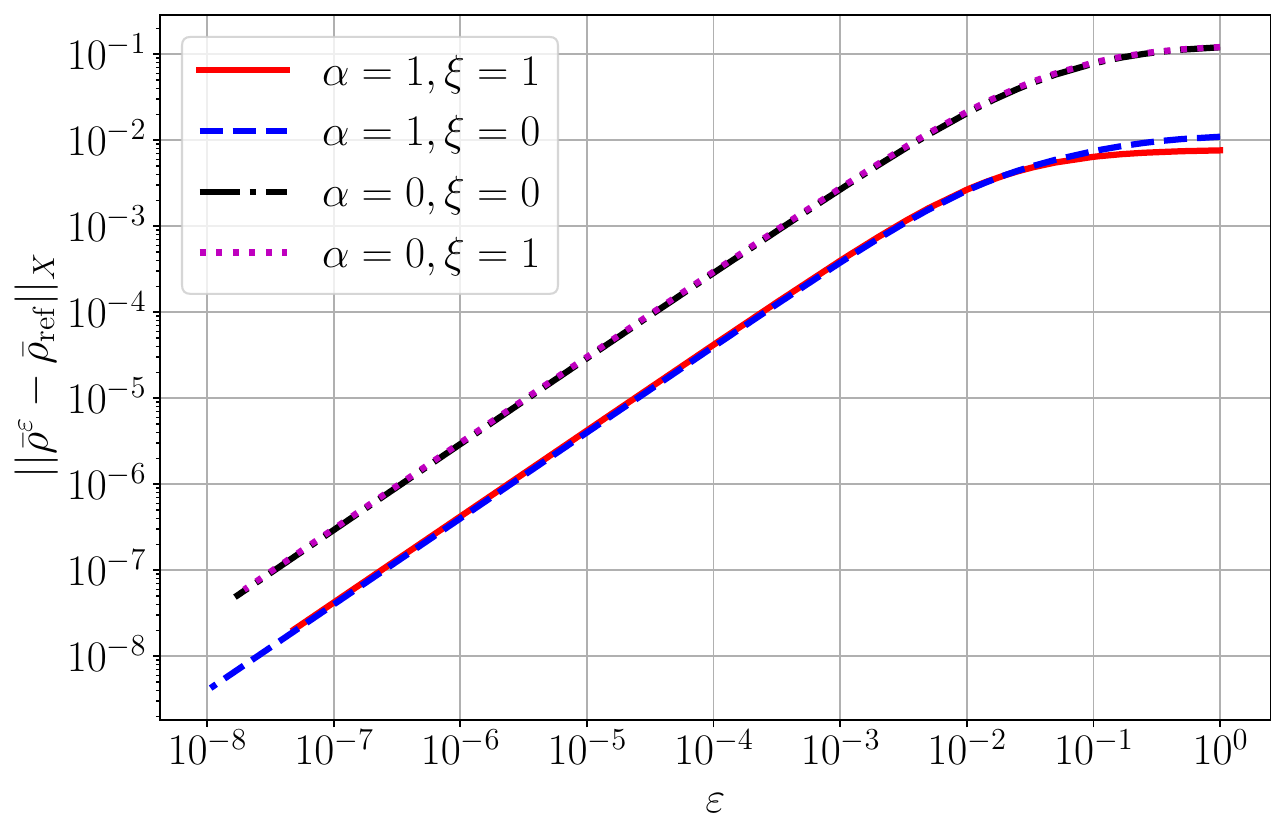}
            \label{subfig:rho_norm_err_lamb_seq}
    \end{subfigure} 	
    \begin{subfigure}[b]{0.9\linewidth}
            \centering
            \caption{}
            \includegraphics[width=0.98\linewidth]{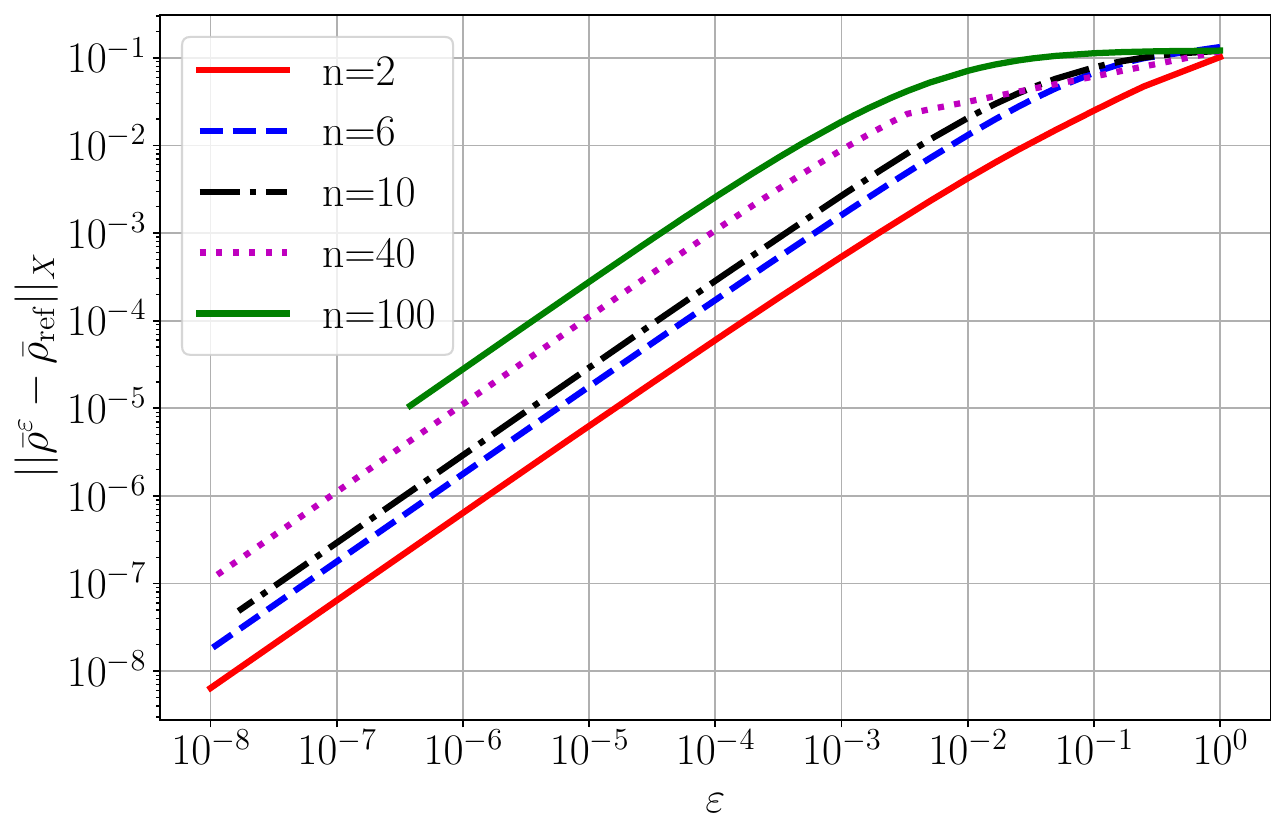}
            \label{subfig:rho_norm_err_N_seq}
    \end{subfigure} 	
    \caption{Distance between the proximal density, $\rho^{\varepsilon}$, and the reference density, $\rhoref$, as a function of the regularisation parameter $\varepsilon$. In (a) the reference density has $n=10$ electrons and each line style correspond to a different parameter values in the model functional $F^{\mathrm{mod}}_{\alpha,\xi}$. In (b) the number of electrons is varied for the model functional with $\alpha=\xi=0$ and the reference density is scaled accordingly.}
\end{figure}

\begin{figure}[h]
    \begin{subfigure}[b]{0.48\linewidth}
            \centering
            \caption{}
            \includegraphics[width=0.98\linewidth]{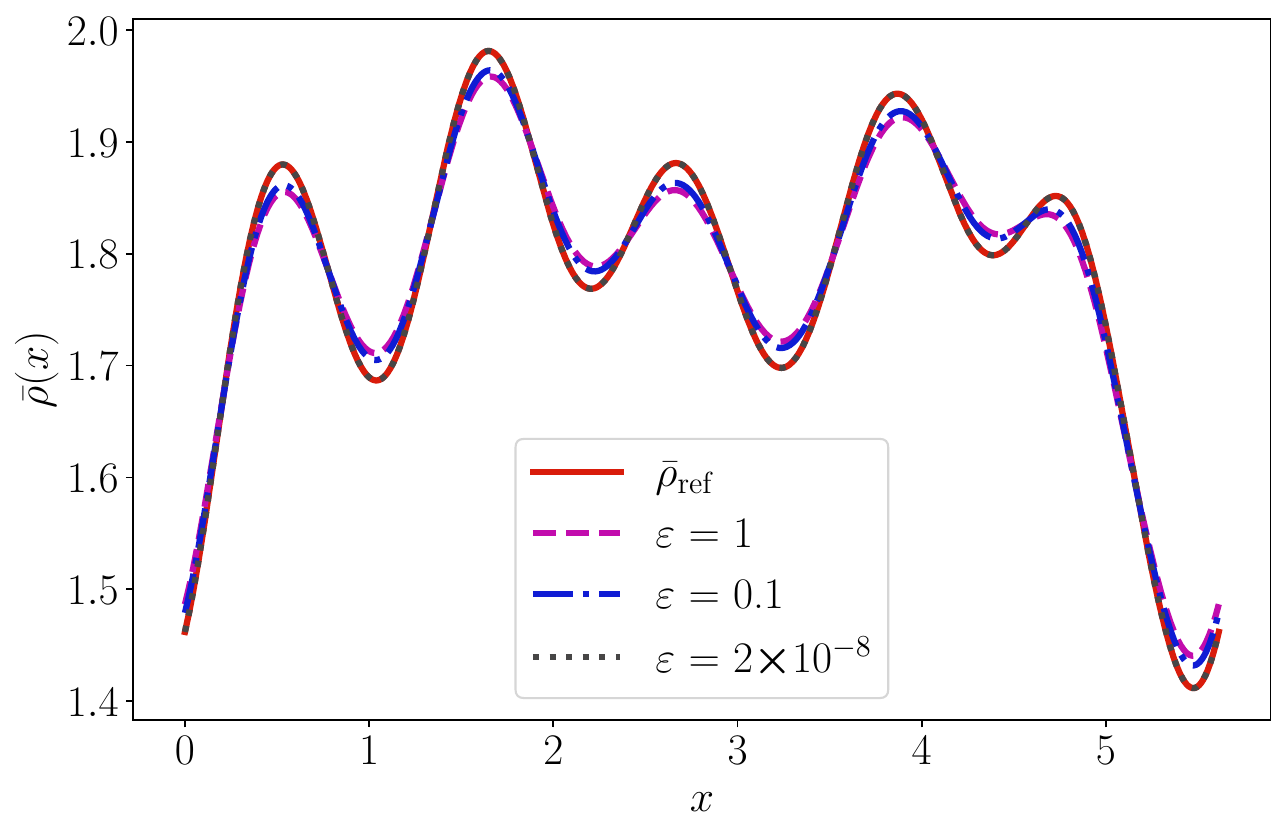}
            \label{subfig:rho_x_X_N10}
    \end{subfigure} 	
    \begin{subfigure}[b]{0.48\linewidth}
            \centering
            \caption{}
            \includegraphics[width=0.98\linewidth]{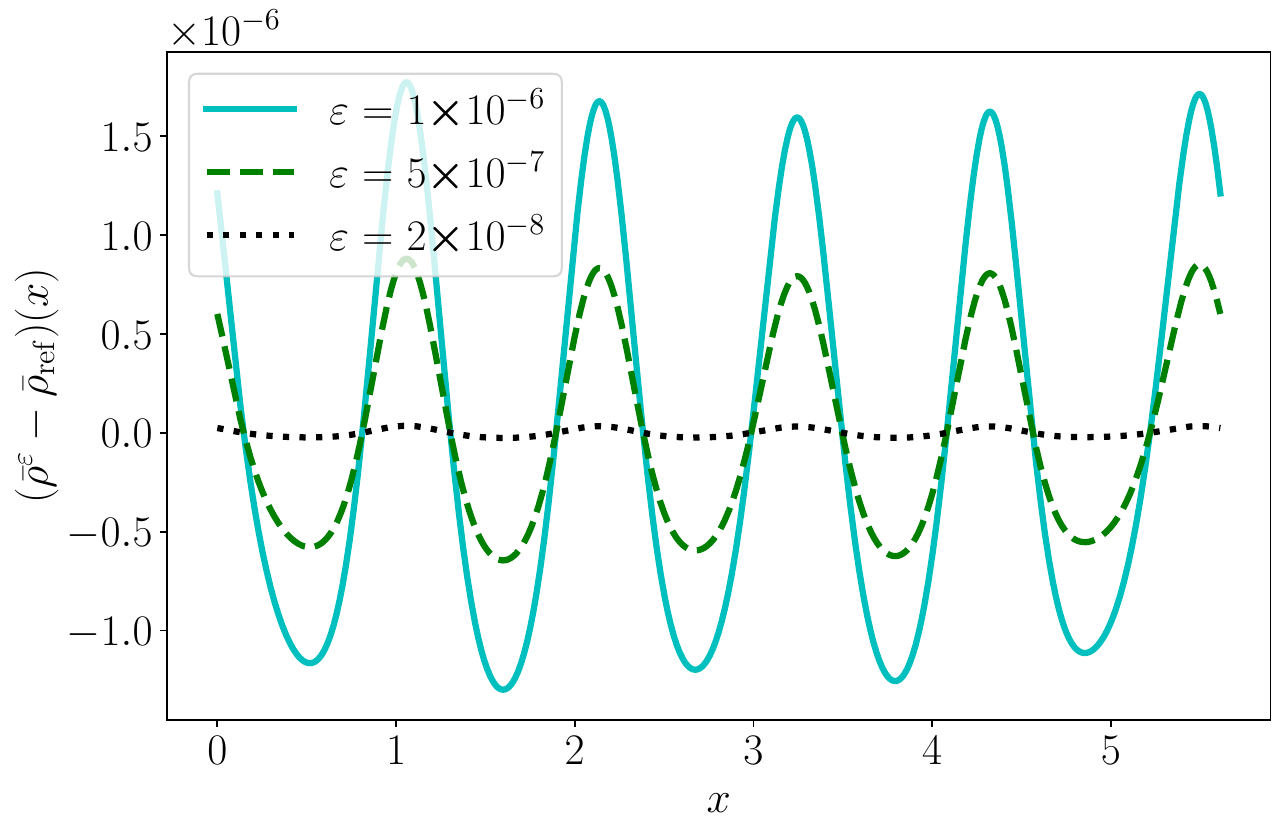}
            \label{subfig:rho_err_x_X_N10}
    \end{subfigure} 	
    \begin{subfigure}[b]{0.48\linewidth}
            \centering
            \caption{}
            \includegraphics[width=0.98\linewidth]{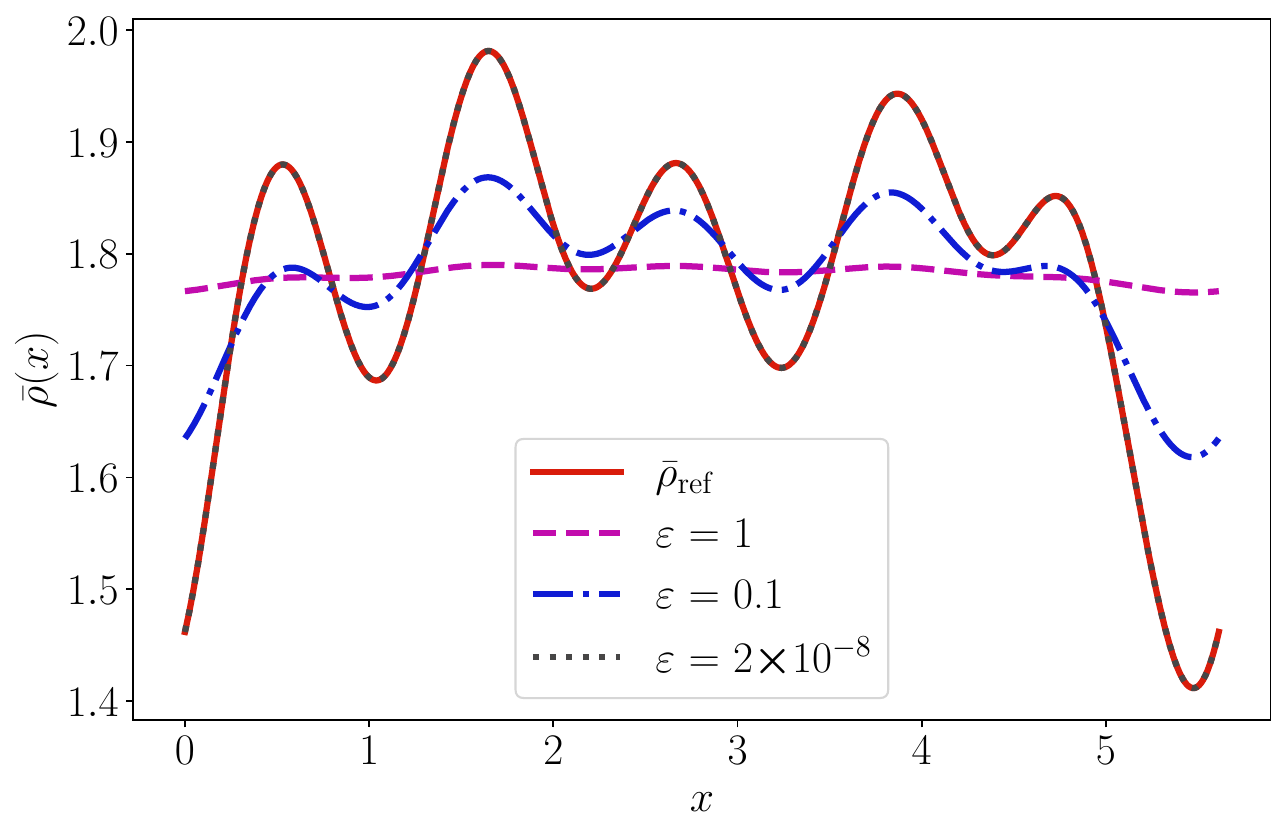}
            \label{subfig:rho_x_vHX_N10}
    \end{subfigure}
    \begin{subfigure}[b]{0.48\linewidth}
            \centering
            \caption{}
            \includegraphics[width=0.98\linewidth]{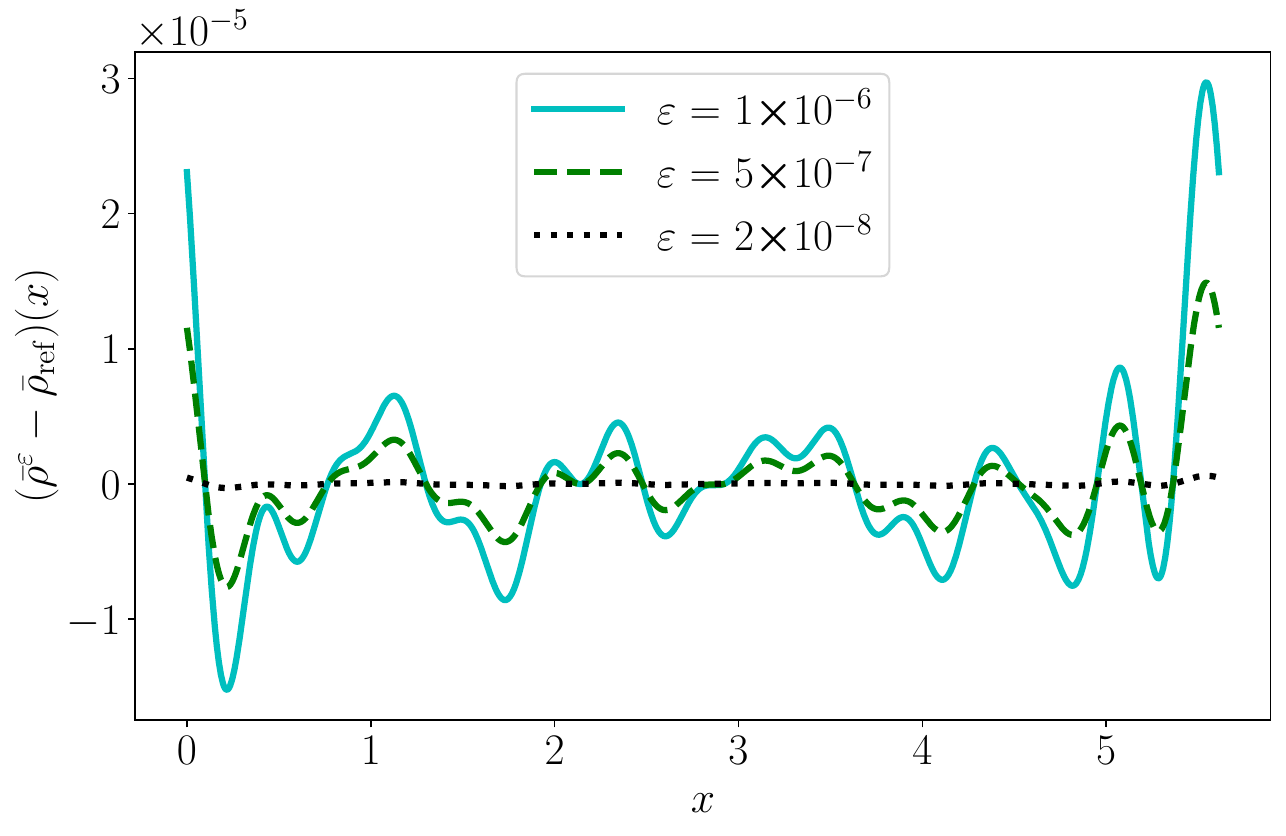}
            \label{subfig:rho_err_x_vHX_N10}
    \end{subfigure}
    \caption{\label{fig:dens_plots_and_dens_err_N10} In (a) and (c), the reference density and proximal densities are displayed as functions of position within a unit cell. The model functional with $\alpha=\xi=1$ was used in (a) and $\alpha=\xi=0$ in (c). In (b) and (d), the corresponding density errors as functions of position are shown for a different set of regularisation parameters.}    
\end{figure}

\begin{figure}[h]
        \begin{subfigure}[b]{0.9\linewidth}
                \centering
                \caption{}
                \includegraphics[width=0.98\linewidth]{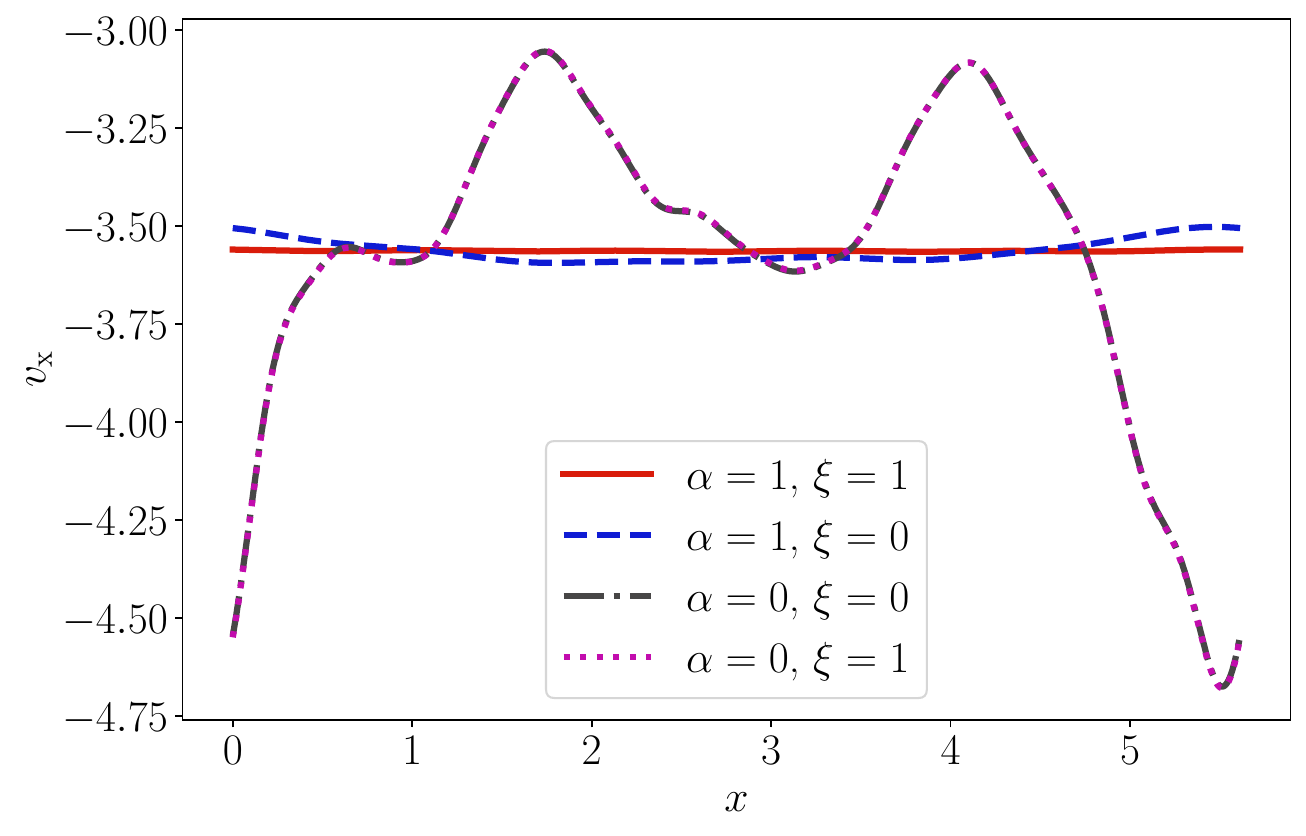}
                \label{subfig:vx_conv_00_N10} 
        \end{subfigure} 	
        \begin{subfigure}[b]{0.9\linewidth}
                \centering
                \caption{}
                \includegraphics[width=0.98\linewidth]{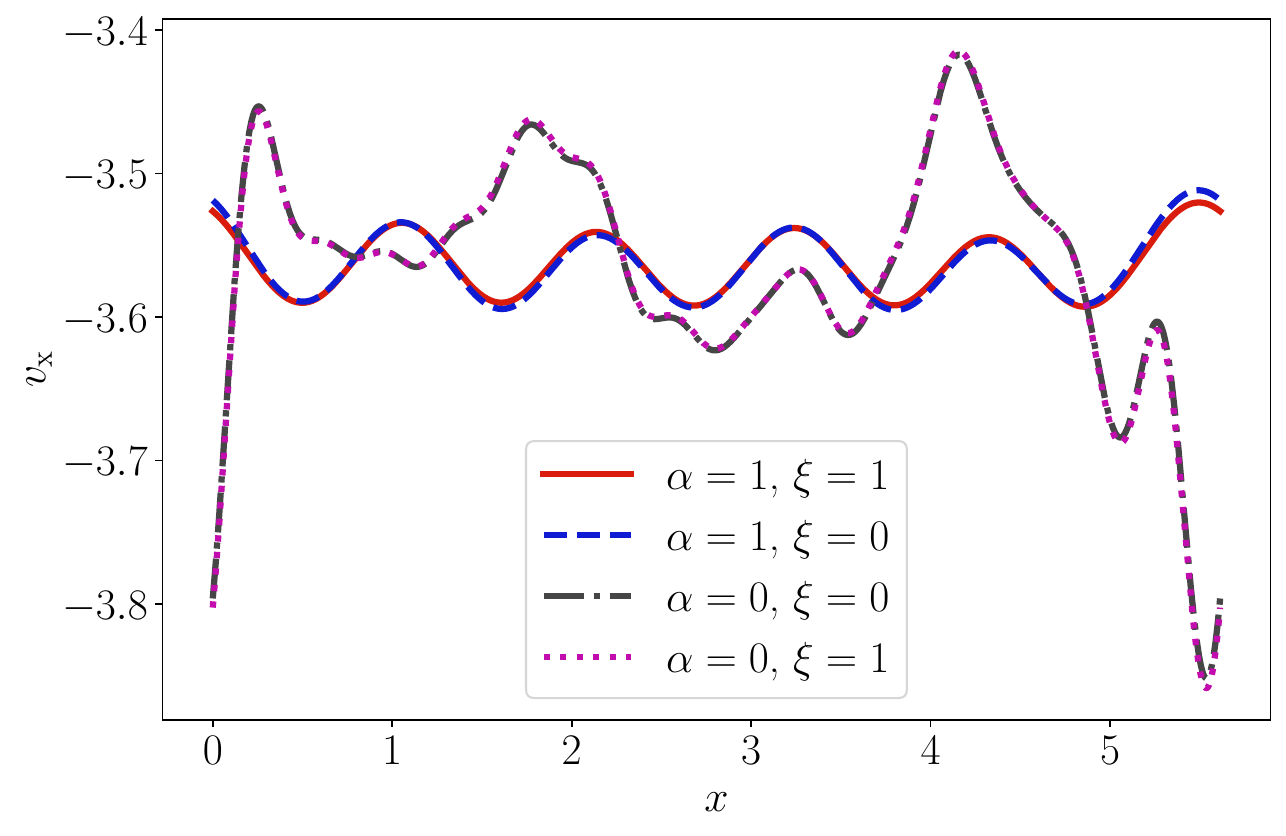}
                \label{subfig:vx_conv_02_N10} 
        \end{subfigure} 	
        \begin{subfigure}[b]{0.9\linewidth}
                \centering
                \caption{}
                \includegraphics[width=0.98\linewidth]{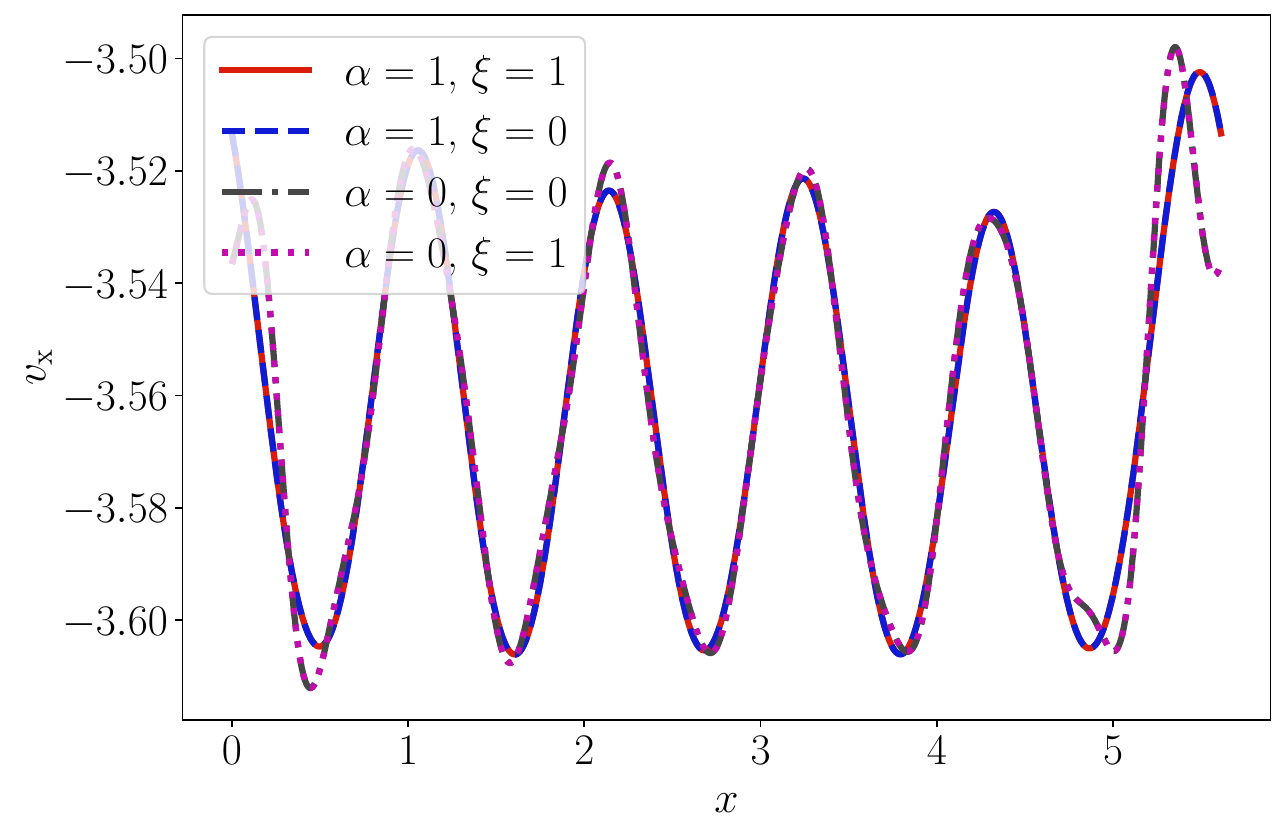}
                \label{subfig:vx_conv_04_N10} 
        \end{subfigure}
        \begin{subfigure}[b]{0.9\linewidth}
                \centering
                \caption{}
                \includegraphics[width=0.98\linewidth]{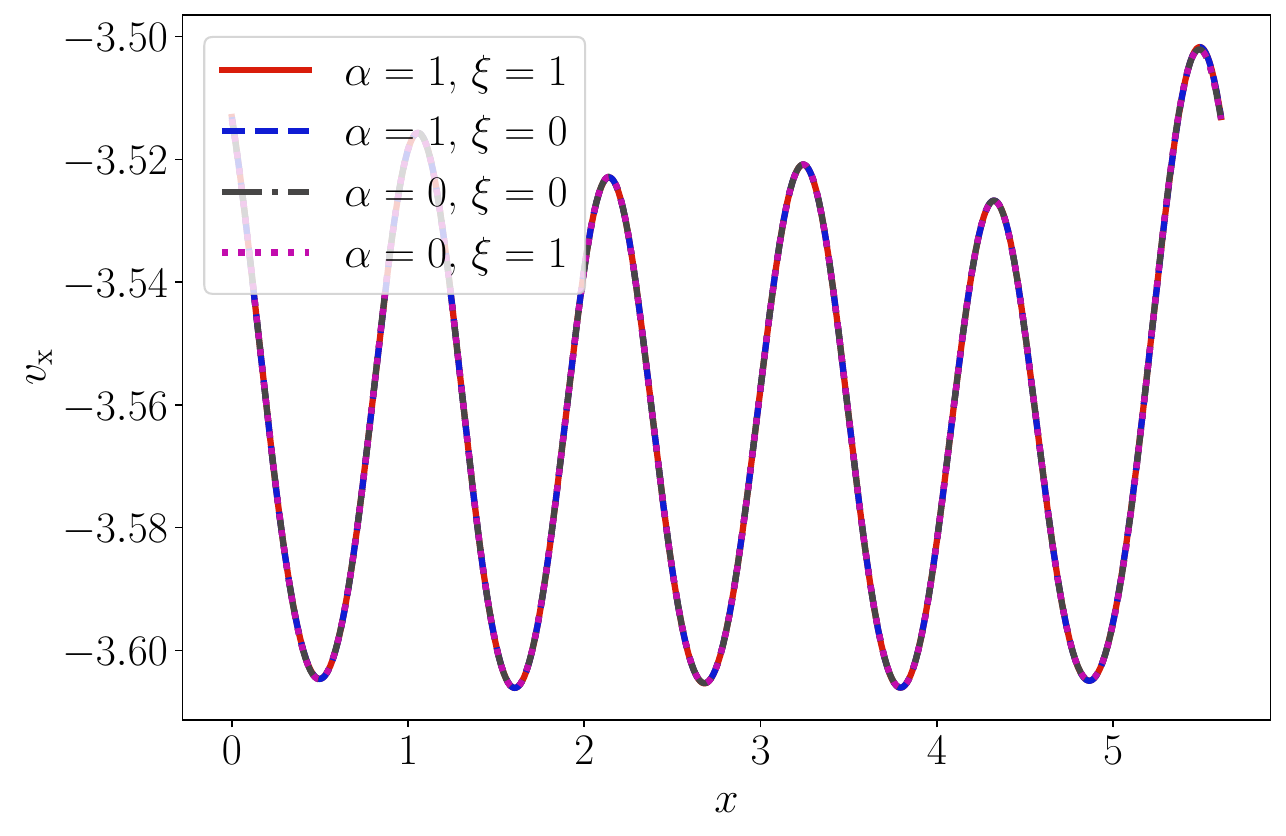}
                \label{subfig:vx_conv_06_N10} 
        \end{subfigure}
        \caption{\label{fig:vx_conv_N10}  Local exact exchange potentials $v_{\mathrm{x}}^{\varepsilon}$ obtained with four different choices of model functionals. In (a) $\varepsilon = 1$, (b) $\varepsilon = 0.01$, (c) $\varepsilon = 9.8\times10^{-5}$ and (d) $\varepsilon = 10^{-6}$, which is small enough to obtain converged results.}        
\end{figure}

\begin{figure}[h]
    \includegraphics[width=0.45\textwidth]{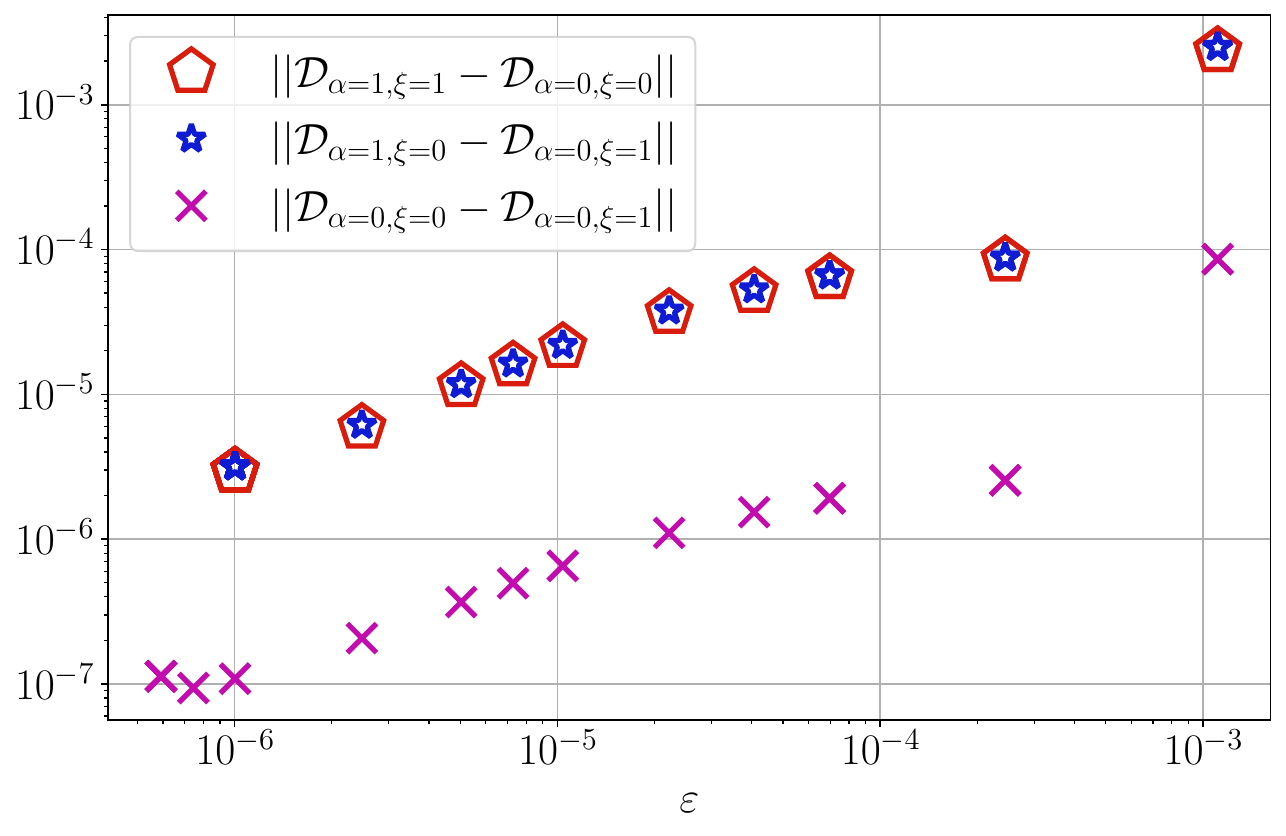}
    \caption{\label{fig:D_conv} The Frobenius norm of the density matrix different for each pair of different model functionals $F^{\mathrm{mod}}_{\alpha,\xi}$. The underlying reference density is the $n=10$ electron Hartree--Fock ground-state density for the external potential in Fig.~\ref{fig:vext_and_rho_twin_N10}.}
\end{figure}

\begin{figure}[h]
    \includegraphics[width=0.49\textwidth]{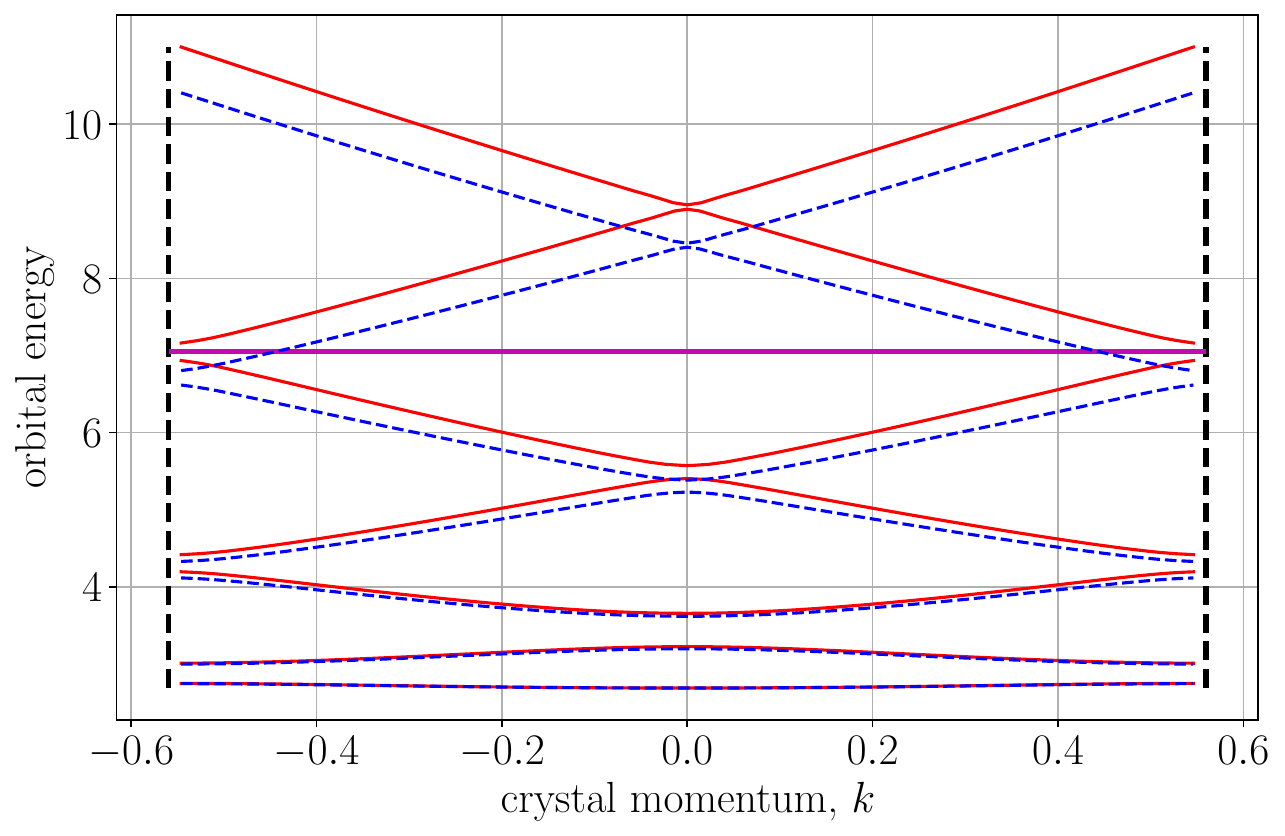}
    \caption{Hartree--Fock and Kohn--Sham band structure of the $n=10$ electron system with external potential as in Fig.~\ref{fig:vext_and_rho_twin_N10} displayed with solid red lines and blue dashed lines, respectively. The Kohn-Sham band structure was obtained with $\alpha=0$ and $\xi=0$ and the regularisation parameter $\varepsilon = 1.6\times10^{-8}$. The horizontal, purple line is the Fermi level of the Hartree--Fock system and the dashed vertical lines are the boundaries of the first Brillouin zone.}
    \label{fig:band_struc}
\end{figure}

How small a regularisation parameter value is needed to achieve a given accuracy depends on the parameters in the model functional $F^{\mathrm{mod}}_{\alpha,\xi}$.
To illustrate this dependence, we take as $\bar{\rho}_{\mathrm{ref}}$ the Hartree--Fock ground-state density for the external potential shown in Fig.~\ref{fig:vext_and_rho_twin_N10}. The number of electrons per unit cell was fixed to $n=10$. The resulting convergence of $\bar{\rho}^{\varepsilon}$ to $\bar{\rho}_{\mathrm{ref}}$ for different parameter values $\alpha,\xi$ is shown in Fig.~\ref{subfig:rho_norm_err_lamb_seq}. For this system, the inclusion ($\alpha=1$) of the external potential in the model functional has a very large effect and reduces the error by an order of magnitude for a given $\varepsilon$. By contrast, the inclusion ($\xi=1$) of the Hartree potential has only a negligible effect on the accuracy. Several resulting proximal densities are shown Fig.~\ref{fig:dens_plots_and_dens_err_N10} for the cases where both the external and Hartree potentials are included ($\alpha=\xi=1$) and excluded ($\alpha=\xi=0$), respectively. In the former case, even a large value of $\varepsilon=1$ qualitatively reproduces the reference density, though the agreement is not yet quantitative (see Fig.~\ref{subfig:rho_x_X_N10}).  In the latter case, the large value $\varepsilon=1$ results in a qualitatively incorrect proximal density and $\varepsilon \lesssim 0.1$ is needed to reproduce qualitative features (see Fig.~\ref{subfig:rho_x_vHX_N10}). Turning to the pointwise density errors in Figs.~\ref{subfig:rho_err_x_X_N10} and \ref{subfig:rho_err_x_vHX_N10}, one sees that $\varepsilon = 10^{-6}$ yields density errors on the order of $10^{-6}$ when the external potential is included and an order of magnitude larger when it is excluded. Moreover, the former density error have slowly oscillating character, where the latter density error has the rapidly oscillating pointwise density errors, indicating that Fourier components with large wave vectors have yet to reach optimal values. We interpret this result as the external potential introducing Fourier modes into the density with large wave vectors $\myvec{G}$. To reproduce these modes using the MY penalty term requires very small values of $\varepsilon$.

The convergence of the computed exchange potentials $v_{\mathrm{x}}^{\varepsilon}$ to a common limit is shown in Fig.~\ref{fig:vx_conv_N10}. When the external potential is included in the model, Fig.~\ref{subfig:vx_conv_02_N10} shows that the exchange potential is quantitatively accurate already for $\varepsilon = 10^{-2}$, and the bottom panel, Fig.~\ref{subfig:vx_conv_06_N10}, shows that every model eventually becomes accurate for small enough $\varepsilon$.
Fig.~\ref{fig:D_conv} shows the pairwise distance, measured with the Frobenius norm, between the density matrices produced for all four combinations of $\alpha,\xi\in\{0,1\}$. To within numerical noise, all distances vanish in the limit $\varepsilon \to 0^+$, indicating that not only the densities, but also the Kohn--Sham orbitals converge to the same limit. This result is expected since the Kohn--Sham orbitals are determined by the Kohn--Sham potential, which in turn is determined by the exchange potential in the present setting. Finally, Fig.~\ref{fig:band_struc} illustrates the effect on the band structure, where the HF system has a band gap of $0.227$ [a.u.] and the KS system has a band gap of $0.187$ [a.u.] hence, the local exchange reduced the band gap by $0.04$ [a.u.] and lowering the unoccupied bands compared to the nonlocal exact exchange.
 
\begin{figure}[h]
    \includegraphics[width=0.45\textwidth]{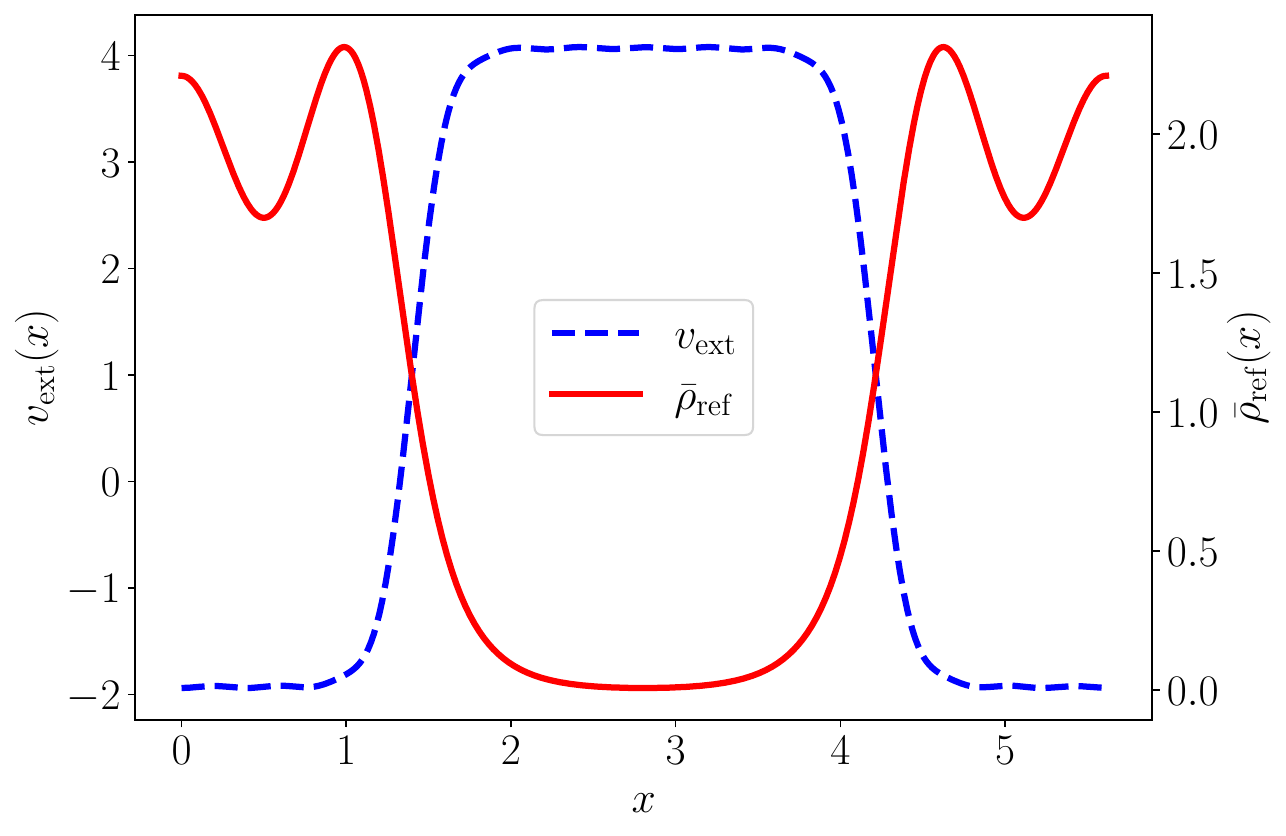}
    \caption{\label{fig:vext_and_rho_twin_N6} A confining external potential, biasing the system to part of the unit cell, and the corresponding $n=6$ electron Hartree--Fock ground-state density.}    
\end{figure}

\begin{figure}[h]
    \includegraphics[width=0.98\linewidth]{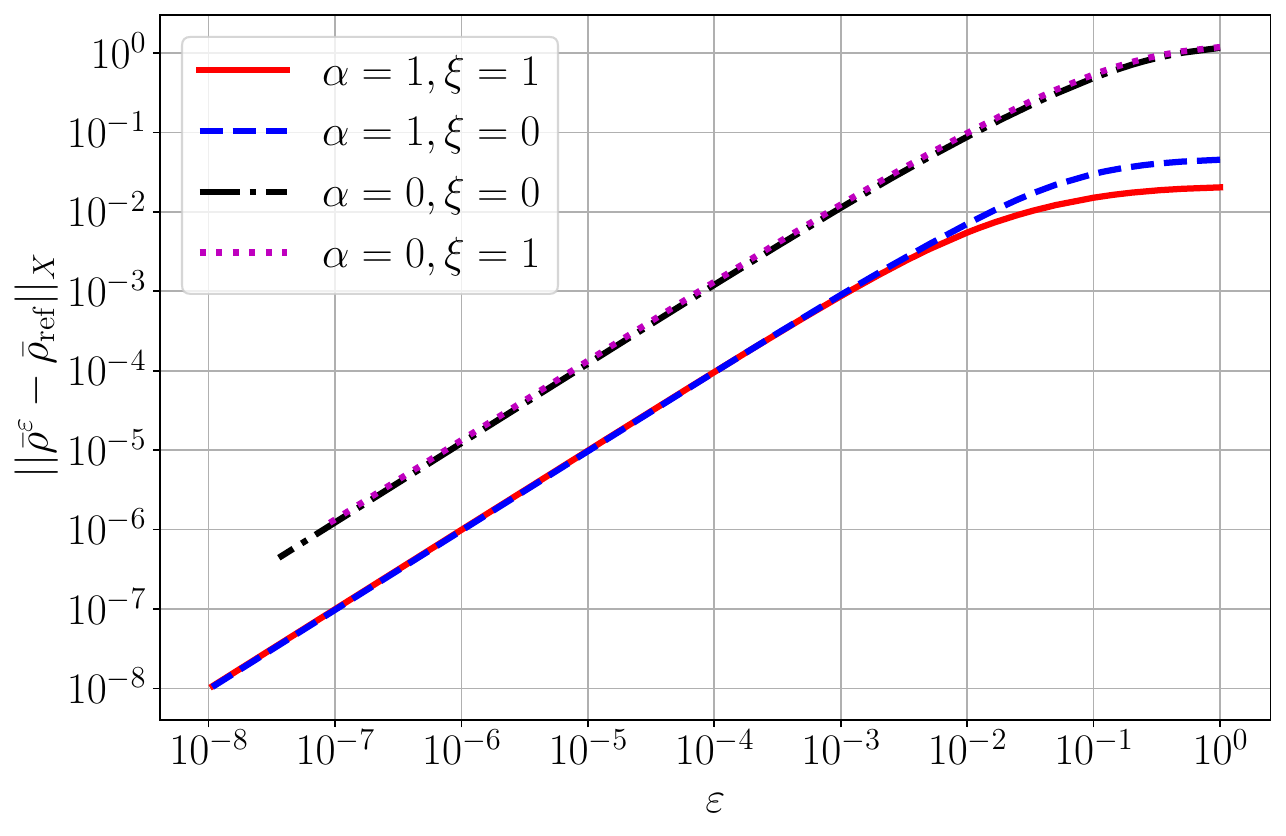}    \caption{\label{fig:rho_norm_err_lamb_seq_N6} Distance between the proximal density and the reference density obtained as the ground-state in a confining potential.}    
\end{figure}

\begin{figure}[h]
    \begin{subfigure}[b]{0.75\linewidth}
            \centering
            \caption{}
            \includegraphics[width=0.98\linewidth]{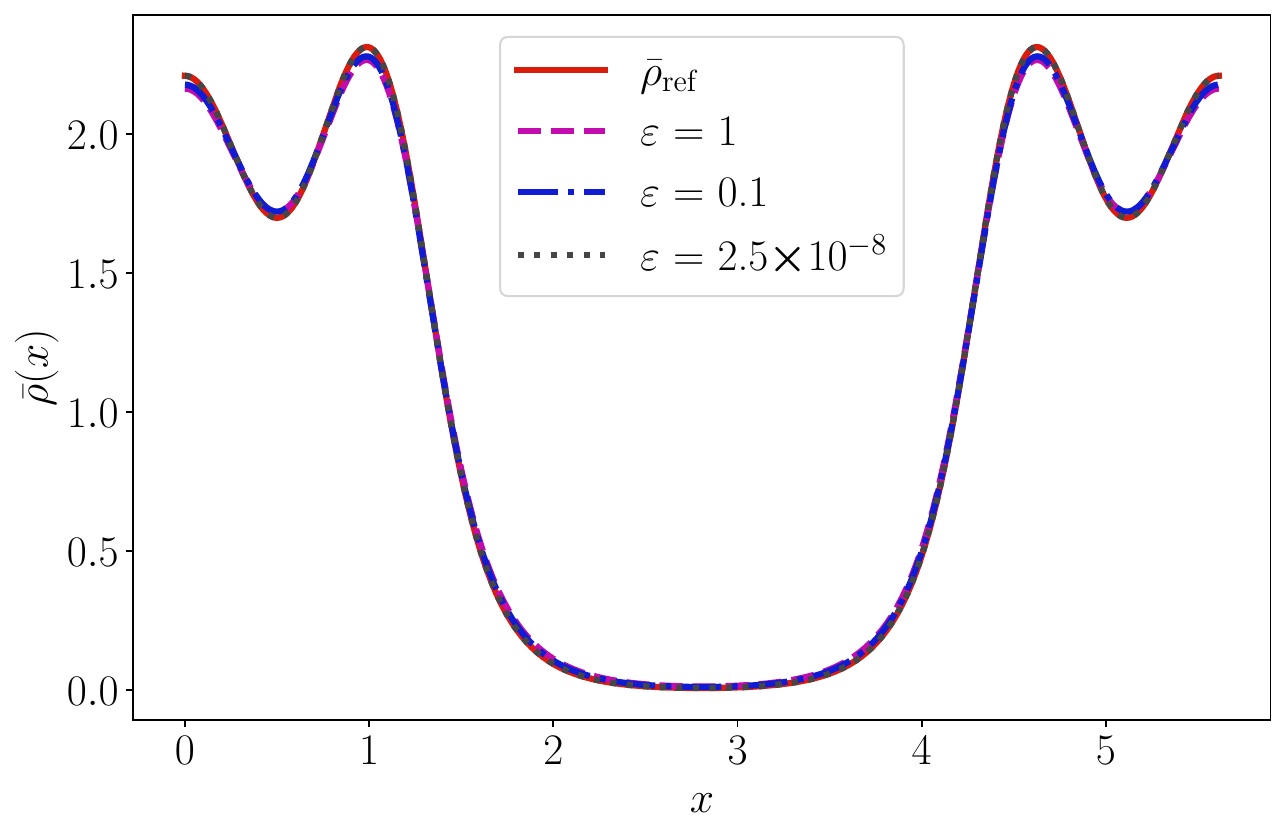}
            \label{subfig:rho_x_X_N6}
    \end{subfigure} 	
    \begin{subfigure}[b]{0.75\linewidth}
            \centering
            \caption{}
            \includegraphics[width=0.98\linewidth]{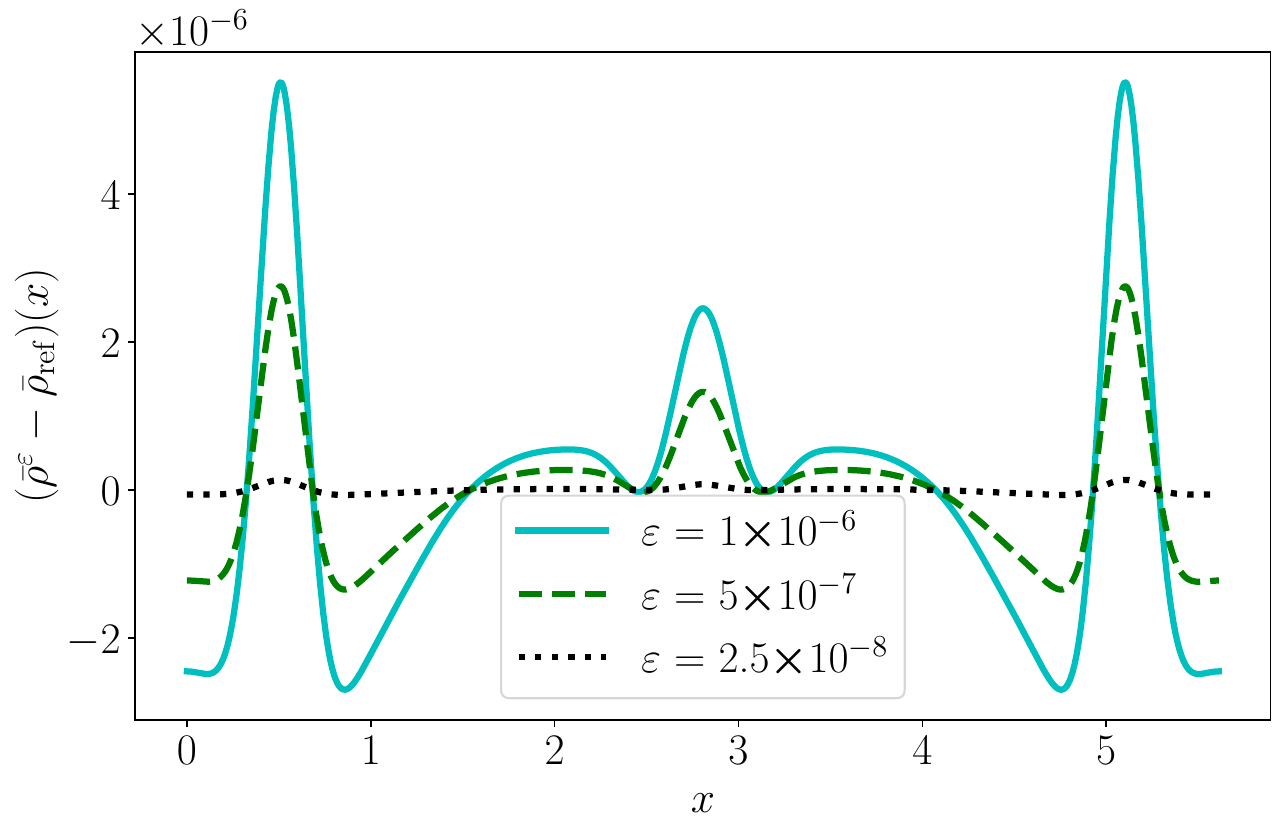}
            \label{subfig:rho_err_x_X_N6}
    \end{subfigure} 	
    \begin{subfigure}[b]{0.75\linewidth}
            \centering
            \caption{}
            \includegraphics[width=0.98\linewidth]{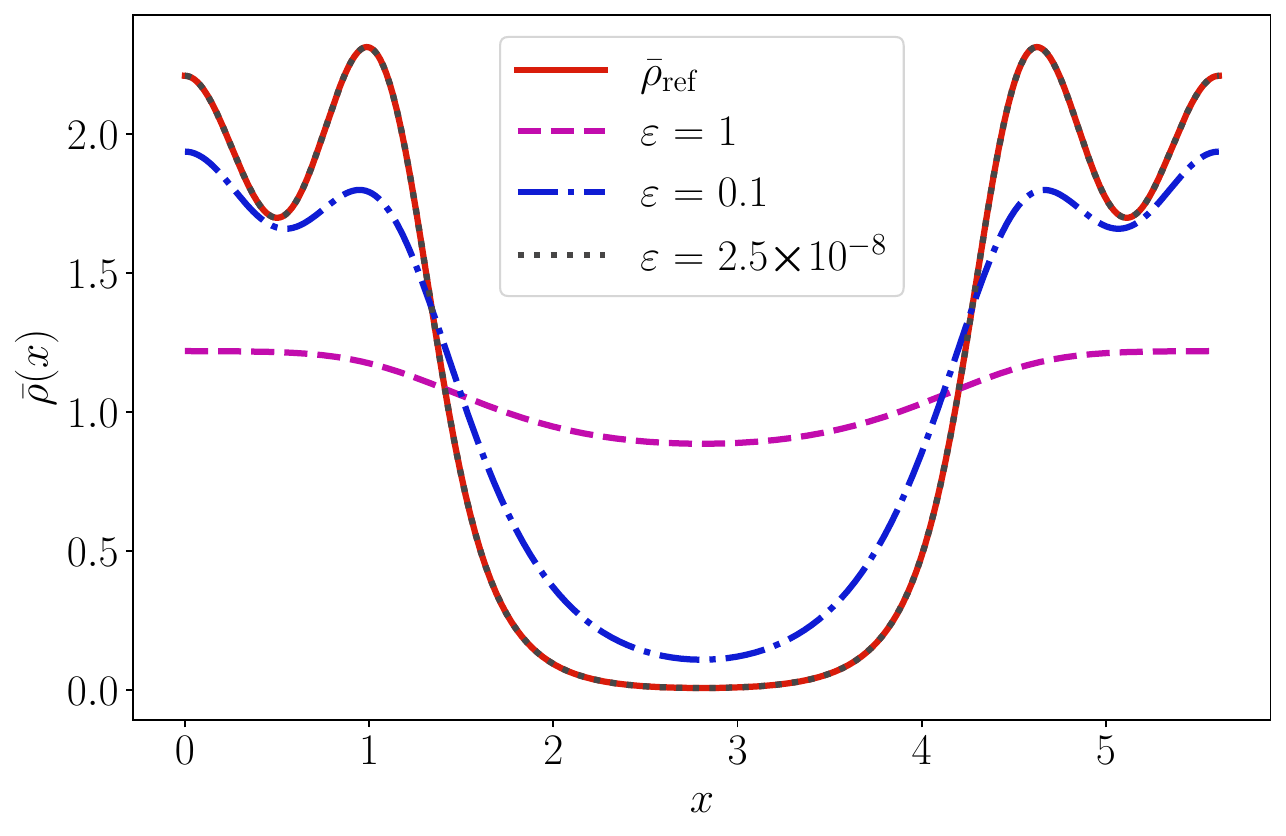}
            \label{subfig:rho_x_vHX_N6}
    \end{subfigure}
    \begin{subfigure}[b]{0.75\linewidth}
            \centering
            \caption{}
            \includegraphics[width=0.98\linewidth]{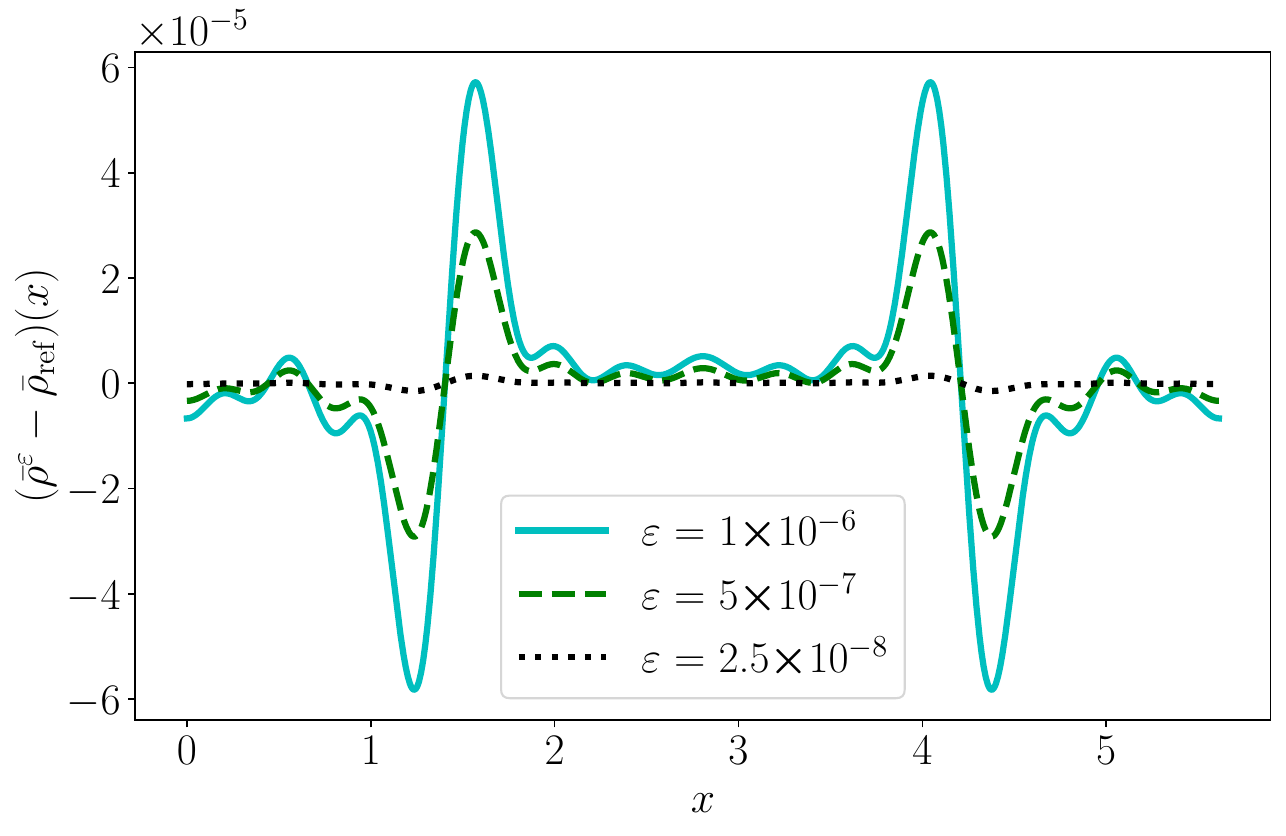}
            \label{subfig:rho_err_x_vHX_N6}
    \end{subfigure}
    \caption{\label{fig:dens_plots_and_dens_err_N6} Panels (a) and (c) show the the reference density and proximal densities as functions of position within a unit cell. The model functional with $\alpha=\xi=1$ (a) and $\alpha=\xi=0$ (c) was used. Panels (b) and (d) show the corresponding pointwise deviations from the reference density for a different set of regularisation parameter values.} 
\end{figure}

As a second numerical example, we choose $v_{\mathrm{ext}}$ to be the confining potential in Fig.~\ref{fig:vext_and_rho_twin_N6}. Using the resulting $n=6$ electron Hartree--Fock ground-state density as the reference density, $\bar{\rho}_{\mathrm{ref}}$, which is also displayed in Fig.~\ref{fig:vext_and_rho_twin_N6}, we investigated the convergence of $\bar{\rho}^{\varepsilon} \to \bar{\rho}_{\mathrm{ref}}$ for four different parameter values $\alpha,\xi \in \{0,1\}$. Again, as seen in Fig.~\ref{fig:rho_norm_err_lamb_seq_N6}, the inclusion of the external potential in the model functional makes a dramatic difference, reducing the density error by an order of magnitude for a given value of $\varepsilon$. By contrast, the inclusion of the Hartree potential has no visible effects except for the largest values of $\varepsilon$. Several resulting proximal densities $\bar{\rho}^{\varepsilon}$ are shown in Fig.~\ref{fig:dens_plots_and_dens_err_N6}: When both the external and Hartree potentials are included in the model functional ($\alpha=\xi=1$), the agreement with the reference density is excellent already for the very large $\varepsilon=1$ (see Fig.~\ref{subfig:rho_x_X_N6}). When the external and the Hartree potential are excluded ($\alpha=\xi=0$), the large value $\varepsilon$ results in very poor reproduction of the reference density as seen in Fig.~\ref{subfig:rho_x_vHX_N6}.

Similar to the first example, $\varepsilon = 10^{-6}$ yields pointwise density errors on the order of $10^{-6}$ and $10^{-5}$ (see Figs.~\ref{subfig:rho_err_x_X_N6} and \ref{subfig:rho_err_x_vHX_N6}) when the external potential is included and excluded, respectively. In the latter case the density error also has a more oscillatory character. Hence also this external potential introduces modes in the ground-state density with relatively large wave vectors that are challenging to recover with the MY-type penalty term alone. However, for sufficiently small $\varepsilon$ our results illustrate that even these modes are eventually recovered. 

\begin{figure}[h]
        \begin{subfigure}[b]{0.49\linewidth}
                \centering
                \caption{}
                \includegraphics[width=0.98\linewidth]{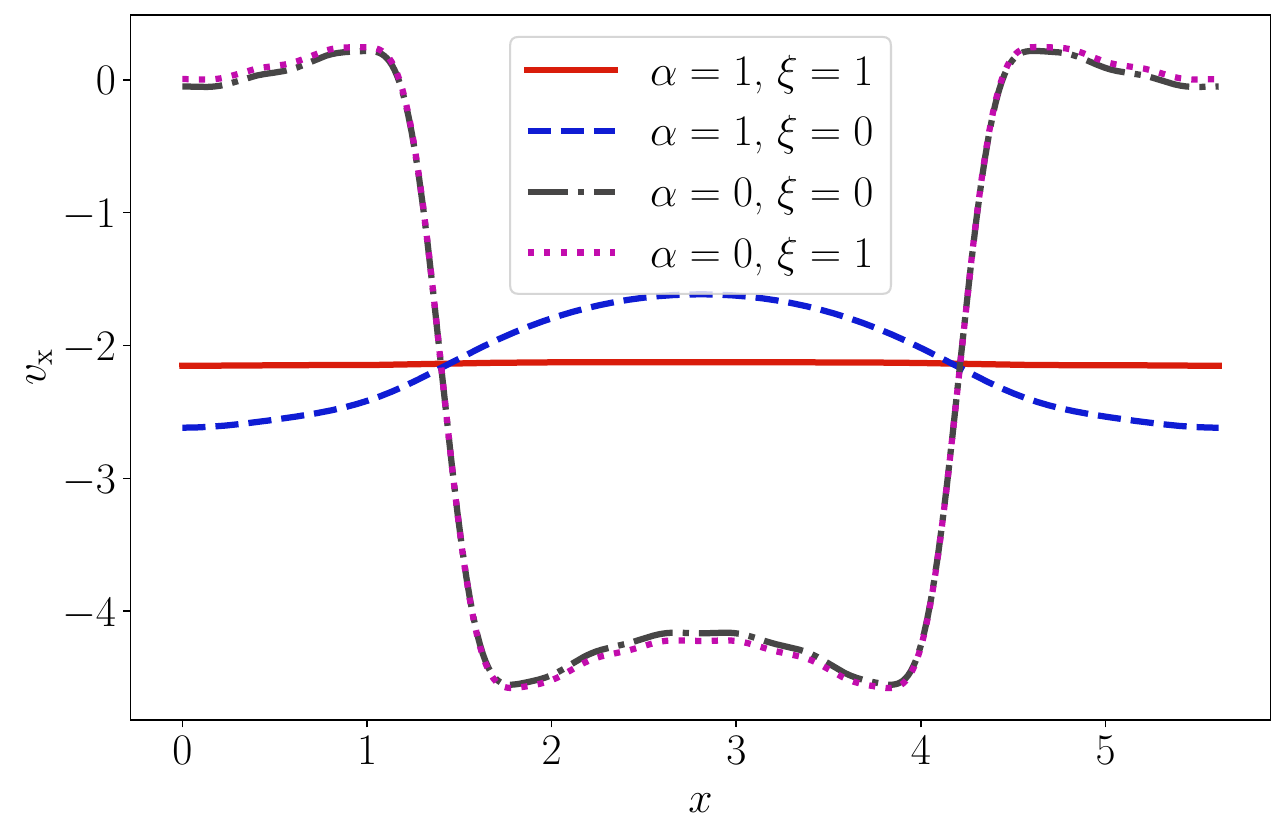}
                \label{subfig:vx_conv_00_N6} 
        \end{subfigure} 	
        \begin{subfigure}[b]{0.49\linewidth}
                \centering
                \caption{}
                \includegraphics[width=0.98\linewidth]{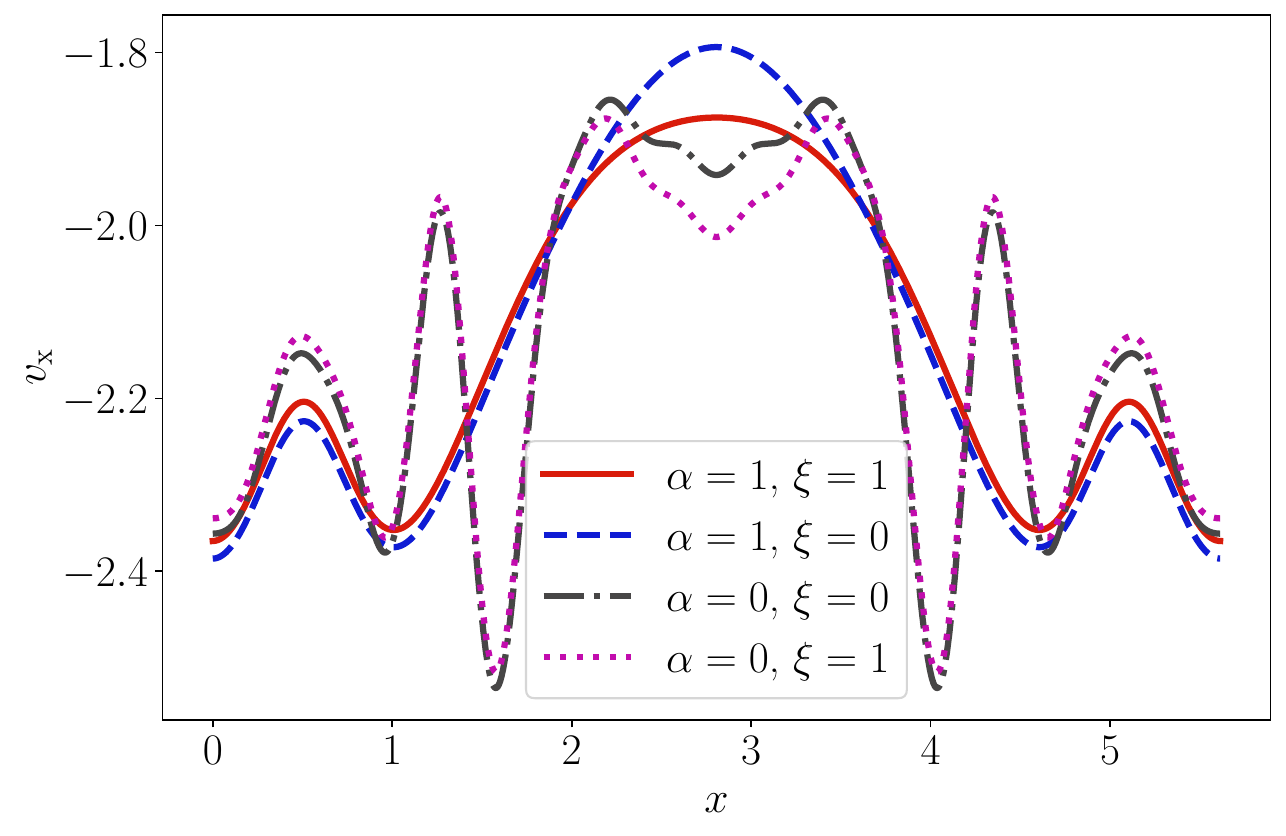}
                \label{subfig:vx_conv_04_N6} 
        \end{subfigure}
        \begin{subfigure}[b]{0.49\linewidth}
                \centering
                \caption{}
                \includegraphics[width=0.98\linewidth]{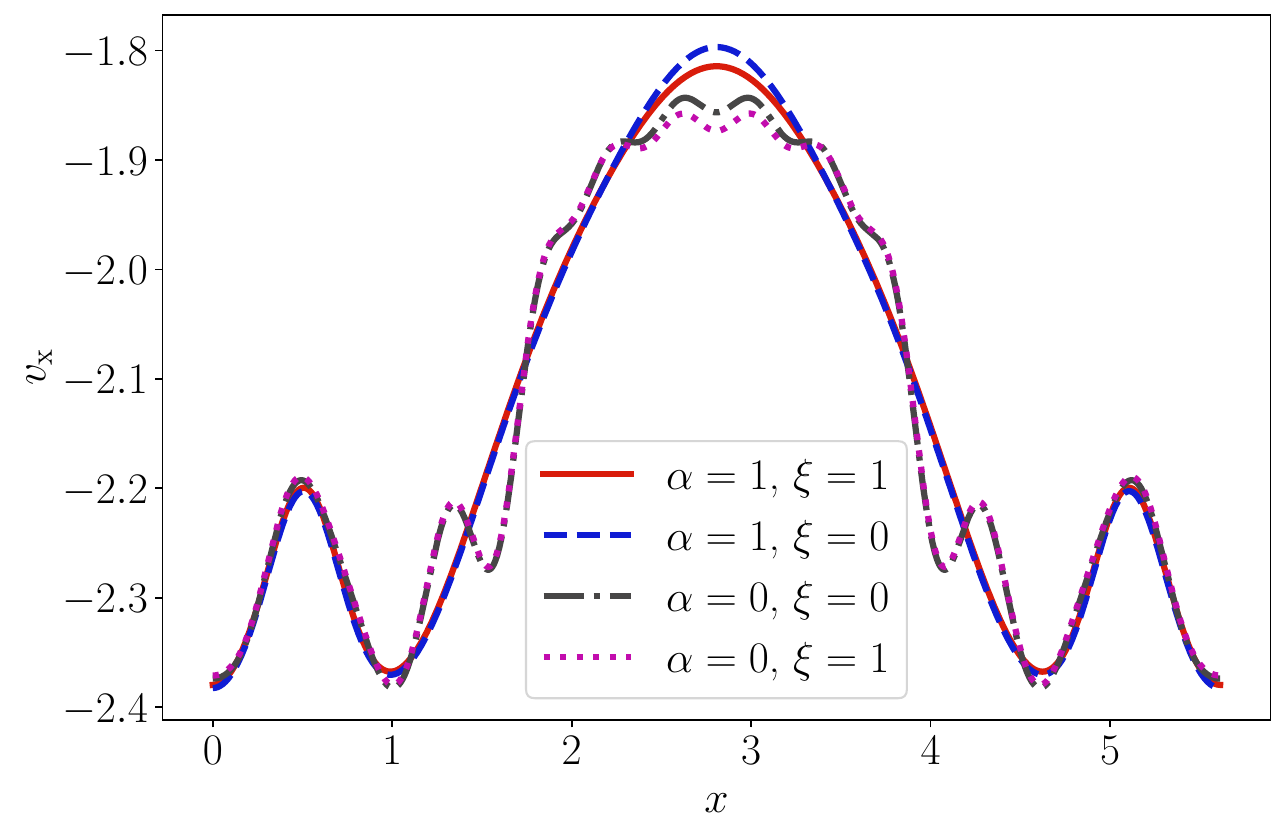}
                \label{subfig:vx_conv_06_N6} 
        \end{subfigure}
        \begin{subfigure}[b]{0.49\linewidth}
                \centering
                \caption{}
                \includegraphics[width=0.98\linewidth]{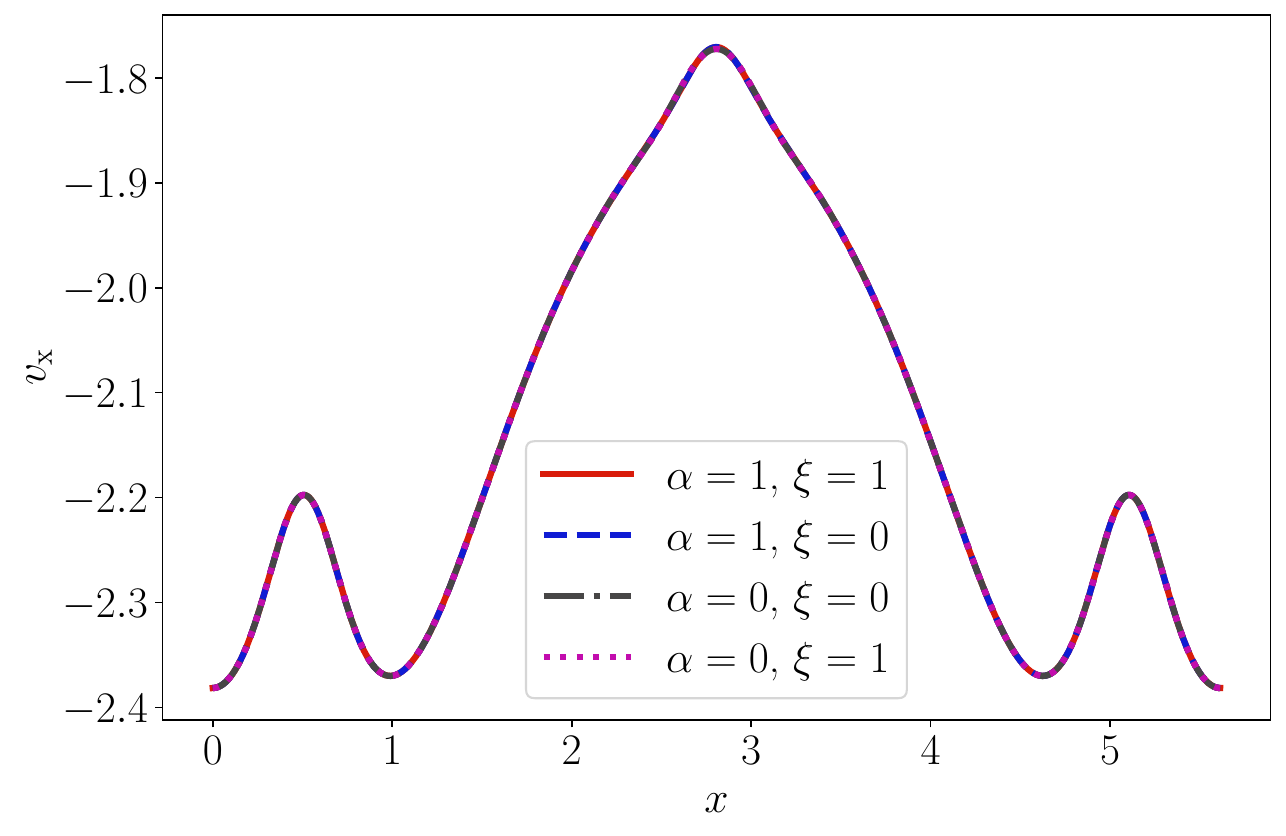}
                \label{subfig:vx_conv_07_N6} 
        \end{subfigure}
        \caption{\label{fig:vx_conv_N6} The local exact exchange potential as a function of position in the unit cell for different model functionals $F^{\mathrm{mod}}_{\alpha,\xi}$. The values of the regularisation parameter are (a) $\varepsilon=1$, (b) $\varepsilon = 9.8\times 10^{-4}$, (c) $\varepsilon = 9.8\times 10^{-5}$, and (d) $\varepsilon = 10^{-7}$. The last $\varepsilon $ value corresponds to the exact exchange potential being numerically converged for all the choice of $\alpha$ and $\xi$.}       
\end{figure}

The convergence of the computed local exchange potentials $v_{\mathrm{x}}^{\varepsilon}$ are exhibited in Fig.~\ref{fig:vx_conv_N6}. The converged potentials in the last panel show that the local exchange mostly reflects the confinement (see Fig.~\ref{fig:vext_and_rho_twin_N6}) in that $v_{\mathrm{x}}$ is minimal where the reference density $\bar{\rho}_{\mathrm{ref}}$ is maximal.


\subsubsection{Dependence on the number of electrons}
  
Next, we consider the variation with the number of electrons. As the external potential was constructed to give a potential energy $\pairing{v_{\mathrm{ext}}}{\rho}$ of similar magnitude as the Hartree energy for a $n=10$ electron system, we scale the external potential proportionally to $n$. Hence, both $v_{\mathrm{ext}}$ and $\bar{\rho}_{\mathrm{ref}}$ vary with $n$. Results are shown in Fig.~\ref{subfig:rho_norm_err_N_seq}, where the convergence of $\bar{\rho}^{\varepsilon}$ to $\bar{\rho}_{\mathrm{ref}}$ is apparent. Larger $n$ require substantially smaller $\varepsilon$ to reach a given accuracy. At a fixed $\varepsilon$, the calculation with $n=2$ electrons achieves an accuracy that is two orders of magnitude better than for $n=100$ electrons per cell. This is likely due to the penalty term having a progressively smaller effect higher up in the Kohn--Sham spectrum.

\subsubsection{Dependence on the Yukawa screening parameter}
\label{subsec:Yukawa_screen_param}

\begin{figure}[h]
    \includegraphics[width=0.45\textwidth]{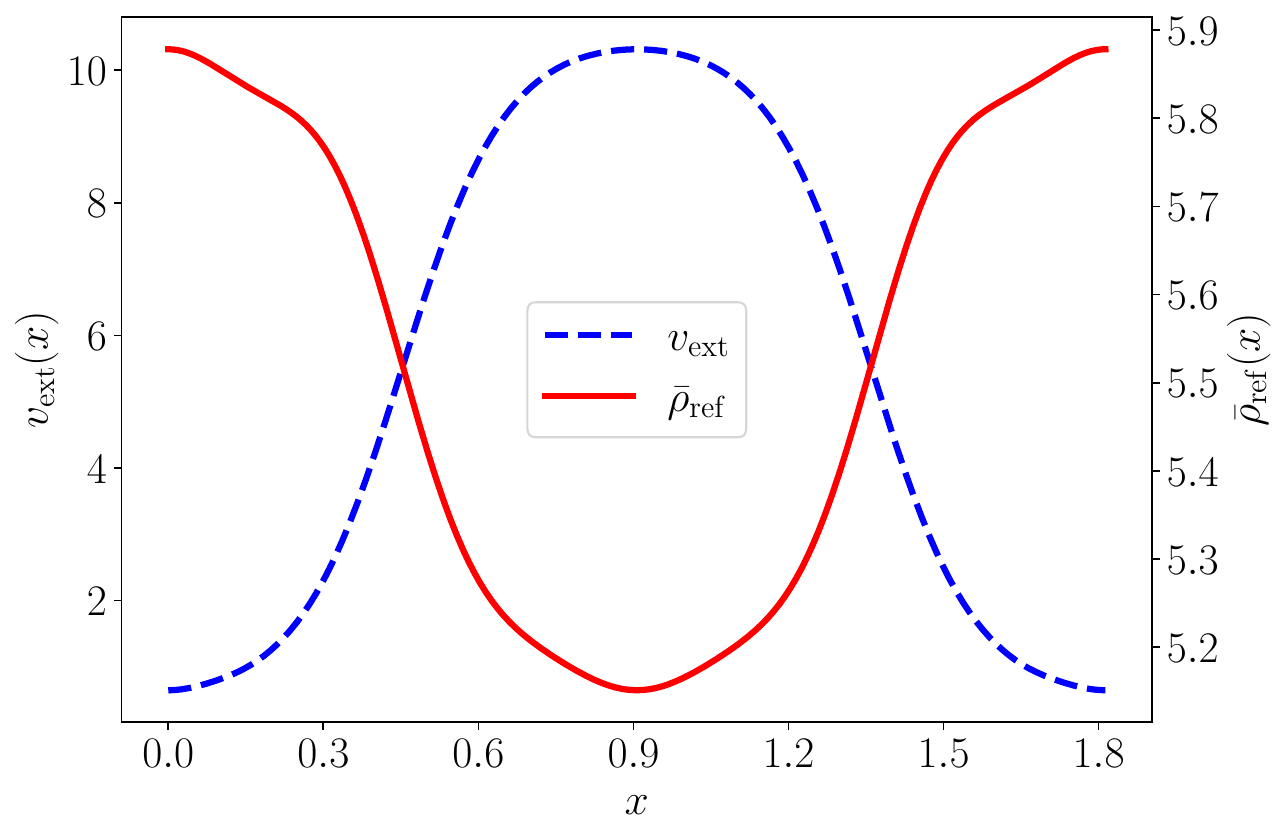}     \caption{\label{fig:vext_and_rho_twin_gamma_seq} An external potential $v_{\mathrm{ext}}$ and the corresponding Hartree--Fock ground-state density, $\rhoref$, with $n=10$ electrons. Note the twin vertical axis.}
\end{figure}

To study the the variation with the Yukawa screening parameter, we fix $n=10$ electrons per cell and a lattice constant of $a=1.813$. The reference density was fixed to the Hartree--Fock ground-state density in the external potential, as shown in Fig.~\ref{fig:vext_and_rho_twin_gamma_seq}, and with electron-electron interactions corresponding to $\gamma=1$. Then, for this fixed reference density, the density-potential inversion was performed with values $\gamma=0.1,0.5,1,5,10$ for the screening parameter. For the median value, $\gamma=1$, the Yukawa force is substantial also in the nearest neighbor cell, but only $e^{-\gamma a} \approx 3\%$ in the second-nearest neighbor cell. For the smallest $\gamma$ value, the Yukawa force extends over several unit cells and for the largest value, the force is negligible even between nearest neighbors.

\begin{figure}[h]
        \begin{subfigure}[b]{0.9\linewidth}
                \centering
                \caption{}
                \includegraphics[width=0.98\linewidth]{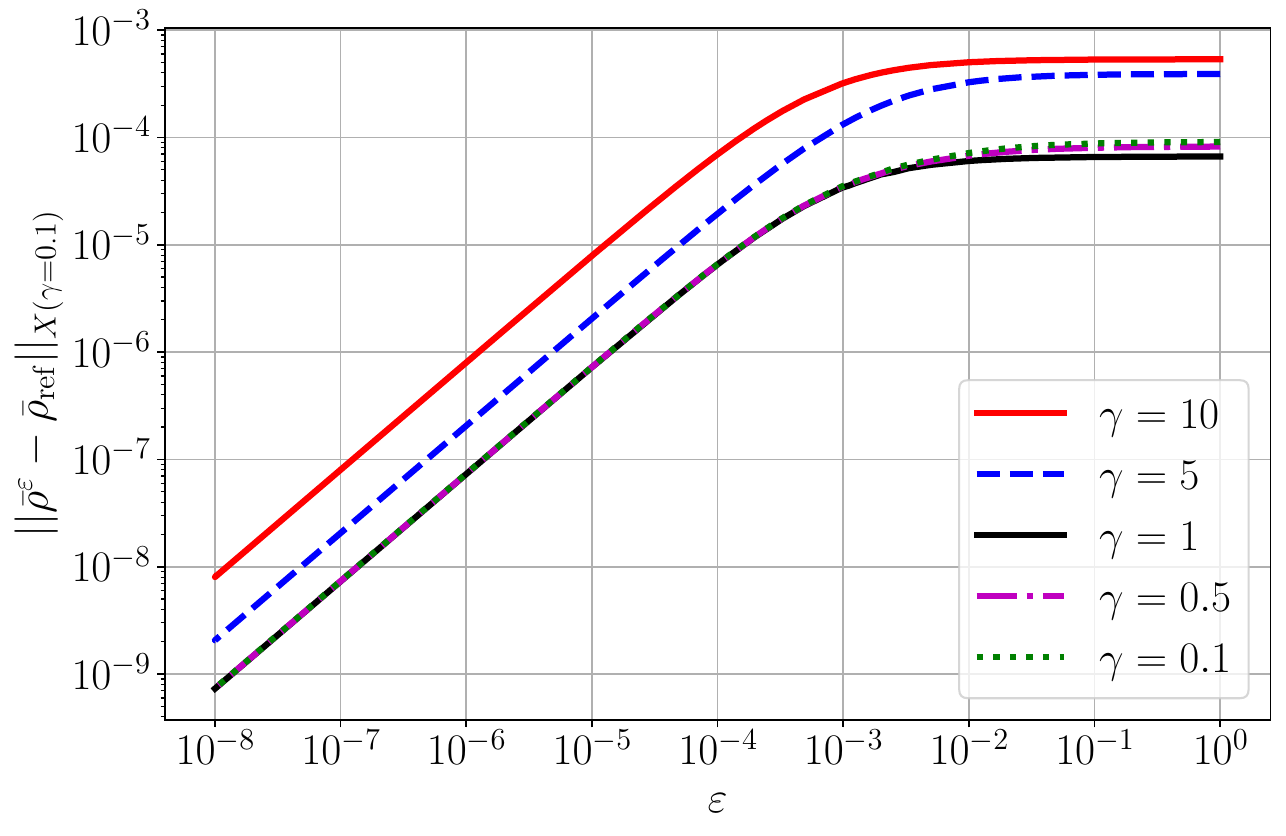}
                \label{subfig:G_norm_dens_err_X01} 
        \end{subfigure} 	
        \begin{subfigure}[b]{0.9\linewidth}
                \centering
                \caption{}
                \includegraphics[width=0.98\linewidth]{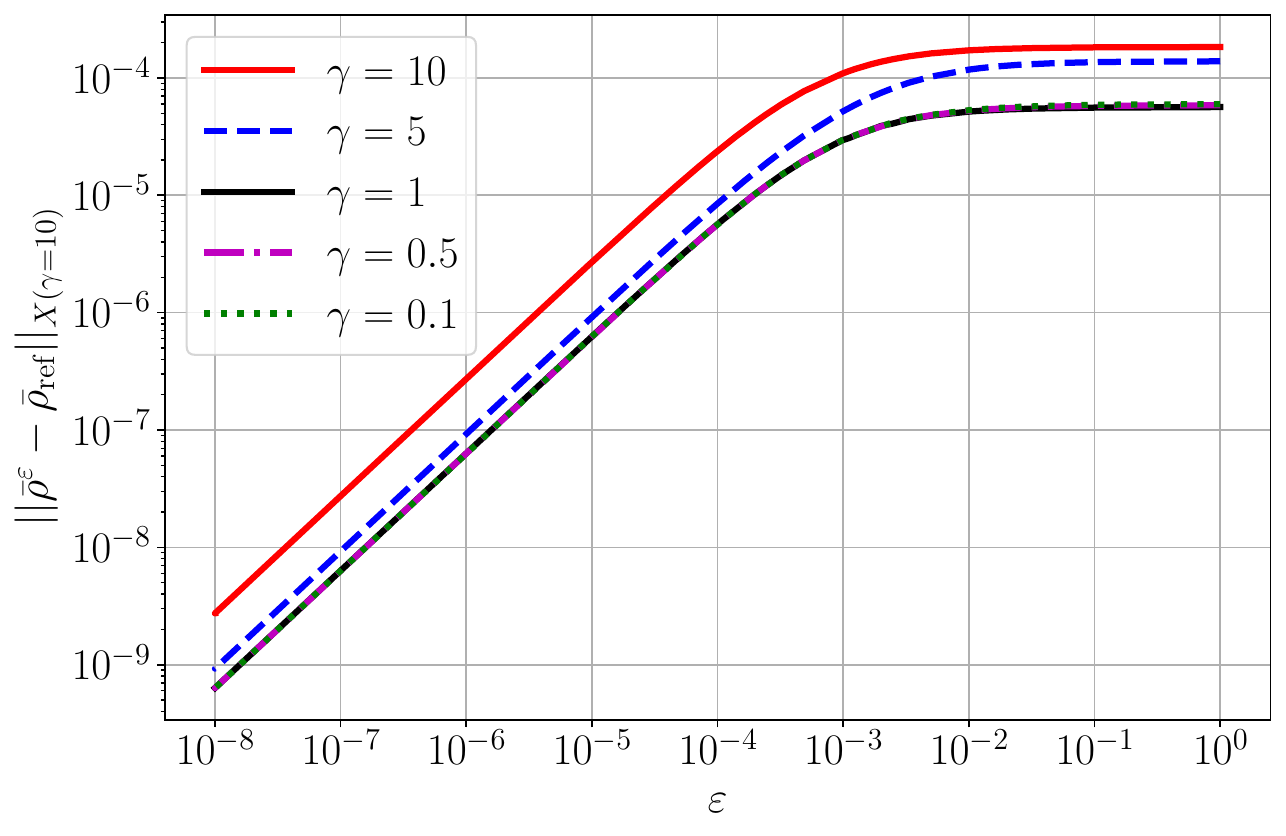}
                \label{subfig:G_norm_dens_err_X10} 
        \end{subfigure}
        \begin{subfigure}[b]{0.9\linewidth}
                \centering
                \caption{}
                \includegraphics[width=0.98\linewidth]{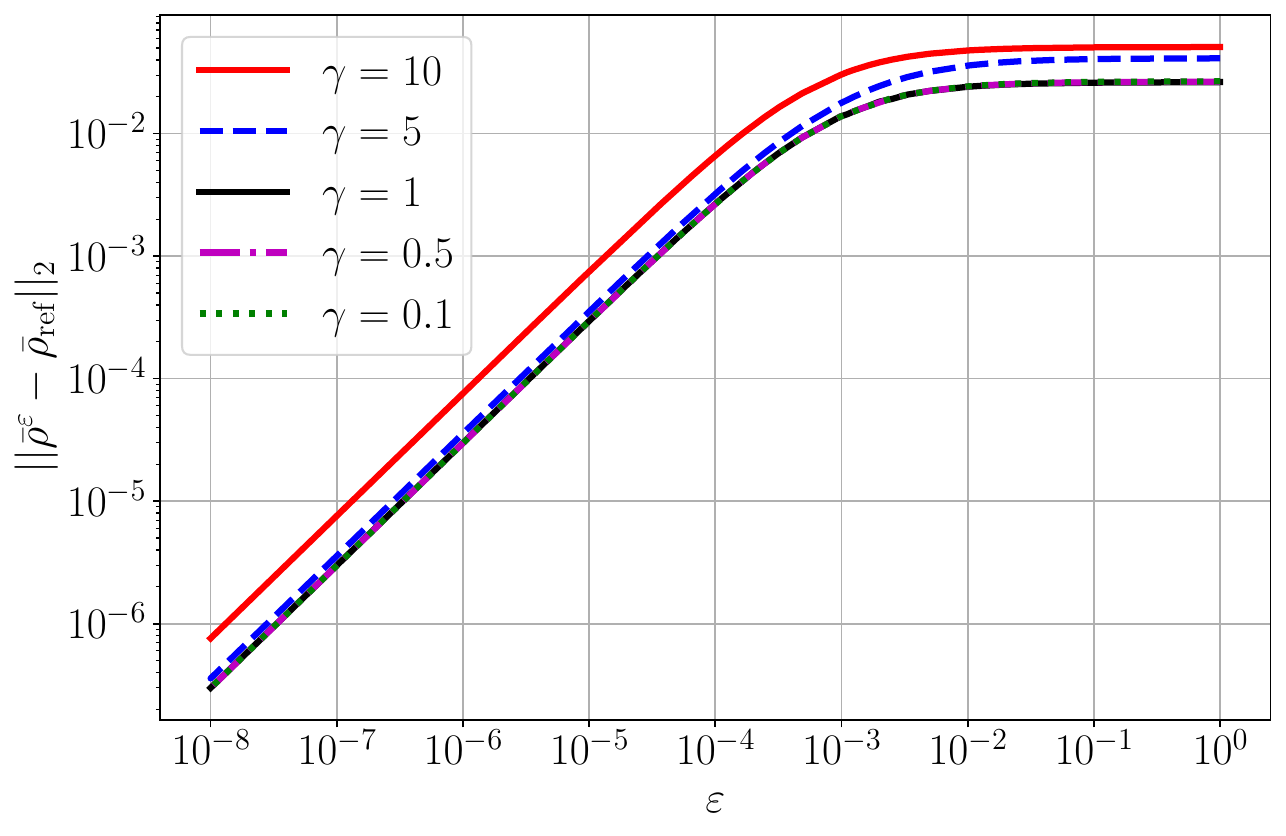}
                \label{subfig:G_norm_dens_err_L2} 
        \end{subfigure}
        \caption{\label{fig:G_norm_err} Distances measured with three different norms between proximal densities and a reference density. The different curves correspond to different values of the Yukawa parameters used in the optimisation of the proximal density. Post-optimisation, the distances were measured using (a) the Sobolev norm with $\gamma = 0.1$, (b) the Sobolev norm with $\gamma=10$, and (c) the $L^2$ norm.}
\end{figure}

For each $\gamma$ value, we performed a MY-based density-potential inversion for a range of $\varepsilon$ values. Since the function space of densities and the proximal density depend on the screening parameter, we denote them $X(\gamma)$ and $\bar{\rho}^{\epsilon}(\gamma)$, respectively. Then we measured the distance $\|\bar{\rho}^{\varepsilon}(\gamma)-\bar{\rho}_{\mathrm{ref}}\|_{X(\gamma_{\mathrm{meas}})}$ using the norm corresponding the largest and smallest values $\gamma_{\mathrm{meas}} \in \{0.1,10\}$. This illustrates the effect of measuring the density error using a different parameter value than the one the density $\bar{\rho}^{\varepsilon}$ was optimised for. Additionally, we also measured the distance using the standard $L^2$ norm. The results are shown in Fig.~\ref{fig:G_norm_err}. Intriguingly, optimising using the smallest value, $\gamma=0.1$, always yields the best fit, even when measured post-optimisation using the largest value, $\gamma_{\mathrm{meas}} = 10$ as seen in Fig.~\ref{subfig:G_norm_dens_err_X10}.

\subsection{Behavior for non-$N$-representable densities}
\label{secNonNRep}

\begin{figure}[h]
        \begin{subfigure}[b]{0.9\linewidth}
                \centering
                \caption{}
                \includegraphics[width=0.98\linewidth]{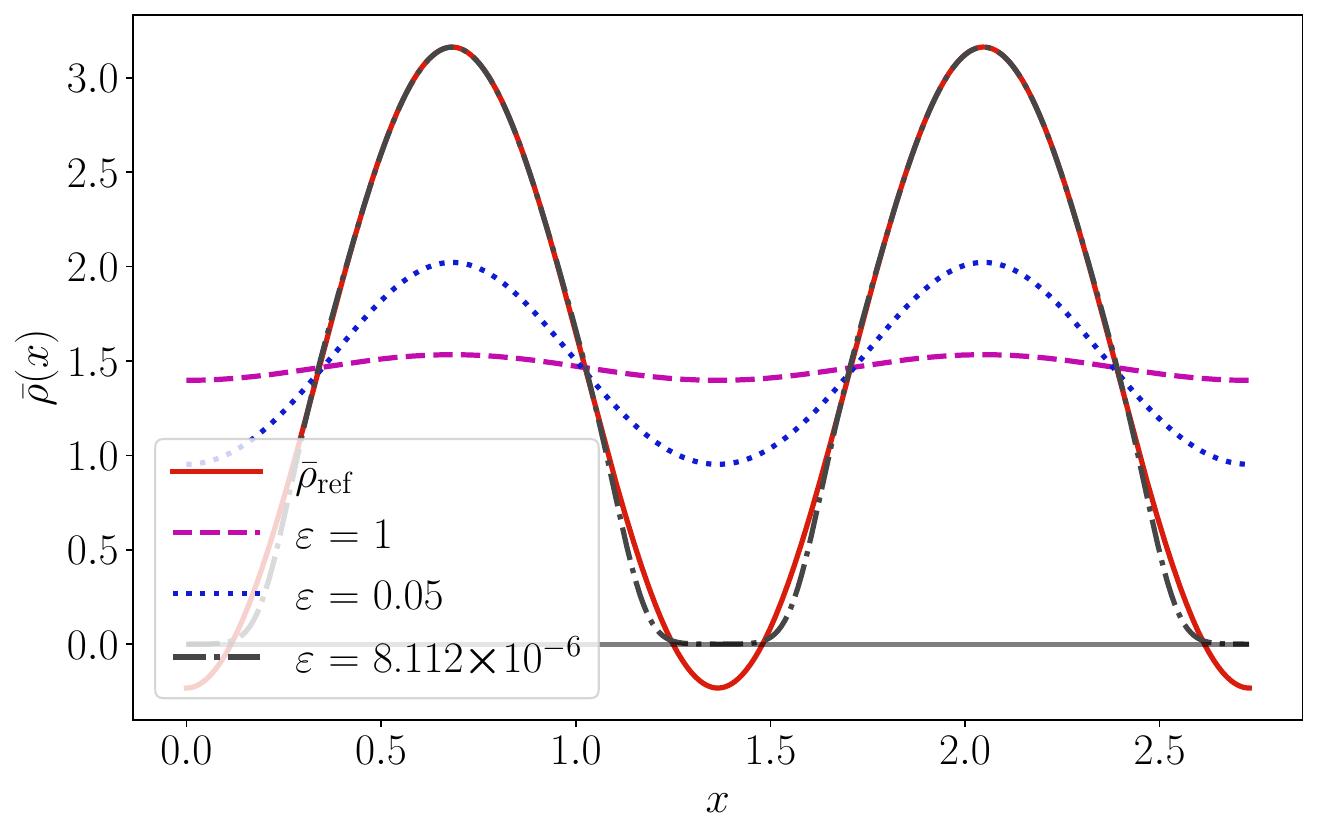}
                \label{subfig:dens_non_N_rep}
        \end{subfigure} 	
        \begin{subfigure}[b]{0.9\linewidth}
                \centering
                \caption{}
                \includegraphics[width=0.98\linewidth]{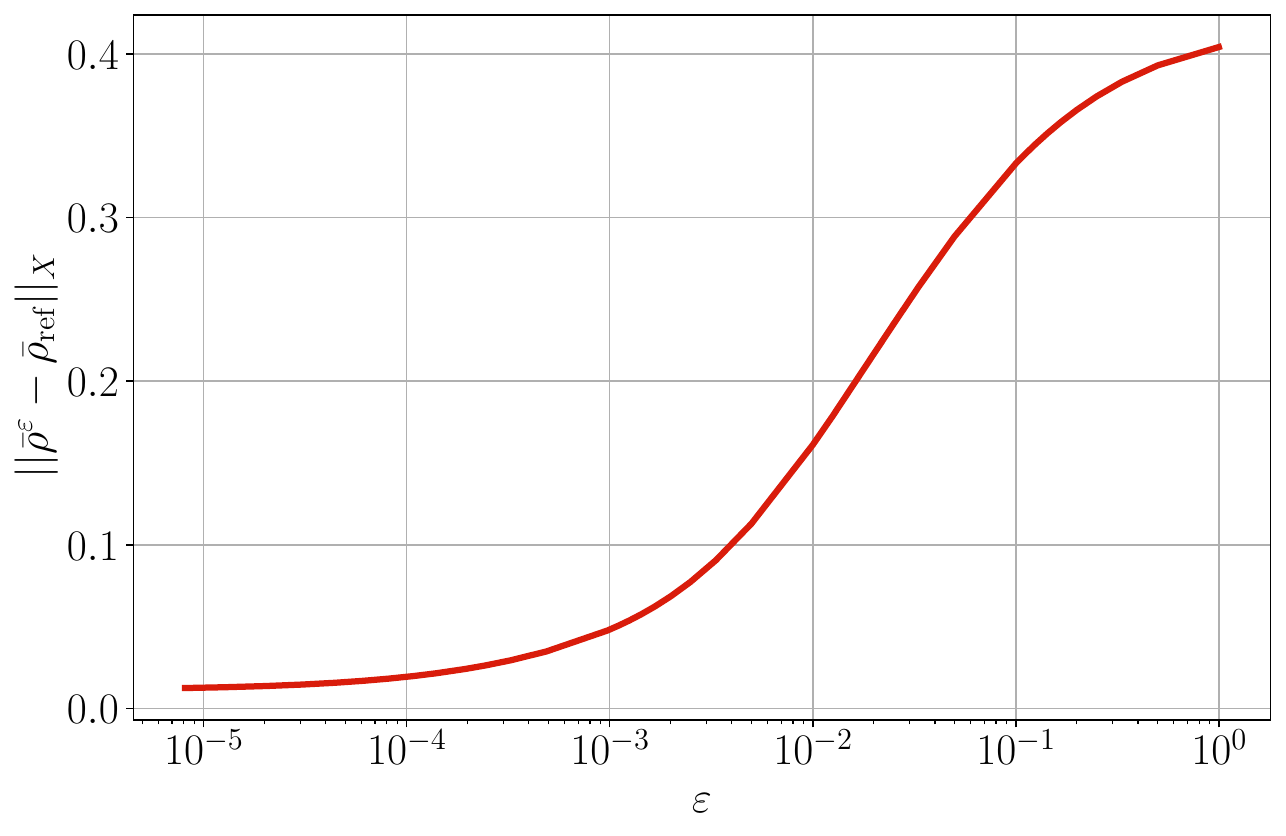}
                \label{subfig:dens_norm_non_N_rep}
        \end{subfigure}        \caption{\label{fig:dens_non_N_rep} Panel (a) shows a a non-$N$-representable reference density and three proximal densities. Panel (b) shows the distance between the proximal density and the reference density as a function of the regularisation parameter $\varepsilon$.}
\end{figure}

Our MY regularisation approach works whether or not the reference density is $N$-representable density.
By tuning the Fourier components to produce a region of a negative density, we constructed a manifestly non-$N$-representable reference density.
We then performed density-potential inversion calculations with $n=4$ electrons per cell for a series of $\varepsilon$ values, were $a=2.73$. Neither the external potential nor the Hartree term contributed to the model functional used, i.e.\ we set $\alpha=\xi=0$ in these calculations.
The resulting sequence of densities is shown together with the reference density in Fig.~\ref{subfig:dens_non_N_rep}. 
This illustrates graphically that the MY-based inversion method in this case produces the best $N$-representable approximation (in some least squares sense) to the non-$N$-representable reference density.
In Fig.~\eqref{subfig:dens_norm_non_N_rep} we show the corresponding behavior of the distance $\|\bar{\rho}^{\varepsilon} - \bar{\rho}_{\mathrm{ref}}\|_X$, which is seen to plateau and tend to a non-zero constant as $\varepsilon \to 0$.
This is the natural consequence of there not existing any nearby $N$-representable density to our constructed reference density.

\subsection{Error analysis}
\label{secErrorAnalysis}

In this section we study the propagation of errors, or perturbations, from the reference density to the proximal density. For simplicity, we focus on the first system discussed in Sec.~\ref{sec:AnsatzDep} above. To be able to reach SCF convergence with as small $\varepsilon$ as possible, we use the model functional that includes both the external potential and the Hartree energy. As a simple, deterministic way of generating perturbed reference densities $\tilde{\rho}_{\mathrm{ref}}$, we zero out Fourier coefficients for the largest wave vectors,
\begin{equation}
    \hat{\tilde{\rho}}_{\mathrm{ref}} (\myvec{G}) = \begin{cases} \hat{\bar{\rho}}_{\mathrm{ref}}(\myvec{G}), & \text{if}\ |\myvec{G}| \leq G_{\mathrm{trunc}}, \\
      0, & \text{otherwise},
      \end{cases}
\end{equation}    
where $G_{\mathrm{trunc}}$ is chosen large enough to give a reasonable order of magnitude of $\delta\bar{\rho} := \tilde{\rho}_{\mathrm{ref}} - \bar{\rho}_{\mathrm{ref}}$. Next we denote by $\bar{\rho}^{\varepsilon}$ and $\tilde{\rho}^{\varepsilon}$ the proximal densities determined with the unperturbed $\bar{\rho}_{\mathrm{ref}}$ and the perturbed density $\tilde{\rho}_{\mathrm{ref}}$, respectively, as the reference density.

\begin{figure}[h]
    \begin{subfigure}[b]{0.9\linewidth}
            \centering
            \caption{}
            \includegraphics[width=0.98\linewidth]{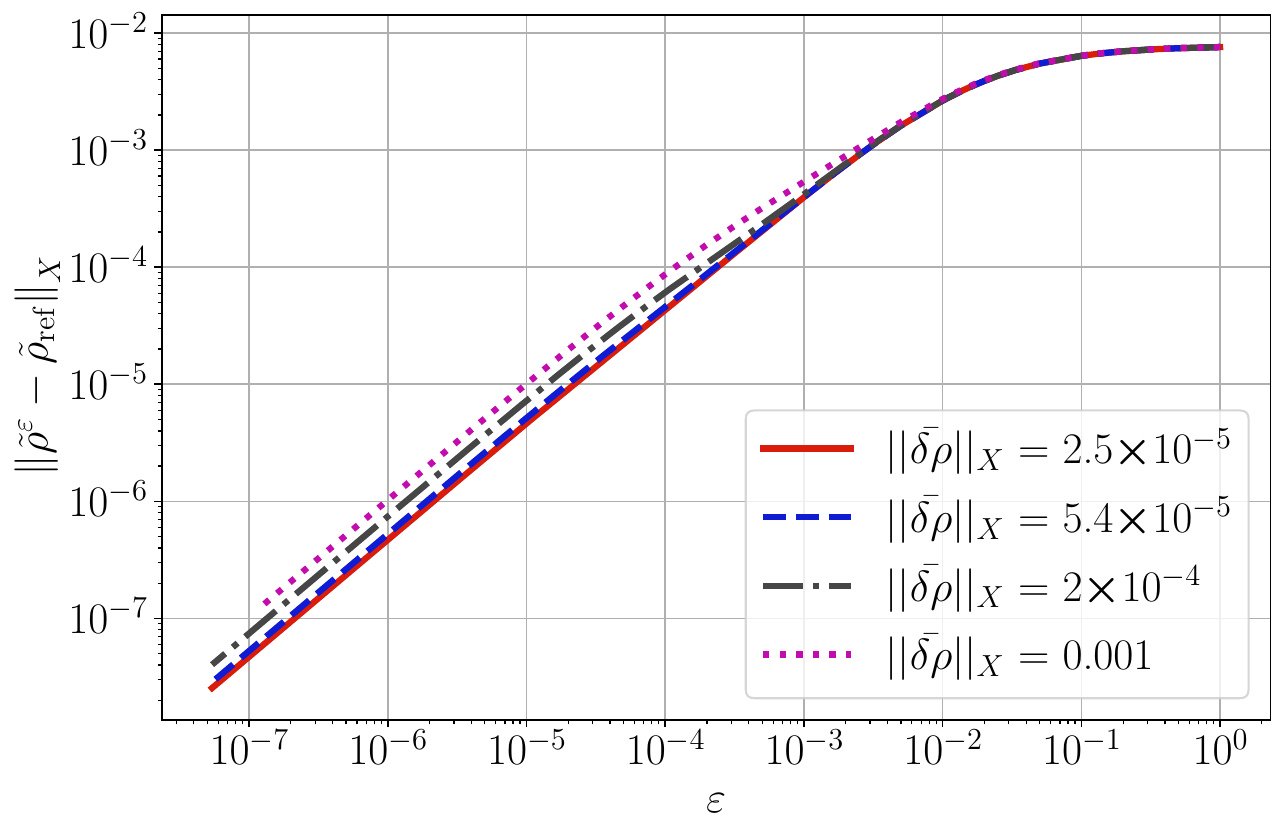}
            \label{subfig:norm_dens_err_with_same_eps} 
    \end{subfigure}
    \begin{subfigure}[b]{0.9\linewidth}
            \centering
            \caption{}
            \includegraphics[width=0.98\linewidth]{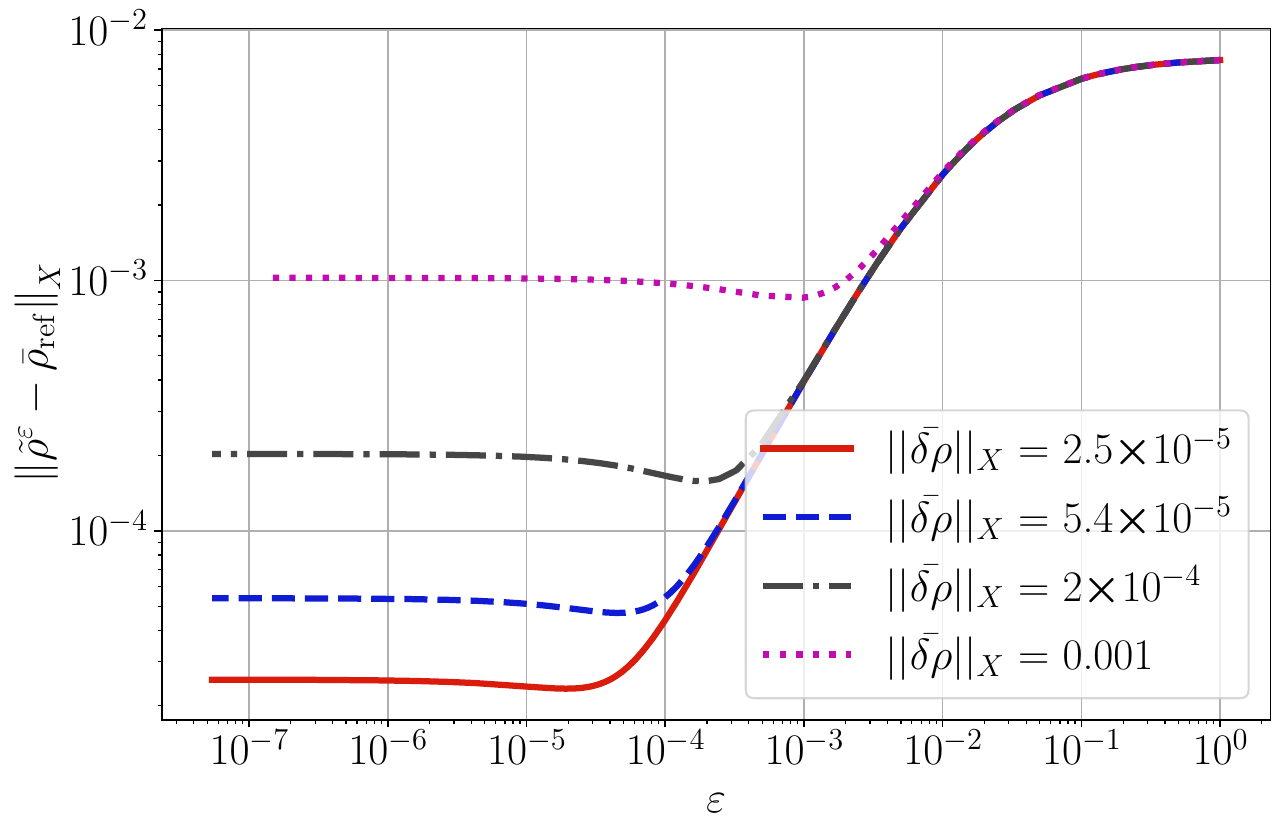}
            \label{subfig:norm_dens_err_with_non_trunc_dens_ref} 
    \end{subfigure} 	    
        \caption{\label{fig:dens_err_analy_norm_err} The top panel (a) shows the distance between the perturbed proximal density and the perturbed reference density. The bottom panel (b) shows the distance between the perturbed proximal density and the unperturbed reference density.}
\end{figure}

For the perturbed reference densities, Fig.~\ref{subfig:norm_dens_err_with_same_eps} illustrates the convergence $\tilde{\rho}^{\varepsilon} \to \tilde{\rho}_{\mathrm{ref}}$ as $\varepsilon \to 0$. The perturbations are sufficiently small that these curves are nearly identical to the unperturbed curves in Fig.~\ref{subfig:rho_norm_err_lamb_seq}. In both these figures, the slope tends asymptotically to $+1$ in the limit of small $\varepsilon$, i.e.\ $\|\bar{\rho}^{\varepsilon} - \bar{\rho}_{\mathrm{ref}}\|_X$ and $\|\tilde{\rho}^{\varepsilon} - \tilde{\rho}_{\mathrm{ref}}\|_X$ are on the order of $\varepsilon$.  In the opposite limit of large $\varepsilon$, we have $u^{\varepsilon} \to 0$ and the proximal densities $\bar{\rho}^{\varepsilon}$ and $\tilde{\rho}^{\varepsilon}$ both tend to the noninteracting ground-state density $\bar{\rho}^{\infty}$ of the potential $\alpha v_{\mathrm{ext}} + \xi v_{\mathrm{H}}$, which is also reflected in the hint of an asymptotic plateau in Fig.~\ref{subfig:rho_norm_err_lamb_seq} with value $\|\bar{\rho}^{\infty}-\tilde{\rho}_{\mathrm{ref}}\|_X$.

In Fig.~\ref{subfig:norm_dens_err_with_non_trunc_dens_ref}, we visualise the deviation $\|\tilde{\rho}^{\varepsilon}-\bar{\rho}_{\mathrm{ref}}\|_X$ between the perturbed proximal density and the unperturbed reference density. Using Eq.~\eqref{eqUpotFromInv}, we find
\begin{align}
   \label{eqRhoBarFromInv}
   \bar{\rho}^{\varepsilon} & = \bar{\rho}_{\mathrm{ref}} + \varepsilon \Jdual^{-1}(u^{\varepsilon}),
  \\
    \label{eqRhoTildeFromInv}
  \tilde{\rho}^{\varepsilon} & = \tilde{\rho}_{\mathrm{ref}} + \varepsilon \Jdual^{-1}(\tilde{u}^{\varepsilon})  = \bar{\rho}_{\mathrm{ref}} + \delta\bar{\rho} + \varepsilon \Jdual^{-1}(\tilde{u}^{\varepsilon})
\end{align}
and thus $\|\tilde{\rho}^{\varepsilon}-\bar{\rho}_{\mathrm{ref}}\|_X = \|\delta\bar{\rho} + \varepsilon \Jdual^{-1}(\tilde{u}^{\varepsilon})\|_X$, which tends to $\|\delta\bar{\rho}\|_X$ in the limit of vanishing $\varepsilon$. In the limit of large $\varepsilon$, we have
\begin{equation}
  \|\tilde{\rho}^{\varepsilon}-\bar{\rho}_{\mathrm{ref}}\|_X \to \|\bar{\rho}^{\infty} - \bar{\rho}_{\mathrm{ref}}\|_X \approx \|\bar{\rho}^{\infty} - \bar{\rho}_{\mathrm{ref}} - \delta\bar{\rho}\|_X.
\end{equation}
Hence, Figs.~\ref{subfig:norm_dens_err_with_same_eps} and \ref{subfig:norm_dens_err_with_non_trunc_dens_ref}, and by analogous reasoning also Fig.~\ref{subfig:rho_norm_err_lamb_seq} ($\alpha=\xi=1$), have approximately the same horizontal asymptote for large $\varepsilon$.

\begin{figure}[h]
        \begin{subfigure}[b]{0.9\linewidth}
                \centering
                \caption{}
                \includegraphics[width=0.98\linewidth]{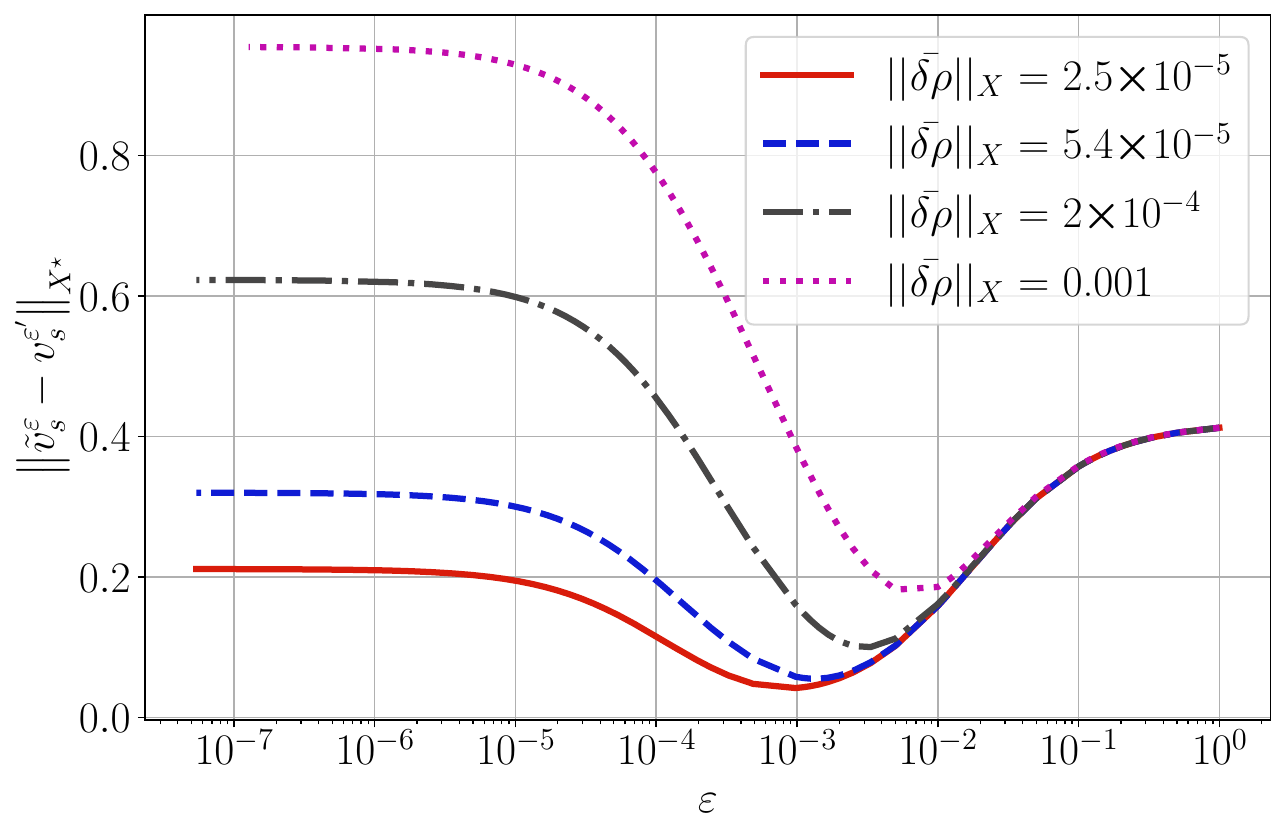}
                \label{subfig:norm_pot_err_with_best_v_eps} 
        \end{subfigure} 	
        \begin{subfigure}[b]{0.9\linewidth}
                \centering
                \caption{}
                \includegraphics[width=0.98\linewidth]{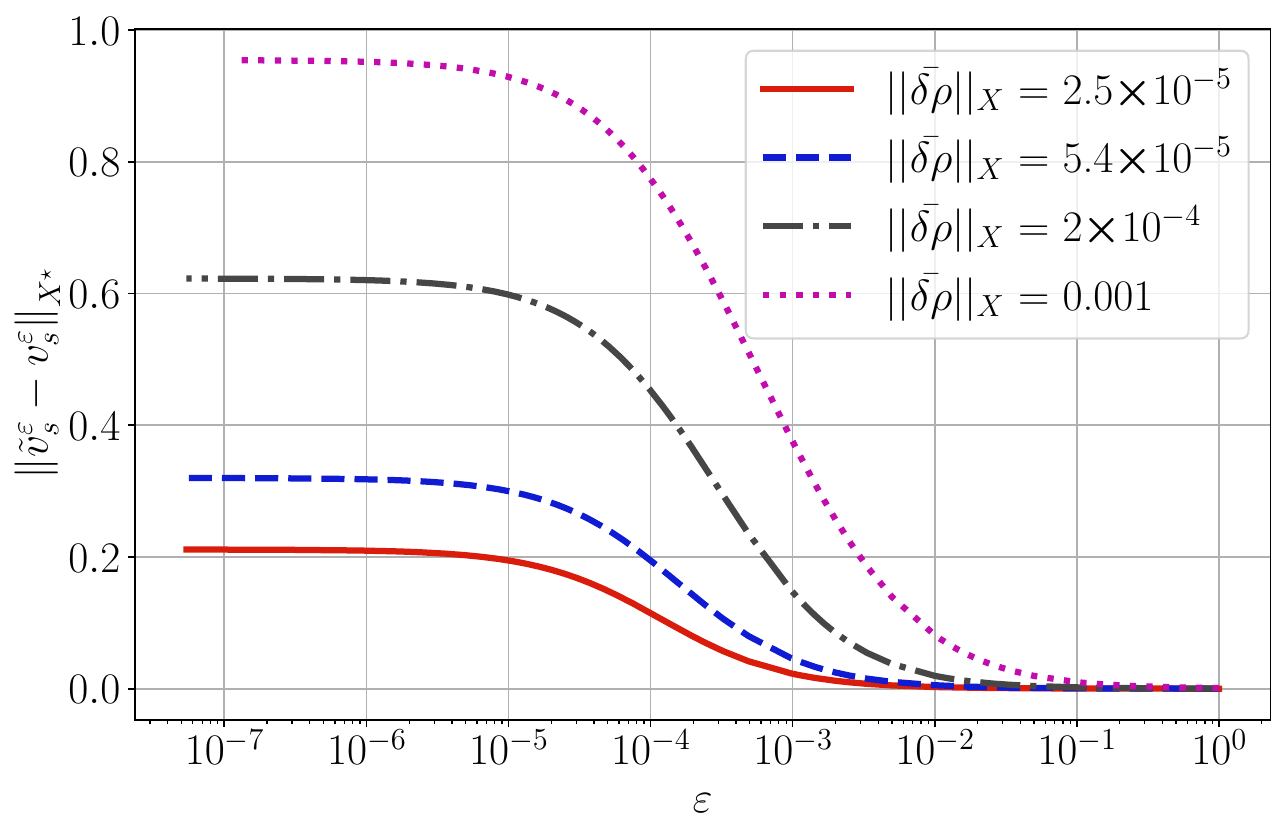}
                \label{subfig:norm_pot_err_with_same_eps} 
        \end{subfigure}        \caption{\label{fig:pot_err_analy_norm_err} The top panel (a) shows the distance between our best estimate $v^{\varepsilon'}_{\mathrm{s}}$ of the Kohn--Sham potential and estimates $v^{\varepsilon}_{\mathrm{s}}$ produced with larger regularisation parameters. The bottom panel (b) shows the distance between the perturbed and unperturbed estimated Kohn--Sham potentials.}
\end{figure}

Next we turn to the difference between the converged Kohn--Sham potential and approximations produced with larger $\varepsilon$. Using the potential $v^{\varepsilon'}_{\mathrm{s}}$ obtained with the smallest regularisation parameter ($\varepsilon' = 5.4\times 10^{-8}$ in this set of calculations) as an approximation of the $\varepsilon\to 0$ limit, we have
\begin{equation}
  \begin{split}
    \|\tilde{v}_{\mathrm{s}}^{\varepsilon} - v_{\mathrm{s}}^{\varepsilon'}\|_{X^*} & = \|\tilde{u}^{\varepsilon} - u^{\varepsilon'}\|_{X^*} \\
    & = \Big\|\frac{\Jdual(\tilde{\rho}^{\varepsilon} - \tilde{\rho}_{\mathrm{ref}})}{\varepsilon} - \frac{\Jdual(\bar{\rho}^{\varepsilon'} - \bar{\rho}_{\mathrm{ref}})}{\varepsilon'}\Big\|_{X^*} \\
    & = \Big\|\frac{\tilde{\rho}^{\varepsilon} - \tilde{\rho}_{\mathrm{ref}}}{\varepsilon} - \frac{\bar{\rho}^{\varepsilon'} - \bar{\rho}_{\mathrm{ref}}}{\varepsilon'}\Big\|_{X},
  \end{split}
\end{equation}
where we used that the duality mapping is linear for our choice of density space $X$. This quantity is visualised in Fig.~\ref{subfig:norm_pot_err_with_best_v_eps}. In the limit of vanishing $\varepsilon$, this quantity tends to a constant, which judging from the plot scales sublinearly with $\|\delta\bar{\rho}\|_X$. However, it should be recalled that the $\delta\bar{\rho}$'s with different norms are not simply proportional to each other as they were produced through truncation, so these numerical results are not definitive by themselves. In the limit of large $\varepsilon$, one has $\tilde{u}^{\varepsilon} \to 0$ and the quantity thus tends to $\|u^{\varepsilon'}\|_{X^*}$. An interesting transition region occurs around $\varepsilon \sim 10 \|\delta\bar{\rho}\|_X$, where the effects of optimisation with respect to the perturbed density become noticeable and the accuracy of $\tilde{v}_{\mathrm{s}}^{\varepsilon}$ as a reproduction of the unperturbed $v_{\mathrm{s}}^{\varepsilon'}$ is saturated.

In Fig.~\ref{subfig:norm_pot_err_with_same_eps}, we display the quantity
\begin{equation}
  \label{eqPreRIdentity}
  \begin{split}    
    \|\tilde{v}_{\mathrm{s}}^{\varepsilon} - v_{\mathrm{s}}^{\varepsilon}\|_{X^*} & = \|\tilde{u}^{\varepsilon} - u^{\varepsilon}\|_{X^*} = \frac{\|\tilde{\rho}^{\varepsilon} - \bar{\rho}^{\varepsilon} - \delta\bar{\rho}\|_{X}}{\varepsilon},
  \end{split}
\end{equation}
where the potentials are now evaluated for the same value of $\varepsilon$. This has, as expected, no effect on the limit of small $\varepsilon$. However, for  large $\varepsilon$ we now more directly see the manifestation of the shared limits $u^{\varepsilon}, \tilde{u}^{\varepsilon} \to 0$ and $v_{\mathrm{s}}^{\varepsilon}, \tilde{v}_{\mathrm{s}}^{\varepsilon} \to \alpha v_{\mathrm{ext}} + \xi v_{\mathrm{H}}$ as well as $\tilde{v}_{\mathrm{s}}^{\varepsilon} - v_{\mathrm{s}}^{\varepsilon} \to 0$.

\begin{figure}[h]
        \begin{subfigure}[b]{0.9\linewidth}
                \centering
                \caption{}
                \includegraphics[width=0.98\linewidth]{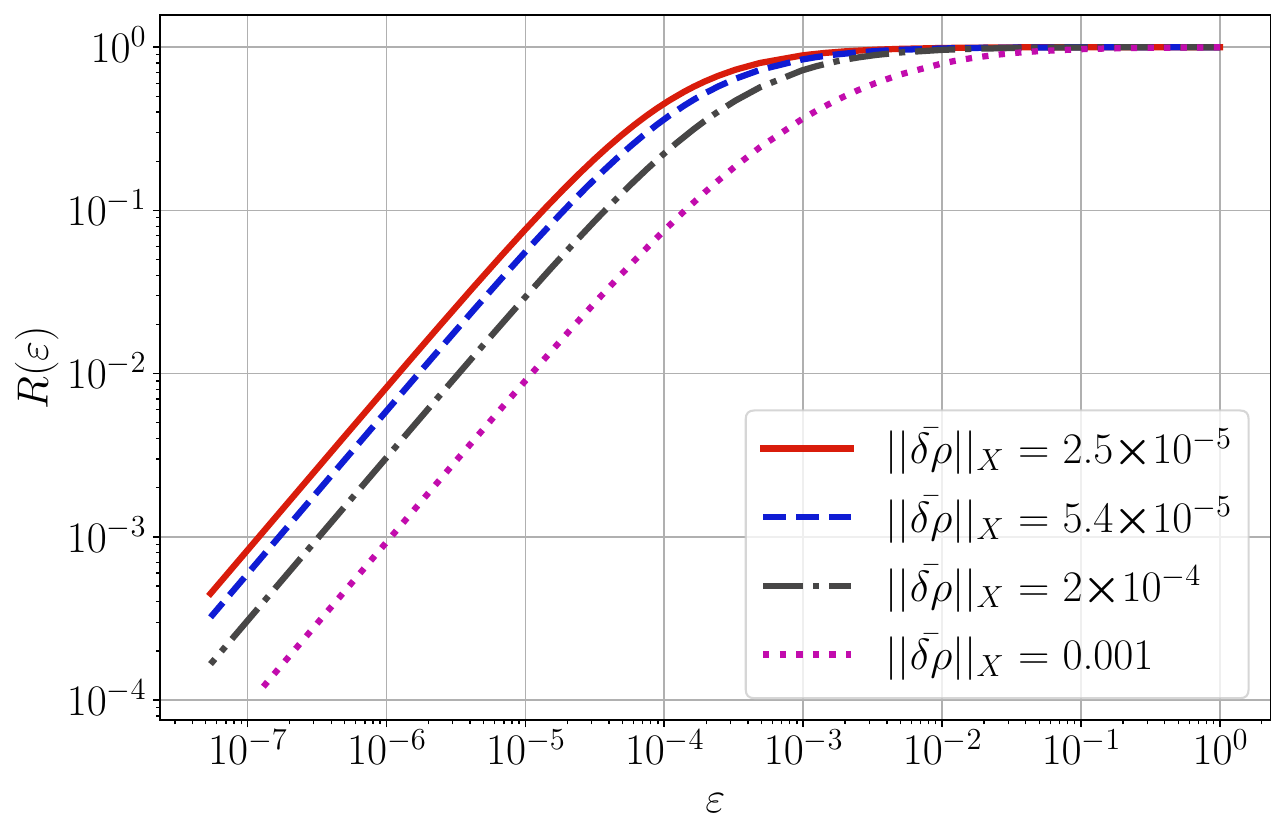}
                \label{subfig:err_analy_R} 
        \end{subfigure} 	
        \begin{subfigure}[b]{0.9\linewidth}
                \centering
                \caption{}
                \includegraphics[width=0.98\linewidth]{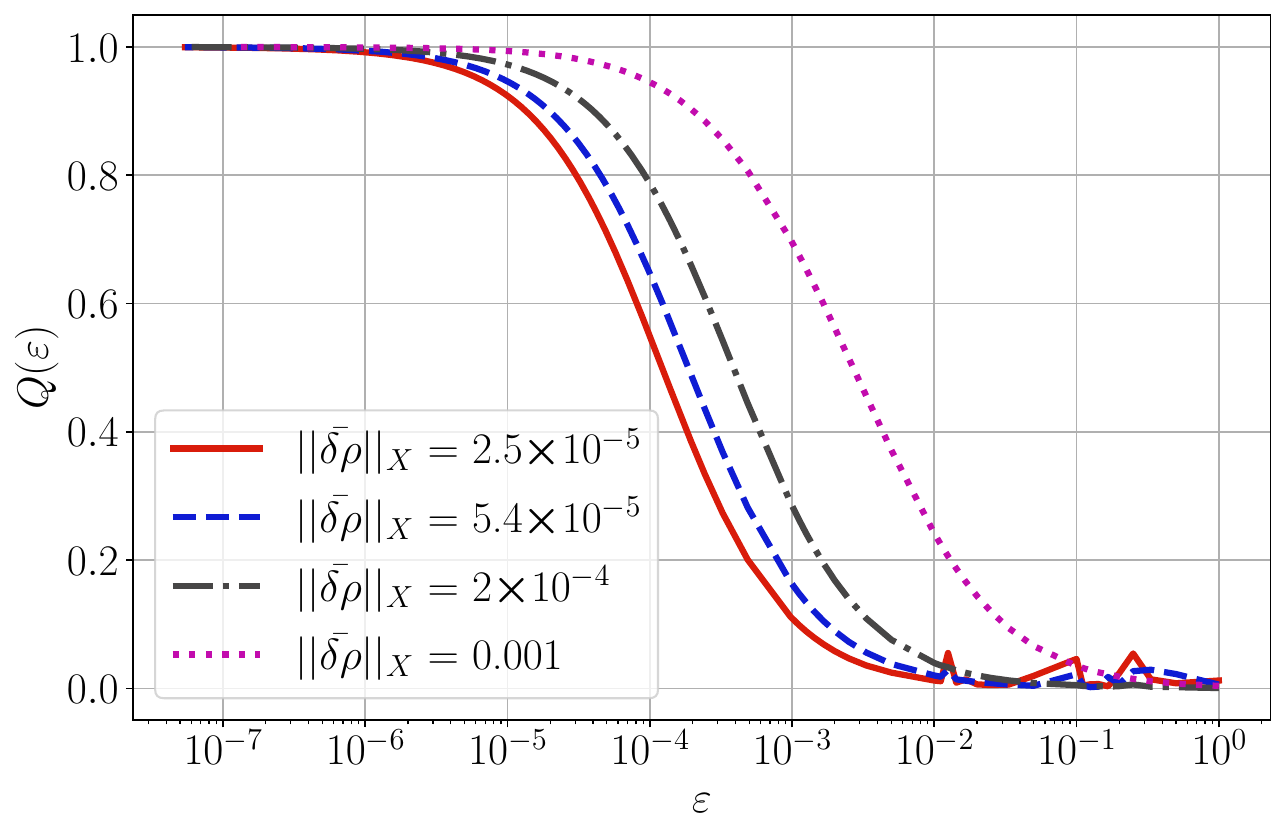}
                \label{subfig:err_analy_Q} 
        \end{subfigure}
        \begin{subfigure}[b]{0.9\linewidth}
                \centering
                \caption{}
                \includegraphics[width=0.98\linewidth]{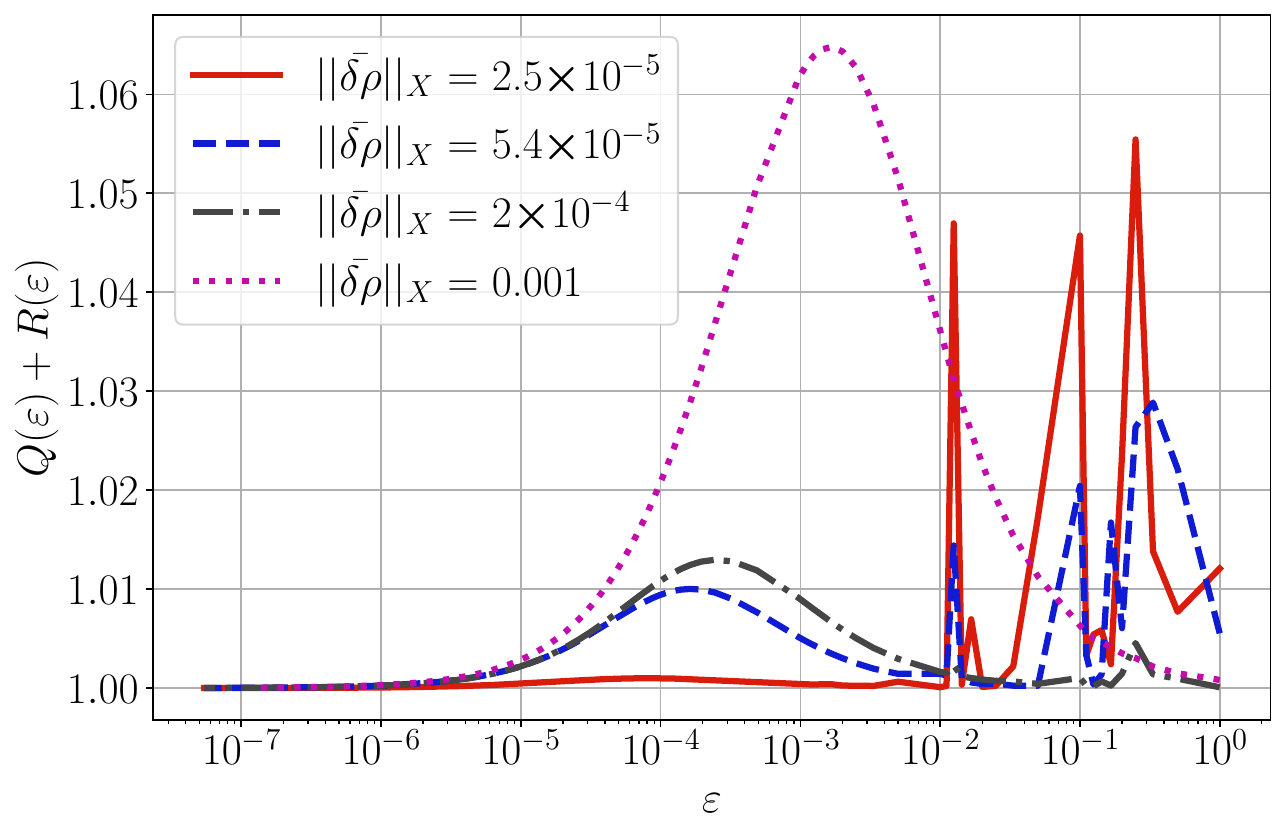}
                \label{subfig:err_analy_Q_plus_R} 
        \end{subfigure}
        \caption{\label{fig:S_and_Q_err_analy} The first two panels show (a) the measure $R(\varepsilon)$ of sensitivity of the Kohn--Sham potential and (b) the measure $Q(\varepsilon)$ of sensitivity of the proximal density to reference density perturbations. The bottom panel shows the sum $Q(\varepsilon) + R(\varepsilon)$, which is bounded from below by 1.}
\end{figure}

The mathematical fact that $\bar{\rho}_{\mathrm{ref}} \mapsto \bar{\rho}^{\varepsilon}$ is a non-expansive mapping means that we have the inequality $\|\tilde{\rho}^{\varepsilon}-\bar{\rho}^{\varepsilon}\|_X \leq \|\tilde{\rho}_{\mathrm{ref}}-\bar{\rho}_{\mathrm{ref}}\|_X = \|\delta\bar{\rho}\|_X $. In Fig.~\ref{subfig:err_analy_Q} we therefore plot the ratio
\begin{equation}
  \label{eqQdef}
  Q(\varepsilon) = \frac{\|\tilde{\rho}^{\varepsilon}-\bar{\rho}^{\varepsilon}\|_X}{\|\delta\bar{\rho}\|_X},
\end{equation}
which is seen to not exceed 1 in our numerical example, consistent with non-expansivity. Moreover, the approach to 1 for small $\varepsilon$ seen in the plot is generic: Using Eqs.~\eqref{eqRhoBarFromInv} and \eqref{eqRhoTildeFromInv} we have~\cite{Herbst2025}
\begin{equation}
  \label{eqQalt}
  Q(\varepsilon) = \frac{\|\delta\bar{\rho} + \varepsilon(\Jdual^{-1}(\tilde{u}^{\varepsilon}) - \Jdual^{-1}(u^{\varepsilon}))\|_X}{\|\delta\bar{\rho}\|_X},
\end{equation}
which tends to 1 as $\varepsilon \to 0$.
That $Q(\varepsilon)$ vanishes in the limit of large $\varepsilon$ is again a manifestation of both $\bar{\rho}^{\varepsilon}$ and $\tilde{\rho}^{\varepsilon}$ converging to the same density -- the ground-state density of $\alpha v_{\mathrm{ext}} + \xi v_{\mathrm{H}}$ -- irrespective of the reference density. Comparing Eqs.~\eqref{eqPreRIdentity} and \eqref{eqQdef}, we also calculate the related quantity~\cite{Herbst2025}
\begin{equation}
  \label{eqRdef}
  R(\varepsilon) = \frac{\|\tilde{\rho}^{\varepsilon}-\bar{\rho}^{\varepsilon} - \delta\bar{\rho}\|_X}{\|\delta\bar{\rho}\|_X} = \frac{\varepsilon \|\tilde{v}_{\mathrm{s}}^{\varepsilon} - v_{\mathrm{s}}^{\varepsilon}\|_{X^*}}{\|\delta\bar{\rho}\|_X},
\end{equation}
which is plotted in Fig.~\ref{subfig:err_analy_R}. For small $\varepsilon$, the difference $\tilde{v}_{\mathrm{s}}^{\varepsilon} - v_{\mathrm{s}}^{\varepsilon}$ converges some constant $\tilde{v}_{\mathrm{s}}^{0} - v_{\mathrm{s}}^{0}$ and we have that $R(\varepsilon)$ is proportional $\varepsilon$. This is also exemplified in Fig.~\ref{subfig:err_analy_R}, where the slope on the log-log plot is $+1$ for $\varepsilon \leq 10^{-5}$. In the limit of large $\varepsilon$, $R(\varepsilon)$ tends to 1 since $\tilde{\rho}^{\varepsilon}-\bar{\rho}^{\varepsilon}$ tends to 0 (since the penalty term vanishes in this limit).

Finally, we exploit the fact that for our choice of density space $X$, the norm $\|\cdot\|_X$ is given by a quadratic form in Fourier space, which gives that $\|\bar{\rho}+\bar{\sigma}\|_X^2 = \|\bar{\rho}\|_X^2 + 2 \pairing{\Jdual(\bar{\rho})}{\bar{\sigma}}+ \|\bar{\sigma}\|_X^2$. Hence, we may write
\begin{equation}
  \begin{split}
    R(\varepsilon)^2 & = \frac{\|\tilde{\rho}^{\varepsilon}-\bar{\rho}^{\varepsilon}\|_X^2 - 2\pairing{\Jdual(\tilde{\rho}^{\varepsilon}-\bar{\rho}^{\varepsilon})}{\delta\bar{\rho}} + \|\delta\bar{\rho}\|_X^2}{ \|\delta\bar{\rho}\|_X^2} \\
    & = Q(\varepsilon)^2 - 2Q(\varepsilon) \cos(\theta)+ 1
  \end{split}
\end{equation}
with
\begin{equation}
  \label{}
  \begin{split}
    \cos(\theta) & = \frac{\pairing{\Jdual(\tilde{\rho}^{\varepsilon}-\bar{\rho}^{\varepsilon})}{\delta\bar{\rho}}}{ \|\tilde{\rho}^{\varepsilon}-\bar{\rho}^{\varepsilon}\|_X \cdot \|\delta\bar{\rho}\|_X}
    \\
   & = \frac{\|\delta\bar{\rho}\|_X^2 + \varepsilon \pairing{\tilde{u}^{\varepsilon}-u^{\varepsilon}}{\delta\bar{\rho}}}{ \|\delta\bar{\rho} + \varepsilon\Jdual^{-1}(\tilde{u}^{\varepsilon}-u^{\varepsilon})\|_X \cdot \|\delta\bar{\rho}\|_X}
   \\
   & = \frac{\|\delta\bar{\rho}\|_X^2 + \varepsilon \pairing{\tilde{u}^{\varepsilon}-u^{\varepsilon}}{\delta\bar{\rho}}}{ \|\delta\bar{\rho}\|_X \sqrt{\|\delta\bar{\rho}\|_X^2 + 2\varepsilon\pairing{\tilde{u}^{\varepsilon}-u^{\varepsilon}}{\delta\bar{\rho}} + \varepsilon^2 \|\tilde{u}^{\varepsilon}-u^{\varepsilon}\|_{X^*}^2}}
   \\
    & = \frac{1 + \frac{\varepsilon}{\|\delta\bar{\rho}\|_X^2} \pairing{\tilde{u}^{\varepsilon}-u^{\varepsilon}}{\delta\bar{\rho}}}{ \sqrt{1 + 2\frac{\varepsilon}{\|\delta\bar{\rho}\|_X^2} \pairing{\tilde{u}^{\varepsilon}-u^{\varepsilon}}{\delta\bar{\rho}} + \frac{\varepsilon^2}{\|\delta\bar{\rho}\|_X^2} \|\tilde{u}^{\varepsilon}-u^{\varepsilon}\|_{X^*}^2}}.
  \end{split}
\end{equation}
Alternatively, the the triangle inequality and the reverse triangle inequality, combined with the fact that $Q(\varepsilon) \leq 1$ (i.e., non-expansivity), directly yield the following lower and upper bounds~\cite{Herbst2025}
\begin{equation}\label{eq:QRbound}
  1 - Q(\varepsilon) \leq R(\varepsilon) \leq 1 + Q(\varepsilon) \leq 2.
\end{equation}
These bounds are equivalent to\cite{Herbst2025}
\begin{equation}
 \begin{split}
    & \frac{\|\delta\bar{\rho}\|_X - \|\tilde{\rho}^{\varepsilon}-\bar{\rho}^{\varepsilon}\|_X}{\varepsilon} \leq \|\tilde{v}_{\mathrm{s}}^{\varepsilon}-v_{\mathrm{s}}^{\varepsilon}\|_{X^*} 
    \\
    & \qquad \leq \frac{\|\delta\bar{\rho}\|_X + \|\tilde{\rho}^{\varepsilon}-\bar{\rho}^{\varepsilon}\|_X}{\varepsilon} \leq \frac{2\|\delta\bar{\rho}\|_X}{\varepsilon}    . 
 \end{split}
\end{equation}
The quantity $Q(\varepsilon) + R(\varepsilon)$ is plotted in Fig.~\ref{subfig:err_analy_Q_plus_R}. (The discontinuous jumps in the curves representing the smallest perturbations $\|\delta\bar{\rho}\|_X$ are due to numerical noise below the gradient norm criterion used in the SCF optimisation---$\|[\mathcal{F},D]\|_F < 10^{-6}$. These discontinuities are also visible in Fig.~\ref{subfig:err_analy_Q} but become amplified on the vertical axis used in Fig.~\ref{subfig:err_analy_Q_plus_R}. With a smaller threshold relative to $\|\delta\bar{\rho}\|_X$ these discontinuities disappear.) For our numerical example, we observe that $Q(\varepsilon) + R(\varepsilon) \approx 1$, i.e.\ the lower bound in Eq.~\eqref{eq:QRbound} above turns out to be very tight. This can happen if $\cos(\theta) \approx 1$ in the entire range of $\varepsilon$ for which both $Q(\varepsilon)$ and $R(\varepsilon)$ are non-negligible, so that $R(\varepsilon) \approx \sqrt{Q(\varepsilon)^2 - 2Q(\varepsilon) + 1} = 1-Q(\varepsilon)$. Writing $\delta u^{\varepsilon} = \tilde{u}^{\varepsilon} - u^{\varepsilon}$ and expanding to second order in $\varepsilon$ yields
\begin{equation}
  \begin{split}
    \cos(\theta) & = 1 + \frac 1 2 \frac{\varepsilon^2}{\|\delta\bar{\rho}\|_X^4} |\pairing{\delta u^{\varepsilon}}{\delta\bar{\rho}}|^2 - \frac{\varepsilon^2}{2\|\delta\bar{\rho}\|_X^2} \|\delta u^{\varepsilon}\|_{X^*}^2 + \mathcal{O}(\varepsilon^3)
    \\
    & \geq 1 -  \frac{\varepsilon^2  \|\delta u^{\varepsilon}\|_{X^*}^2  }{\|\delta\bar{\rho}\|_X^2}  + \mathcal{O}(\varepsilon^3).
  \end{split}
\end{equation}
Hence, as long as $\varepsilon \ll \|\delta\bar{\rho}\|_X$, one typically has $\cos(\theta) \approx 1$. Consequently, we believe the lower bound $1-Q(\varepsilon) \leq R(\varepsilon)$ is typically tight in this regime.

\section{Summary and Conclusion}
\label{sec:concl}

We have presented a general formalism for density-functional theory for periodic systems. Born--von K\'arm\'an periodic boundary conditions are applied to wave functions, while potentials are subject to simple periodic boundary conditions, along an arbitrary number of dimensions. As a consequence, the natural dual quantity within a rigorous convex analysis-based formulation that parallels that of Lieb~\cite{Lieb1983} is not the conventional electron density, but the density averaged over all translations within the Born--von K\'arm\'an zone. The electron-electron interactions were taken to be screened Yukawa interactions as this choice is naturally compatible with the function spaces for wave functions and densities. Within this formalism, we formulated a Moreau--Yosida based method for density-potential inversion. The theory was illustrated using a one-dimensional Hartree--Fock implementation that was used to obtain Kohn--Sham potentials and local exact exchange potentials. This was done by first computing a ground-state Hartree--Fock density for a given $v_\mathrm{ext}$. This density was then inverted in a regularised iKS procedure where we allow for various degree of guidance by the same external potential and Hartree term. While our ground-state density is determined at the uncorrelated Hartree-Fock level, the inversion procedure do allow us to translate exact exchange energy to a local exchange potential. Within this one-dimensional set-up, we expect the Kohn--Sham potential to provide a good local description of the effects of the nonlocal exact exchange. The inversion of ground-state densities from correlated methods is left for future work where we hope to investigate the regularised iKS methods ability to also capture correlation in an effective potential obtained in the limit of the Moreau--Yosida regularisation parameter $\varepsilon$ to zero.

Our numerical results demonstrate that the method can be used to reliably compute approximate Kohn--Sham potentials corresponding to regularisation parameters $\varepsilon \sim 10^{-8}-10^{-6}$, which is sufficient for converged results. Moreover, we demonstrate the behavior of the method when the reference density is not $N$-representable, in which case the closest $N$-representable density is produced instead. As discussed in Sec.~\ref{sec:AnsatzDep}, the results also illustrate that convergence can be accelerated by a suitable choice of a model functional subjected to regularisation (cf. Eq.~\eqref{eq:mod_func_alp_xi}) that incorporates some features of the Kohn--Sham potential. 

From our experience, the large amount of Hartree energy contributed by the penalty term makes SCF convergence by simple methods infeasible and represents a practical implementation challenge. Nonetheless, this challenge can be overcome with more robust SCF optimisation methods and good initial guesses.

Remarkably, our initial results indicate that optimisation with the smallest studied Yukawa screening parameter ($\gamma=0.1$) provides the best reproduction of the reference density irrespective of which screening parameter value is used to measure the fit. It is an avenue for future work to investigate this further and also address the effects of an electron-electron interaction that harmonises less well with the underlying function spaces of densities and wave functions.

Finally, the numerical results in Sec.~\ref{secErrorAnalysis} illustrate the effects of perturbations under the \emph{regularised} (non-interacting) Hohenberg--Kohn mapping $\bar{\rho}_{\mathrm{ref}} \mapsto v^{\varepsilon}_{\mathrm{s}}$ as well as under the proximal mapping $\bar{\rho}_{\mathrm{ref}} \mapsto \bar{\rho}^{\varepsilon}$. A crucial ingredient in our error analysis is the non-expansiveness of the proximal mapping, i.e. a perturbation of the reference density propagates to a (typically) smaller (but never larger) perturbation in the proximal densities. The propagation to the Kohn--Sham potential can be rigorously bounded from above and below in terms of the regularisation parameter $\varepsilon$, the proximal density perturbation $\|\tilde{\rho}^{\varepsilon}-\bar{\rho}^{\varepsilon}\|_X$, and the reference density perturbation $\|\delta\bar{\rho}_{\mathrm{ref}}\|_X$. In a common parameter regime, the lower bound $\|\delta\bar{\rho}_{\mathrm{ref}}\|_X \leq \|\tilde{\rho}^{\varepsilon}-\bar{\rho}^{\varepsilon}\|_X + \varepsilon \|\tilde{v}^{\varepsilon}_{\mathrm{s}} - v^{\varepsilon}_{\mathrm{s}}\|_{X^*}$ is tight to within a few percent.

\begin{acknowledgements}
AL and ML have received funding from the ERC-2021-STG under grant agreement No.~101041487 REGAL. AL, OB and EIT were funded by the Research Council of Norway through CoE Hylleraas Centre for Quantum Molecular Sciences Grant No.~262695.
AL has also received funding from the Research Council of Norway through CCerror Grant No.~287906. The work also received support from the UNINETT Sigma2, the National Infrastructure for High Performance Computing and Data Storage, through a grant of computer time (Grant No. NN4654K). This project has received funding from the European Union's Horizon 2020 research and innovation programme under the Marie Sk\l{}odowska-Curie grant agreement No.~945371. The authors thank Vegard Falm\aa r for useful discussions.
\end{acknowledgements}

\clearpage

\onecolumngrid

%

\begin{thebibliography}{62}%
\makeatletter
\providecommand \@ifxundefined [1]{%
 \@ifx{#1\undefined}
}%
\providecommand \@ifnum [1]{%
 \ifnum #1\expandafter \@firstoftwo
 \else \expandafter \@secondoftwo
 \fi
}%
\providecommand \@ifx [1]{%
 \ifx #1\expandafter \@firstoftwo
 \else \expandafter \@secondoftwo
 \fi
}%
\providecommand \natexlab [1]{#1}%
\providecommand \enquote  [1]{``#1''}%
\providecommand \bibnamefont  [1]{#1}%
\providecommand \bibfnamefont [1]{#1}%
\providecommand \citenamefont [1]{#1}%
\providecommand \href@noop [0]{\@secondoftwo}%
\providecommand \href [0]{\begingroup \@sanitize@url \@href}%
\providecommand \@href[1]{\@@startlink{#1}\@@href}%
\providecommand \@@href[1]{\endgroup#1\@@endlink}%
\providecommand \@sanitize@url [0]{\catcode `\\12\catcode `\$12\catcode
  `\&12\catcode `\#12\catcode `\^12\catcode `\_12\catcode `\%12\relax}%
\providecommand \@@startlink[1]{}%
\providecommand \@@endlink[0]{}%
\providecommand \url  [0]{\begingroup\@sanitize@url \@url }%
\providecommand \@url [1]{\endgroup\@href {#1}{\urlprefix }}%
\providecommand \urlprefix  [0]{URL }%
\providecommand \Eprint [0]{\href }%
\providecommand \doibase [0]{https://doi.org/}%
\providecommand \selectlanguage [0]{\@gobble}%
\providecommand \bibinfo  [0]{\@secondoftwo}%
\providecommand \bibfield  [0]{\@secondoftwo}%
\providecommand \translation [1]{[#1]}%
\providecommand \BibitemOpen [0]{}%
\providecommand \bibitemStop [0]{}%
\providecommand \bibitemNoStop [0]{.\EOS\space}%
\providecommand \EOS [0]{\spacefactor3000\relax}%
\providecommand \BibitemShut  [1]{\csname bibitem#1\endcsname}%
\let\auto@bib@innerbib\@empty
\bibitem [{\citenamefont {Kohn}\ and\ \citenamefont {Sham}(1965)}]{KS1965}%
  \BibitemOpen
  \bibfield  {author} {\bibinfo {author} {\bibfnamefont {W.}~\bibnamefont
  {Kohn}}\ and\ \bibinfo {author} {\bibfnamefont {L.~J.}\ \bibnamefont
  {Sham}},\ }\bibfield  {title} {\enquote {\bibinfo {title} {{Self-Consistent
  Equations Including Exchange and Correlation Effects}},}\ }\href
  {https://doi.org/10.1103/PhysRev.140.A1133} {\bibfield  {journal} {\bibinfo
  {journal} {Phys. Rev.}\ }\textbf {\bibinfo {volume} {140}},\ \bibinfo {pages}
  {A1133--A1138} (\bibinfo {year} {1965})}\BibitemShut {NoStop}%
\bibitem [{\citenamefont {Burke}(2012)}]{Burke2012}%
  \BibitemOpen
  \bibfield  {author} {\bibinfo {author} {\bibfnamefont {K.}~\bibnamefont
  {Burke}},\ }\bibfield  {title} {\enquote {\bibinfo {title} {Perspective on
  density functional theory},}\ }\href {http://dx.doi.org/10.1063/1.4704546}
  {\bibfield  {journal} {\bibinfo  {journal} {J. Chem. Phys.}\ }\textbf
  {\bibinfo {volume} {136}} (\bibinfo {year} {2012})}\BibitemShut {NoStop}%
\bibitem [{\citenamefont {Verma}\ and\ \citenamefont
  {Truhlar}(2020)}]{Verma_2020}%
  \BibitemOpen
  \bibfield  {author} {\bibinfo {author} {\bibfnamefont {P.}~\bibnamefont
  {Verma}}\ and\ \bibinfo {author} {\bibfnamefont {D.~G.}\ \bibnamefont
  {Truhlar}},\ }\bibfield  {title} {\enquote {\bibinfo {title} {{Status and
  Challenges of Density Functional Theory}},}\ }\href
  {https://doi.org/https://doi.org/10.1016/j.trechm.2020.02.005} {\bibfield
  {journal} {\bibinfo  {journal} {Trends in Chemistry}\ }\textbf {\bibinfo
  {volume} {2}},\ \bibinfo {pages} {302--318} (\bibinfo {year}
  {2020})}\BibitemShut {NoStop}%
\bibitem [{\citenamefont {Teale~et al.}(2022)}]{Teale2022}%
  \BibitemOpen
  \bibfield  {author} {\bibinfo {author} {\bibfnamefont {A.~M.}\ \bibnamefont
  {Teale~et al.}},\ }\bibfield  {title} {\enquote {\bibinfo {title} {{DFT}
  exchange: Sharing perspectives on the workhorse of quantum chemistry and
  materials science},}\ }\href {https://doi.org/10.1039/d2cp02827a} {\bibfield
  {journal} {\bibinfo  {journal} {Phys. Chem. Chem. Phys.}\ }\textbf {\bibinfo
  {volume} {24}},\ \bibinfo {pages} {28700--28781} (\bibinfo {year}
  {2022})}\BibitemShut {NoStop}%
\bibitem [{\citenamefont {Hohenberg}\ and\ \citenamefont
  {Kohn}(1964)}]{Hohenberg1964}%
  \BibitemOpen
  \bibfield  {author} {\bibinfo {author} {\bibfnamefont {P.}~\bibnamefont
  {Hohenberg}}\ and\ \bibinfo {author} {\bibfnamefont {W.}~\bibnamefont
  {Kohn}},\ }\bibfield  {title} {\enquote {\bibinfo {title} {{Inhomogeneous
  Electron Gas}},}\ }\href {https://doi.org/10.1103/PhysRev.136.B864}
  {\bibfield  {journal} {\bibinfo  {journal} {Phys. Rev.}\ }\textbf {\bibinfo
  {volume} {136}},\ \bibinfo {pages} {B864--B871} (\bibinfo {year}
  {1964})}\BibitemShut {NoStop}%
\bibitem [{\citenamefont {Percus}(1978)}]{PERCUS_IJQC13_89}%
  \BibitemOpen
  \bibfield  {author} {\bibinfo {author} {\bibfnamefont {J.~K.}\ \bibnamefont
  {Percus}},\ }\bibfield  {title} {\enquote {\bibinfo {title} {{The role of
  model systems in the few-body reduction of the $N$-fermion problem}},}\
  }\href@noop {} {\bibfield  {journal} {\bibinfo  {journal} {Int. J. Quantum
  Chem.}\ }\textbf {\bibinfo {volume} {13}},\ \bibinfo {pages} {89--124}
  (\bibinfo {year} {1978})}\BibitemShut {NoStop}%
\bibitem [{\citenamefont {Levy}(1979)}]{levy1979}%
  \BibitemOpen
  \bibfield  {author} {\bibinfo {author} {\bibfnamefont {M.}~\bibnamefont
  {Levy}},\ }\bibfield  {title} {\enquote {\bibinfo {title} {Universal
  variational functionals of electron densities, first-order density matrices,
  and natural spin-orbitals and solution of the v-representability problem},}\
  }\href@noop {} {\bibfield  {journal} {\bibinfo  {journal} {Proc. Natl. Acad.
  Sci. USA}\ }\textbf {\bibinfo {volume} {76}},\ \bibinfo {pages} {6062--6065}
  (\bibinfo {year} {1979})}\BibitemShut {NoStop}%
\bibitem [{\citenamefont {Lieb}(1983)}]{Lieb1983}%
  \BibitemOpen
  \bibfield  {author} {\bibinfo {author} {\bibfnamefont {E.~H.}\ \bibnamefont
  {Lieb}},\ }\bibfield  {title} {\enquote {\bibinfo {title} {Density
  functionals for coulomb systems},}\ }\href
  {https://doi.org/10.1002/qua.560240302} {\bibfield  {journal} {\bibinfo
  {journal} {Int. J. Quantum Chem.}\ }\textbf {\bibinfo {volume} {24}},\
  \bibinfo {pages} {243--277} (\bibinfo {year} {1983})}\BibitemShut {NoStop}%
\bibitem [{\citenamefont {Penz}\ \emph {et~al.}(2023)\citenamefont {Penz},
  \citenamefont {Tellgren}, \citenamefont {Csirik}, \citenamefont
  {Ruggenthaler},\ and\ \citenamefont {Laestadius}}]{PenzPartI2023}%
  \BibitemOpen
  \bibfield  {author} {\bibinfo {author} {\bibfnamefont {M.}~\bibnamefont
  {Penz}}, \bibinfo {author} {\bibfnamefont {E.~I.}\ \bibnamefont {Tellgren}},
  \bibinfo {author} {\bibfnamefont {M.~A.}\ \bibnamefont {Csirik}}, \bibinfo
  {author} {\bibfnamefont {M.}~\bibnamefont {Ruggenthaler}},\ and\ \bibinfo
  {author} {\bibfnamefont {A.}~\bibnamefont {Laestadius}},\ }\bibfield  {title}
  {\enquote {\bibinfo {title} {{The Structure of Density-Potential Mapping.
  Part I: Standard Density--Functional Theory}},}\ }\href
  {https://doi.org/10.1021/acsphyschemau.2c00069} {\bibfield  {journal}
  {\bibinfo  {journal} {ACS Physical Chemistry Au}\ }\textbf {\bibinfo {volume}
  {3}},\ \bibinfo {pages} {334–347} (\bibinfo {year} {2023})}\BibitemShut
  {NoStop}%
\bibitem [{\citenamefont {Garrigue}(2021)}]{GarrigueCMP386_1803}%
  \BibitemOpen
  \bibfield  {author} {\bibinfo {author} {\bibfnamefont {L.}~\bibnamefont
  {Garrigue}},\ }\bibfield  {title} {\enquote {\bibinfo {title} {{Some
  Properties of the Potential-to-Ground State Map in Quantum Mechanics}},}\
  }\href {https://doi.org/10.1007/s00220-021-04140-9} {\bibfield  {journal}
  {\bibinfo  {journal} {Comm. Math. Phys.}\ }\textbf {\bibinfo {volume}
  {386}},\ \bibinfo {pages} {1803--1844} (\bibinfo {year} {2021})}\BibitemShut
  {NoStop}%
\bibitem [{\citenamefont {Aryasetiawan}\ and\ \citenamefont
  {Stott}(1988)}]{Aryasetiawan1988}%
  \BibitemOpen
  \bibfield  {author} {\bibinfo {author} {\bibfnamefont {F.}~\bibnamefont
  {Aryasetiawan}}\ and\ \bibinfo {author} {\bibfnamefont {M.~J.}\ \bibnamefont
  {Stott}},\ }\bibfield  {title} {\enquote {\bibinfo {title} {Effective
  potentials in density-functional theory},}\ }\href
  {https://doi.org/10.1103/PhysRevB.38.2974} {\bibfield  {journal} {\bibinfo
  {journal} {Phys. Rev. B}\ }\textbf {\bibinfo {volume} {38}},\ \bibinfo
  {pages} {2974--2987} (\bibinfo {year} {1988})}\BibitemShut {NoStop}%
\bibitem [{\citenamefont {Knorr}\ and\ \citenamefont
  {Godby}(1992)}]{Knorr-Godby-1992}%
  \BibitemOpen
  \bibfield  {author} {\bibinfo {author} {\bibfnamefont {W.}~\bibnamefont
  {Knorr}}\ and\ \bibinfo {author} {\bibfnamefont {R.~W.}\ \bibnamefont
  {Godby}},\ }\bibfield  {title} {\enquote {\bibinfo {title} {Investigating
  exact density-functional theory of a model semiconductor},}\ }\href
  {https://doi.org/10.1103/PhysRevLett.68.639} {\bibfield  {journal} {\bibinfo
  {journal} {Phys. Rev. Lett.}\ }\textbf {\bibinfo {volume} {68}},\ \bibinfo
  {pages} {639--641} (\bibinfo {year} {1992})}\BibitemShut {NoStop}%
\bibitem [{\citenamefont {G\"orling}(1992)}]{Gorling1992}%
  \BibitemOpen
  \bibfield  {author} {\bibinfo {author} {\bibfnamefont {A.}~\bibnamefont
  {G\"orling}},\ }\bibfield  {title} {\enquote {\bibinfo {title} {{Kohn--Sham
  potentials and wave functions from electron densities}},}\ }\href
  {https://doi.org/10.1103/PhysRevA.46.3753} {\bibfield  {journal} {\bibinfo
  {journal} {Phys. Rev. A}\ }\textbf {\bibinfo {volume} {46}},\ \bibinfo
  {pages} {3753--3757} (\bibinfo {year} {1992})}\BibitemShut {NoStop}%
\bibitem [{\citenamefont {Zhao}\ and\ \citenamefont {Parr}(1993)}]{ZP1993}%
  \BibitemOpen
  \bibfield  {author} {\bibinfo {author} {\bibfnamefont {Q.}~\bibnamefont
  {Zhao}}\ and\ \bibinfo {author} {\bibfnamefont {R.~G.}\ \bibnamefont
  {Parr}},\ }\bibfield  {title} {\enquote {\bibinfo {title} {Constrained-search
  method to determine electronic wave functions from electronic densities},}\
  }\href {https://doi.org/10.1063/1.465093} {\bibfield  {journal} {\bibinfo
  {journal} {J. Chem. Phys.}\ }\textbf {\bibinfo {volume} {98}},\ \bibinfo
  {pages} {543--548} (\bibinfo {year} {1993})}\BibitemShut {NoStop}%
\bibitem [{\citenamefont {Wang}\ and\ \citenamefont {Parr}(1993)}]{Wang1993}%
  \BibitemOpen
  \bibfield  {author} {\bibinfo {author} {\bibfnamefont {Y.}~\bibnamefont
  {Wang}}\ and\ \bibinfo {author} {\bibfnamefont {R.~G.}\ \bibnamefont
  {Parr}},\ }\bibfield  {title} {\enquote {\bibinfo {title} {{Construction of
  exact Kohn--Sham orbitals from a given electron density}},}\ }\href
  {https://doi.org/10.1103/PhysRevA.47.R1591} {\bibfield  {journal} {\bibinfo
  {journal} {Phys. Rev. A}\ }\textbf {\bibinfo {volume} {47}},\ \bibinfo
  {pages} {R1591--R1593} (\bibinfo {year} {1993})}\BibitemShut {NoStop}%
\bibitem [{\citenamefont {Zhao}, \citenamefont {Morrison},\ and\ \citenamefont
  {Parr}(1994)}]{ZMP1994}%
  \BibitemOpen
  \bibfield  {author} {\bibinfo {author} {\bibfnamefont {Q.}~\bibnamefont
  {Zhao}}, \bibinfo {author} {\bibfnamefont {R.~C.}\ \bibnamefont {Morrison}},\
  and\ \bibinfo {author} {\bibfnamefont {R.~G.}\ \bibnamefont {Parr}},\
  }\bibfield  {title} {\enquote {\bibinfo {title} {From electron densities to
  {K}ohn--{S}ham kinetic energies, orbital energies, exchange-correlation
  potentials, and exchange-correlation energies},}\ }\href
  {https://doi.org/10.1103/PhysRevA.50.2138} {\bibfield  {journal} {\bibinfo
  {journal} {Phys. Rev. A}\ }\textbf {\bibinfo {volume} {50}},\ \bibinfo
  {pages} {2138} (\bibinfo {year} {1994})}\BibitemShut {NoStop}%
\bibitem [{\citenamefont {Knorr}\ and\ \citenamefont
  {Godby}(1994)}]{Knorr_1994}%
  \BibitemOpen
  \bibfield  {author} {\bibinfo {author} {\bibfnamefont {W.}~\bibnamefont
  {Knorr}}\ and\ \bibinfo {author} {\bibfnamefont {R.~W.}\ \bibnamefont
  {Godby}},\ }\bibfield  {title} {\enquote {\bibinfo {title} {{Quantum Monte
  Carlo study of density-functional theory for a semiconducting wire}},}\
  }\href {https://doi.org/10.1103/PhysRevB.50.1779} {\bibfield  {journal}
  {\bibinfo  {journal} {Phys. Rev. B}\ }\textbf {\bibinfo {volume} {50}},\
  \bibinfo {pages} {1779--1791} (\bibinfo {year} {1994})}\BibitemShut {NoStop}%
\bibitem [{\citenamefont {van Leeuwen}\ and\ \citenamefont
  {Baerends}(1994{\natexlab{a}})}]{vanleeuwen1994exchange}%
  \BibitemOpen
  \bibfield  {author} {\bibinfo {author} {\bibfnamefont {R.}~\bibnamefont {van
  Leeuwen}}\ and\ \bibinfo {author} {\bibfnamefont {E.~J.}\ \bibnamefont
  {Baerends}},\ }\bibfield  {title} {\enquote {\bibinfo {title}
  {Exchange-correlation potential with correct asymptotic behavior},}\ }\href
  {https://doi.org/10.1103/PhysRevA.49.2421} {\bibfield  {journal} {\bibinfo
  {journal} {Phys. Rev. A}\ }\textbf {\bibinfo {volume} {49}},\ \bibinfo
  {pages} {2421} (\bibinfo {year} {1994}{\natexlab{a}})}\BibitemShut {NoStop}%
\bibitem [{\citenamefont {Yang}\ and\ \citenamefont {Wu}(2002)}]{Yang_Wu_2002}%
  \BibitemOpen
  \bibfield  {author} {\bibinfo {author} {\bibfnamefont {W.}~\bibnamefont
  {Yang}}\ and\ \bibinfo {author} {\bibfnamefont {Q.}~\bibnamefont {Wu}},\
  }\bibfield  {title} {\enquote {\bibinfo {title} {{Direct Method for Optimized
  Effective Potentials in Density-Functional Theory}},}\ }\href
  {https://doi.org/10.1103/PhysRevLett.89.143002} {\bibfield  {journal}
  {\bibinfo  {journal} {Phys. Rev. Lett.}\ }\textbf {\bibinfo {volume} {89}},\
  \bibinfo {pages} {143002} (\bibinfo {year} {2002})}\BibitemShut {NoStop}%
\bibitem [{\citenamefont {Wu}\ and\ \citenamefont {Yang}(2003)}]{Wu-Yang-2003}%
  \BibitemOpen
  \bibfield  {author} {\bibinfo {author} {\bibfnamefont {Q.}~\bibnamefont
  {Wu}}\ and\ \bibinfo {author} {\bibfnamefont {W.}~\bibnamefont {Yang}},\
  }\bibfield  {title} {\enquote {\bibinfo {title} {A direct optimization method
  for calculating density functionals and exchange-correlation potentials from
  electron densities},}\ }\href {https://doi.org/10.1063/1.1535422} {\bibfield
  {journal} {\bibinfo  {journal} {J. Chem. Phys.}\ }\textbf {\bibinfo {volume}
  {118}},\ \bibinfo {pages} {2498--2509} (\bibinfo {year} {2003})}\BibitemShut
  {NoStop}%
\bibitem [{\citenamefont {Peirs}, \citenamefont {Van~Neck},\ and\ \citenamefont
  {Waroquier}(2003)}]{Peirs2003}%
  \BibitemOpen
  \bibfield  {author} {\bibinfo {author} {\bibfnamefont {K.}~\bibnamefont
  {Peirs}}, \bibinfo {author} {\bibfnamefont {D.}~\bibnamefont {Van~Neck}},\
  and\ \bibinfo {author} {\bibfnamefont {M.}~\bibnamefont {Waroquier}},\
  }\bibfield  {title} {\enquote {\bibinfo {title} {Algorithm to derive exact
  exchange-correlation potentials from correlated densities in atoms},}\ }\href
  {https://doi.org/10.1103/PhysRevA.67.012505} {\bibfield  {journal} {\bibinfo
  {journal} {Phys. Rev. A}\ }\textbf {\bibinfo {volume} {67}},\ \bibinfo
  {pages} {012505} (\bibinfo {year} {2003})}\BibitemShut {NoStop}%
\bibitem [{\citenamefont {Kadantsev}\ and\ \citenamefont
  {Stott}(2004)}]{Kadantsev2004}%
  \BibitemOpen
  \bibfield  {author} {\bibinfo {author} {\bibfnamefont {E.~S.}\ \bibnamefont
  {Kadantsev}}\ and\ \bibinfo {author} {\bibfnamefont {M.~J.}\ \bibnamefont
  {Stott}},\ }\bibfield  {title} {\enquote {\bibinfo {title} {{Variational
  method for inverting the Kohn--Sham procedure}},}\ }\href
  {https://doi.org/10.1103/PhysRevA.69.012502} {\bibfield  {journal} {\bibinfo
  {journal} {Phys. Rev. A}\ }\textbf {\bibinfo {volume} {69}},\ \bibinfo
  {pages} {012502} (\bibinfo {year} {2004})}\BibitemShut {NoStop}%
\bibitem [{\citenamefont {Bulat}\ \emph {et~al.}(2007)\citenamefont {Bulat},
  \citenamefont {Heaton-Burgess}, \citenamefont {Cohen},\ and\ \citenamefont
  {Yang}}]{Bulat2007}%
  \BibitemOpen
  \bibfield  {author} {\bibinfo {author} {\bibfnamefont {F.~A.}\ \bibnamefont
  {Bulat}}, \bibinfo {author} {\bibfnamefont {T.}~\bibnamefont
  {Heaton-Burgess}}, \bibinfo {author} {\bibfnamefont {A.~J.}\ \bibnamefont
  {Cohen}},\ and\ \bibinfo {author} {\bibfnamefont {W.}~\bibnamefont {Yang}},\
  }\bibfield  {title} {\enquote {\bibinfo {title} {{Optimized effective
  potentials from electron densities in finite basis sets}},}\ }\href
  {https://doi.org/10.1063/1.2800021} {\bibfield  {journal} {\bibinfo
  {journal} {J. Chem. Phys.}\ }\textbf {\bibinfo {volume} {127}},\ \bibinfo
  {pages} {174101} (\bibinfo {year} {2007})}\BibitemShut {NoStop}%
\bibitem [{\citenamefont {Gaiduk}, \citenamefont {Ryabinkin},\ and\
  \citenamefont {Staroverov}(2013)}]{Gaiduk2013}%
  \BibitemOpen
  \bibfield  {author} {\bibinfo {author} {\bibfnamefont {A.~P.}\ \bibnamefont
  {Gaiduk}}, \bibinfo {author} {\bibfnamefont {I.~G.}\ \bibnamefont
  {Ryabinkin}},\ and\ \bibinfo {author} {\bibfnamefont {V.~N.}\ \bibnamefont
  {Staroverov}},\ }\bibfield  {title} {\enquote {\bibinfo {title} {{Removal of
  Basis-Set Artifacts in Kohn--Sham Potentials Recovered from Electron
  Densities}},}\ }\href {https://doi.org/10.1021/ct4004146} {\bibfield
  {journal} {\bibinfo  {journal} {J. Chem. Theory Comput.}\ }\textbf {\bibinfo
  {volume} {9}},\ \bibinfo {pages} {3959--3964} (\bibinfo {year}
  {2013})}\BibitemShut {NoStop}%
\bibitem [{\citenamefont {Wagner}\ \emph {et~al.}(2014)\citenamefont {Wagner},
  \citenamefont {Baker}, \citenamefont {Stoudenmire}, \citenamefont {Burke},\
  and\ \citenamefont {White}}]{Wagner2014}%
  \BibitemOpen
  \bibfield  {author} {\bibinfo {author} {\bibfnamefont {L.~O.}\ \bibnamefont
  {Wagner}}, \bibinfo {author} {\bibfnamefont {T.~E.}\ \bibnamefont {Baker}},
  \bibinfo {author} {\bibfnamefont {E.~M.}\ \bibnamefont {Stoudenmire}},
  \bibinfo {author} {\bibfnamefont {K.}~\bibnamefont {Burke}},\ and\ \bibinfo
  {author} {\bibfnamefont {S.~R.}\ \bibnamefont {White}},\ }\bibfield  {title}
  {\enquote {\bibinfo {title} {{Kohn--Sham calculations with the exact
  functional}},}\ }\href {https://doi.org/10.1103/PhysRevB.90.045109}
  {\bibfield  {journal} {\bibinfo  {journal} {Phys. Rev. B}\ }\textbf {\bibinfo
  {volume} {90}},\ \bibinfo {pages} {045109} (\bibinfo {year}
  {2014})}\BibitemShut {NoStop}%
\bibitem [{\citenamefont {Jensen}\ and\ \citenamefont
  {Wasserman}(2018)}]{JensenWasserman2017}%
  \BibitemOpen
  \bibfield  {author} {\bibinfo {author} {\bibfnamefont {D.~S.}\ \bibnamefont
  {Jensen}}\ and\ \bibinfo {author} {\bibfnamefont {A.}~\bibnamefont
  {Wasserman}},\ }\bibfield  {title} {\enquote {\bibinfo {title} {Numerical
  methods for the inverse problem of density functional theory},}\ }\href
  {https://onlinelibrary.wiley.com/doi/abs/10.1002/qua.25425} {\bibfield
  {journal} {\bibinfo  {journal} {Int J Quantum Chem.}\ }\textbf {\bibinfo
  {volume} {118}} (\bibinfo {year} {2018})}\BibitemShut {NoStop}%
\bibitem [{\citenamefont {Zhang}\ and\ \citenamefont
  {Carter}(2018)}]{Zhang2018}%
  \BibitemOpen
  \bibfield  {author} {\bibinfo {author} {\bibfnamefont {X.}~\bibnamefont
  {Zhang}}\ and\ \bibinfo {author} {\bibfnamefont {E.~A.}\ \bibnamefont
  {Carter}},\ }\bibfield  {title} {\enquote {\bibinfo {title} {{Kohn--Sham
  potentials from electron densities using a matrix representation within
  finite atomic orbital basis sets}},}\ }\href
  {https://doi.org/10.1063/1.5005839} {\bibfield  {journal} {\bibinfo
  {journal} {J. Chem. Phys.}\ }\textbf {\bibinfo {volume} {148}},\ \bibinfo
  {pages} {034105} (\bibinfo {year} {2018})}\BibitemShut {NoStop}%
\bibitem [{\citenamefont {Ou}\ and\ \citenamefont {Carter}(2018)}]{Ou2018}%
  \BibitemOpen
  \bibfield  {author} {\bibinfo {author} {\bibfnamefont {Q.}~\bibnamefont
  {Ou}}\ and\ \bibinfo {author} {\bibfnamefont {E.~A.}\ \bibnamefont
  {Carter}},\ }\bibfield  {title} {\enquote {\bibinfo {title} {{Potential
  Functional Embedding Theory with an Improved Kohn--Sham Inversion
  Algorithm}},}\ }\href {https://doi.org/10.1021/acs.jctc.8b00717} {\bibfield
  {journal} {\bibinfo  {journal} {J. Chem. Theory Comput.}\ }\textbf {\bibinfo
  {volume} {14}},\ \bibinfo {pages} {5680--5689} (\bibinfo {year}
  {2018})}\BibitemShut {NoStop}%
\bibitem [{\citenamefont {Kumar}, \citenamefont {Singh},\ and\ \citenamefont
  {Harbola}(2019)}]{kumar2019universal}%
  \BibitemOpen
  \bibfield  {author} {\bibinfo {author} {\bibfnamefont {A.}~\bibnamefont
  {Kumar}}, \bibinfo {author} {\bibfnamefont {R.}~\bibnamefont {Singh}},\ and\
  \bibinfo {author} {\bibfnamefont {M.~K.}\ \bibnamefont {Harbola}},\
  }\bibfield  {title} {\enquote {\bibinfo {title} {Universal nature of
  different methods of obtaining the exact {K}ohn--{S}ham exchange-correlation
  potential for a given density},}\ }\href
  {https://doi.org/10.1088/1361-6455/ab04e8} {\bibfield  {journal} {\bibinfo
  {journal} {J. Phys. B}\ }\textbf {\bibinfo {volume} {52}},\ \bibinfo {pages}
  {075007} (\bibinfo {year} {2019})}\BibitemShut {NoStop}%
\bibitem [{\citenamefont {Kanungo}, \citenamefont {Zimmerman},\ and\
  \citenamefont {Gavini}(2019)}]{Kanungo2019}%
  \BibitemOpen
  \bibfield  {author} {\bibinfo {author} {\bibfnamefont {B.}~\bibnamefont
  {Kanungo}}, \bibinfo {author} {\bibfnamefont {P.~M.}\ \bibnamefont
  {Zimmerman}},\ and\ \bibinfo {author} {\bibfnamefont {V.}~\bibnamefont
  {Gavini}},\ }\bibfield  {title} {\enquote {\bibinfo {title} {Exact
  exchange-correlation potentials from ground-state electron densities},}\
  }\href {https://doi.org/10.1038/s41467-019-12467-0} {\bibfield  {journal}
  {\bibinfo  {journal} {Nat Commun}\ }\textbf {\bibinfo {volume} {10}},\
  \bibinfo {pages} {4497} (\bibinfo {year} {2019})}\BibitemShut {NoStop}%
\bibitem [{\citenamefont {Kumar}\ and\ \citenamefont
  {Harbola}(2020)}]{Kumar2020}%
  \BibitemOpen
  \bibfield  {author} {\bibinfo {author} {\bibfnamefont {A.}~\bibnamefont
  {Kumar}}\ and\ \bibinfo {author} {\bibfnamefont {M.~K.}\ \bibnamefont
  {Harbola}},\ }\bibfield  {title} {\enquote {\bibinfo {title} {A general
  penalty method for density-to-potential inversion},}\ }\href
  {https://onlinelibrary.wiley.com/doi/abs/10.1002/qua.26400} {\bibfield
  {journal} {\bibinfo  {journal} {Int J Quantum Chem.}\ }\textbf {\bibinfo
  {volume} {120}},\ \bibinfo {pages} {e26400} (\bibinfo {year}
  {2020})}\BibitemShut {NoStop}%
\bibitem [{\citenamefont {Garrick}\ \emph {et~al.}(2020)\citenamefont
  {Garrick}, \citenamefont {Natan}, \citenamefont {Gould},\ and\ \citenamefont
  {Kronik}}]{Garrick2020}%
  \BibitemOpen
  \bibfield  {author} {\bibinfo {author} {\bibfnamefont {R.}~\bibnamefont
  {Garrick}}, \bibinfo {author} {\bibfnamefont {A.}~\bibnamefont {Natan}},
  \bibinfo {author} {\bibfnamefont {T.}~\bibnamefont {Gould}},\ and\ \bibinfo
  {author} {\bibfnamefont {L.}~\bibnamefont {Kronik}},\ }\bibfield  {title}
  {\enquote {\bibinfo {title} {{Exact Generalized Kohn--Sham Theory for Hybrid
  Functionals}},}\ }\href {https://doi.org/10.1103/PhysRevX.10.021040}
  {\bibfield  {journal} {\bibinfo  {journal} {Phys. Rev. X}\ }\textbf {\bibinfo
  {volume} {10}},\ \bibinfo {pages} {021040} (\bibinfo {year}
  {2020})}\BibitemShut {NoStop}%
\bibitem [{\citenamefont {Callow}, \citenamefont {Lathiotakis},\ and\
  \citenamefont {Gidopoulos}(2020)}]{Callow2020}%
  \BibitemOpen
  \bibfield  {author} {\bibinfo {author} {\bibfnamefont {T.~J.}\ \bibnamefont
  {Callow}}, \bibinfo {author} {\bibfnamefont {N.~N.}\ \bibnamefont
  {Lathiotakis}},\ and\ \bibinfo {author} {\bibfnamefont {N.~I.}\ \bibnamefont
  {Gidopoulos}},\ }\bibfield  {title} {\enquote {\bibinfo {title}
  {{Density-inversion method for the Kohn--Sham potential: Role of the
  screening density}},}\ }\href {https://doi.org/10.1063/5.0005781} {\bibfield
  {journal} {\bibinfo  {journal} {J. Chem. Phys.}\ }\textbf {\bibinfo {volume}
  {152}},\ \bibinfo {pages} {164114} (\bibinfo {year} {2020})}\BibitemShut
  {NoStop}%
\bibitem [{\citenamefont {Shi}\ and\ \citenamefont
  {Wasserman}(2021)}]{Shi2021}%
  \BibitemOpen
  \bibfield  {author} {\bibinfo {author} {\bibfnamefont {Y.}~\bibnamefont
  {Shi}}\ and\ \bibinfo {author} {\bibfnamefont {A.}~\bibnamefont
  {Wasserman}},\ }\bibfield  {title} {\enquote {\bibinfo {title} {{Inverse
  Kohn--Sham Density Functional Theory: Progress and Challenges}},}\ }\href
  {https://doi.org/10.1021/acs.jpclett.1c00752} {\bibfield  {journal} {\bibinfo
   {journal} {J. Phys. Chem. Lett.}\ }\textbf {\bibinfo {volume} {12}},\
  \bibinfo {pages} {5308--5318} (\bibinfo {year} {2021})}\BibitemShut {NoStop}%
\bibitem [{\citenamefont {Erhard}, \citenamefont {Trushin},\ and\ \citenamefont
  {G{\"o}rling}(2022)}]{Erhard2022}%
  \BibitemOpen
  \bibfield  {author} {\bibinfo {author} {\bibfnamefont {J.}~\bibnamefont
  {Erhard}}, \bibinfo {author} {\bibfnamefont {E.}~\bibnamefont {Trushin}},\
  and\ \bibinfo {author} {\bibfnamefont {A.}~\bibnamefont {G{\"o}rling}},\
  }\bibfield  {title} {\enquote {\bibinfo {title} {{Numerically stable
  inversion approach to construct Kohn--Sham potentials for given electron
  densities within a Gaussian basis set framework}},}\ }\href
  {https://doi.org/10.1063/5.0087356} {\bibfield  {journal} {\bibinfo
  {journal} {J. Chem. Phys.}\ }\textbf {\bibinfo {volume} {156}},\ \bibinfo
  {pages} {204124} (\bibinfo {year} {2022})}\BibitemShut {NoStop}%
\bibitem [{\citenamefont {Gould}(2023)}]{Gould2023}%
  \BibitemOpen
  \bibfield  {author} {\bibinfo {author} {\bibfnamefont {T.}~\bibnamefont
  {Gould}},\ }\bibfield  {title} {\enquote {\bibinfo {title} {{Toward routine
  Kohn--Sham inversion using the ``Lieb-response'' approach}},}\ }\href
  {https://doi.org/10.1063/5.0134330} {\bibfield  {journal} {\bibinfo
  {journal} {J. Chem. Phys.}\ }\textbf {\bibinfo {volume} {158}},\ \bibinfo
  {pages} {064102} (\bibinfo {year} {2023})}\BibitemShut {NoStop}%
\bibitem [{\citenamefont {Foulkes}\ and\ \citenamefont
  {Haydock}(1989)}]{FOULKES_PRB39_12520}%
  \BibitemOpen
  \bibfield  {author} {\bibinfo {author} {\bibfnamefont {W.~M.~C.}\
  \bibnamefont {Foulkes}}\ and\ \bibinfo {author} {\bibfnamefont
  {R.}~\bibnamefont {Haydock}},\ }\bibfield  {title} {\enquote {\bibinfo
  {title} {Tight-binding models and density-functional theory},}\ }\href
  {https://doi.org/10.1103/PhysRevB.39.12520} {\bibfield  {journal} {\bibinfo
  {journal} {Phys. Rev. B}\ }\textbf {\bibinfo {volume} {39}},\ \bibinfo
  {pages} {12520--12536} (\bibinfo {year} {1989})}\BibitemShut {NoStop}%
\bibitem [{\citenamefont {Shi}, \citenamefont {Ch\'{a}vez},\ and\ \citenamefont
  {Wasserman}(2022)}]{Shi2022}%
  \BibitemOpen
  \bibfield  {author} {\bibinfo {author} {\bibfnamefont {Y.}~\bibnamefont
  {Shi}}, \bibinfo {author} {\bibfnamefont {V.~H.}\ \bibnamefont
  {Ch\'{a}vez}},\ and\ \bibinfo {author} {\bibfnamefont {A.}~\bibnamefont
  {Wasserman}},\ }\bibfield  {title} {\enquote {\bibinfo {title} {{n2v: A
  density-to-potential inversion suite. A sandbox for creating, testing, and
  benchmarking density functional theory inversion methods}},}\ }\href
  {https://doi.org/https://doi.org/10.1002/wcms.1617} {\bibfield  {journal}
  {\bibinfo  {journal} {WIREs Comput Mol Sci.}\ }\textbf {\bibinfo {volume}
  {12}},\ \bibinfo {pages} {e1617} (\bibinfo {year} {2022})}\BibitemShut
  {NoStop}%
\bibitem [{\citenamefont {Crisostomo}\ \emph {et~al.}(2023)\citenamefont
  {Crisostomo}, \citenamefont {Pederson}, \citenamefont {Kozlowski},
  \citenamefont {Kalita}, \citenamefont {Cancio}, \citenamefont {Datchev},
  \citenamefont {Wasserman}, \citenamefont {Song},\ and\ \citenamefont
  {Burke}}]{Crisostomo2023}%
  \BibitemOpen
  \bibfield  {author} {\bibinfo {author} {\bibfnamefont {S.}~\bibnamefont
  {Crisostomo}}, \bibinfo {author} {\bibfnamefont {R.}~\bibnamefont
  {Pederson}}, \bibinfo {author} {\bibfnamefont {J.}~\bibnamefont {Kozlowski}},
  \bibinfo {author} {\bibfnamefont {B.}~\bibnamefont {Kalita}}, \bibinfo
  {author} {\bibfnamefont {A.~C.}\ \bibnamefont {Cancio}}, \bibinfo {author}
  {\bibfnamefont {K.}~\bibnamefont {Datchev}}, \bibinfo {author} {\bibfnamefont
  {A.}~\bibnamefont {Wasserman}}, \bibinfo {author} {\bibfnamefont
  {S.}~\bibnamefont {Song}},\ and\ \bibinfo {author} {\bibfnamefont
  {K.}~\bibnamefont {Burke}},\ }\bibfield  {title} {\enquote {\bibinfo {title}
  {Seven useful questions in density functional theory},}\ }\href
  {http://dx.doi.org/10.1007/s11005-023-01665-z} {\bibfield  {journal}
  {\bibinfo  {journal} {Lett Math Phys}\ }\textbf {\bibinfo {volume} {113}}
  (\bibinfo {year} {2023})}\BibitemShut {NoStop}%
\bibitem [{\citenamefont {Wrighton}\ \emph {et~al.}(2023)\citenamefont
  {Wrighton}, \citenamefont {Albavera-Mata}, \citenamefont {Rodr\'{\i}guez},
  \citenamefont {Tan}, \citenamefont {Cancio}, \citenamefont {Dufty},\ and\
  \citenamefont {Trickey}}]{Wrighton2023}%
  \BibitemOpen
  \bibfield  {author} {\bibinfo {author} {\bibfnamefont {J.}~\bibnamefont
  {Wrighton}}, \bibinfo {author} {\bibfnamefont {A.}~\bibnamefont
  {Albavera-Mata}}, \bibinfo {author} {\bibfnamefont {H.~F.}\ \bibnamefont
  {Rodr\'{\i}guez}}, \bibinfo {author} {\bibfnamefont {T.~S.}\ \bibnamefont
  {Tan}}, \bibinfo {author} {\bibfnamefont {A.~C.}\ \bibnamefont {Cancio}},
  \bibinfo {author} {\bibfnamefont {J.~W.}\ \bibnamefont {Dufty}},\ and\
  \bibinfo {author} {\bibfnamefont {S.~B.}\ \bibnamefont {Trickey}},\
  }\bibfield  {title} {\enquote {\bibinfo {title} {Some problems in density
  functional theory},}\ }\href {http://dx.doi.org/10.1007/s11005-023-01649-z}
  {\bibfield  {journal} {\bibinfo  {journal} {Lett Math Phys}\ }\textbf
  {\bibinfo {volume} {113}} (\bibinfo {year} {2023})}\BibitemShut {NoStop}%
\bibitem [{\citenamefont {Teale}, \citenamefont {Coriani},\ and\ \citenamefont
  {Helgaker}(2009)}]{TEALE_JCP130_104111}%
  \BibitemOpen
  \bibfield  {author} {\bibinfo {author} {\bibfnamefont {A.~M.}\ \bibnamefont
  {Teale}}, \bibinfo {author} {\bibfnamefont {S.}~\bibnamefont {Coriani}},\
  and\ \bibinfo {author} {\bibfnamefont {T.}~\bibnamefont {Helgaker}},\
  }\bibfield  {title} {\enquote {\bibinfo {title} {The calculation of
  adiabatic-connection curves from full configuration-interaction densities:
  {T}wo-electron systems},}\ }\href@noop {} {\bibfield  {journal} {\bibinfo
  {journal} {J. Chem. Phys.}\ }\textbf {\bibinfo {volume} {130}},\ \bibinfo
  {pages} {104111} (\bibinfo {year} {2009})}\BibitemShut {NoStop}%
\bibitem [{\citenamefont {Teale}, \citenamefont {Coriani},\ and\ \citenamefont
  {Helgaker}(2010{\natexlab{a}})}]{TEALE_JCP132_164115}%
  \BibitemOpen
  \bibfield  {author} {\bibinfo {author} {\bibfnamefont {A.~M.}\ \bibnamefont
  {Teale}}, \bibinfo {author} {\bibfnamefont {S.}~\bibnamefont {Coriani}},\
  and\ \bibinfo {author} {\bibfnamefont {T.}~\bibnamefont {Helgaker}},\
  }\bibfield  {title} {\enquote {\bibinfo {title} {Accurate calculation and
  modeling of the adiabatic connection in density functional theory},}\
  }\href@noop {} {\bibfield  {journal} {\bibinfo  {journal} {J. Chem. Phys.}\
  }\textbf {\bibinfo {volume} {132}},\ \bibinfo {pages} {164115} (\bibinfo
  {year} {2010}{\natexlab{a}})}\BibitemShut {NoStop}%
\bibitem [{\citenamefont {Teale}, \citenamefont {Coriani},\ and\ \citenamefont
  {Helgaker}(2010{\natexlab{b}})}]{TEALE_JCP133_164112}%
  \BibitemOpen
  \bibfield  {author} {\bibinfo {author} {\bibfnamefont {A.~M.}\ \bibnamefont
  {Teale}}, \bibinfo {author} {\bibfnamefont {S.}~\bibnamefont {Coriani}},\
  and\ \bibinfo {author} {\bibfnamefont {T.}~\bibnamefont {Helgaker}},\
  }\bibfield  {title} {\enquote {\bibinfo {title} {Range-dependent adiabatic
  connections},}\ }\href@noop {} {\bibfield  {journal} {\bibinfo  {journal} {J.
  Chem. Phys.}\ }\textbf {\bibinfo {volume} {133}},\ \bibinfo {pages} {164112}
  (\bibinfo {year} {2010}{\natexlab{b}})}\BibitemShut {NoStop}%
\bibitem [{\citenamefont {Kvaal}\ \emph {et~al.}(2014)\citenamefont {Kvaal},
  \citenamefont {Ekstr{\"o}m}, \citenamefont {Teale},\ and\ \citenamefont
  {Helgaker}}]{Kvaal2014}%
  \BibitemOpen
  \bibfield  {author} {\bibinfo {author} {\bibfnamefont {S.}~\bibnamefont
  {Kvaal}}, \bibinfo {author} {\bibfnamefont {U.}~\bibnamefont {Ekstr{\"o}m}},
  \bibinfo {author} {\bibfnamefont {A.~M.}\ \bibnamefont {Teale}},\ and\
  \bibinfo {author} {\bibfnamefont {T.}~\bibnamefont {Helgaker}},\ }\bibfield
  {title} {\enquote {\bibinfo {title} {Differentiable but exact formulation of
  density-functional theory},}\ }\href {https://doi.org/10.1063/1.4867005}
  {\bibfield  {journal} {\bibinfo  {journal} {J. Chem. Phys.}\ }\textbf
  {\bibinfo {volume} {140}},\ \bibinfo {pages} {18A518} (\bibinfo {year}
  {2014})}\BibitemShut {NoStop}%
\bibitem [{\citenamefont {Bauschke}\ and\ \citenamefont
  {Combettes}(2017)}]{Bauschke_2017}%
  \BibitemOpen
  \bibfield  {author} {\bibinfo {author} {\bibfnamefont {H.~H.}\ \bibnamefont
  {Bauschke}}\ and\ \bibinfo {author} {\bibfnamefont {P.~L.}\ \bibnamefont
  {Combettes}},\ }\href {https://doi.org/10.1007/978-3-319-48311-5} {\emph
  {\bibinfo {title} {Convex Analysis and Monotone Operator Theory in Hilbert
  Spaces}}},\ \bibinfo {edition} {2nd}\ ed.,\ CMS Books in Mathematics\
  (\bibinfo  {publisher} {Springer International Publishing},\ \bibinfo {year}
  {2017})\BibitemShut {NoStop}%
\bibitem [{\citenamefont {Barbu}\ and\ \citenamefont
  {Precupanu}(2012)}]{Barbu-Precepanu}%
  \BibitemOpen
  \bibfield  {author} {\bibinfo {author} {\bibfnamefont {V.}~\bibnamefont
  {Barbu}}\ and\ \bibinfo {author} {\bibfnamefont {T.}~\bibnamefont
  {Precupanu}},\ }\href {https://doi.org/10.1007/978-94-007-2247-7} {\emph
  {\bibinfo {title} {Convexity and Optimization in Banach Spaces}}},\ \bibinfo
  {edition} {4th}\ ed.,\ Springer Monographs in Mathematics\ (\bibinfo
  {publisher} {Springer Netherlands},\ \bibinfo {year} {2012})\BibitemShut
  {NoStop}%
\bibitem [{\citenamefont {Barbu}(2010)}]{Barbu_2010}%
  \BibitemOpen
  \bibfield  {author} {\bibinfo {author} {\bibfnamefont {V.}~\bibnamefont
  {Barbu}},\ }\href {https://doi.org/10.1007/978-1-4419-5542-5} {\emph
  {\bibinfo {title} {Nonlinear Differential Equations of Monotone Types in
  Banach Spaces}}},\ \bibinfo {edition} {1st}\ ed.,\ Springer Monographs in
  Mathematics\ (\bibinfo  {publisher} {Springer New York},\ \bibinfo {year}
  {2010})\BibitemShut {NoStop}%
\bibitem [{\citenamefont {Lammert}(2007)}]{Lammert2007}%
  \BibitemOpen
  \bibfield  {author} {\bibinfo {author} {\bibfnamefont {P.~E.}\ \bibnamefont
  {Lammert}},\ }\bibfield  {title} {\enquote {\bibinfo {title}
  {Differentiability of {L}ieb functional in electronic density functional
  theory},}\ }\href@noop {} {\bibfield  {journal} {\bibinfo  {journal} {Int. J.
  Quantum Chem.}\ }\textbf {\bibinfo {volume} {107}},\ \bibinfo {pages}
  {1943--1953} (\bibinfo {year} {2007})}\BibitemShut {NoStop}%
\bibitem [{\citenamefont {Laestadius}\ \emph {et~al.}(2018)\citenamefont
  {Laestadius}, \citenamefont {Penz}, \citenamefont {Tellgren}, \citenamefont
  {Ruggenthaler}, \citenamefont {Kvaal},\ and\ \citenamefont
  {Helgaker}}]{KSpaper2018}%
  \BibitemOpen
  \bibfield  {author} {\bibinfo {author} {\bibfnamefont {A.}~\bibnamefont
  {Laestadius}}, \bibinfo {author} {\bibfnamefont {M.}~\bibnamefont {Penz}},
  \bibinfo {author} {\bibfnamefont {E.~I.}\ \bibnamefont {Tellgren}}, \bibinfo
  {author} {\bibfnamefont {M.}~\bibnamefont {Ruggenthaler}}, \bibinfo {author}
  {\bibfnamefont {S.}~\bibnamefont {Kvaal}},\ and\ \bibinfo {author}
  {\bibfnamefont {T.}~\bibnamefont {Helgaker}},\ }\bibfield  {title} {\enquote
  {\bibinfo {title} {{Generalized Kohn--Sham iteration on Banach spaces}},}\
  }\href {https://doi.org/10.1063/1.5037790} {\bibfield  {journal} {\bibinfo
  {journal} {J. Chem. Phys.}\ }\textbf {\bibinfo {volume} {149}},\ \bibinfo
  {pages} {164103} (\bibinfo {year} {2018})}\BibitemShut {NoStop}%
\bibitem [{\citenamefont {Laestadius}\ \emph {et~al.}(2019)\citenamefont
  {Laestadius}, \citenamefont {Tellgren}, \citenamefont {Penz}, \citenamefont
  {Ruggenthaler}, \citenamefont {Kvaal},\ and\ \citenamefont
  {Helgaker}}]{MY-CDFTpaper2019}%
  \BibitemOpen
  \bibfield  {author} {\bibinfo {author} {\bibfnamefont {A.}~\bibnamefont
  {Laestadius}}, \bibinfo {author} {\bibfnamefont {E.~I.}\ \bibnamefont
  {Tellgren}}, \bibinfo {author} {\bibfnamefont {M.}~\bibnamefont {Penz}},
  \bibinfo {author} {\bibfnamefont {M.}~\bibnamefont {Ruggenthaler}}, \bibinfo
  {author} {\bibfnamefont {S.}~\bibnamefont {Kvaal}},\ and\ \bibinfo {author}
  {\bibfnamefont {T.}~\bibnamefont {Helgaker}},\ }\bibfield  {title} {\enquote
  {\bibinfo {title} {{K}ohn--{S}ham {T}heory with {P}aramagnetic {C}urrents:
  {C}ompatibility and {F}unctional {D}ifferentiability},}\ }\href
  {https://doi.org/10.1021/acs.jctc.9b00141} {\bibfield  {journal} {\bibinfo
  {journal} {Journal of Chemical Theory and Computation}\ }\textbf {\bibinfo
  {volume} {15}},\ \bibinfo {pages} {4003--4020} (\bibinfo {year} {2019})},\
  \Eprint {https://arxiv.org/abs/https://doi.org/10.1021/acs.jctc.9b00141}
  {https://doi.org/10.1021/acs.jctc.9b00141} \BibitemShut {NoStop}%
\bibitem [{\citenamefont {Penz}\ \emph {et~al.}(2020)\citenamefont {Penz},
  \citenamefont {Laestadius}, \citenamefont {Tellgren}, \citenamefont
  {Ruggenthaler},\ and\ \citenamefont {Lammert}}]{PRLerrata}%
  \BibitemOpen
  \bibfield  {author} {\bibinfo {author} {\bibfnamefont {M.}~\bibnamefont
  {Penz}}, \bibinfo {author} {\bibfnamefont {A.}~\bibnamefont {Laestadius}},
  \bibinfo {author} {\bibfnamefont {E.~I.}\ \bibnamefont {Tellgren}}, \bibinfo
  {author} {\bibfnamefont {M.}~\bibnamefont {Ruggenthaler}},\ and\ \bibinfo
  {author} {\bibfnamefont {P.~E.}\ \bibnamefont {Lammert}},\ }\bibfield
  {title} {\enquote {\bibinfo {title} {{Erratum: Guaranteed Convergence of a
  Regularized Kohn--Sham Iteration in Finite Dimensions [Phys. Rev. Lett. 123,
  037401 (2019)]}},}\ }\href {https://doi.org/10.1103/PhysRevLett.125.249902}
  {\bibfield  {journal} {\bibinfo  {journal} {Phys. Rev. Lett.}\ }\textbf
  {\bibinfo {volume} {125}},\ \bibinfo {pages} {249902(E)} (\bibinfo {year}
  {2020})}\BibitemShut {NoStop}%
\bibitem [{\citenamefont {Penz}, \citenamefont {Csirik},\ and\ \citenamefont
  {Laestadius}(2023)}]{Penz_2023}%
  \BibitemOpen
  \bibfield  {author} {\bibinfo {author} {\bibfnamefont {M.}~\bibnamefont
  {Penz}}, \bibinfo {author} {\bibfnamefont {M.~A.}\ \bibnamefont {Csirik}},\
  and\ \bibinfo {author} {\bibfnamefont {A.}~\bibnamefont {Laestadius}},\
  }\bibfield  {title} {\enquote {\bibinfo {title} {{Density-potential inversion
  from Moreau--Yosida regularization}},}\ }\href
  {https://doi.org/10.1088/2516-1075/acc626} {\bibfield  {journal} {\bibinfo
  {journal} {Electron. Struct.}\ }\textbf {\bibinfo {volume} {5}},\ \bibinfo
  {pages} {014009} (\bibinfo {year} {2023})}\BibitemShut {NoStop}%
\bibitem [{\citenamefont {Penz}\ and\ \citenamefont
  {Laestadius}(2026)}]{Penz2026}%
  \BibitemOpen
  \bibfield  {author} {\bibinfo {author} {\bibfnamefont {M.}~\bibnamefont
  {Penz}}\ and\ \bibinfo {author} {\bibfnamefont {A.}~\bibnamefont
  {Laestadius}},\ }\bibfield  {title} {\enquote {\bibinfo {title} {{Adaptation
  of Moreau--Yosida regularization to the modulus of convexity}},}\ }\href
  {https://doi.org/10.1016/j.jmaa.2025.129956} {\bibfield  {journal} {\bibinfo
  {journal} {Journal of Mathematical Analysis and Applications}\ }\textbf
  {\bibinfo {volume} {554}},\ \bibinfo {pages} {129956} (\bibinfo {year}
  {2026})}\BibitemShut {NoStop}%
\bibitem [{\citenamefont {Herbst}, \citenamefont {Bakkestuen},\ and\
  \citenamefont {Laestadius}(2025)}]{Herbst2025}%
  \BibitemOpen
  \bibfield  {author} {\bibinfo {author} {\bibfnamefont {M.~F.}\ \bibnamefont
  {Herbst}}, \bibinfo {author} {\bibfnamefont {V.~H.}\ \bibnamefont
  {Bakkestuen}},\ and\ \bibinfo {author} {\bibfnamefont {A.}~\bibnamefont
  {Laestadius}},\ }\bibfield  {title} {\enquote {\bibinfo {title} {{Kohn--Sham
  inversion with mathematical guarantees}},}\ }\href
  {https://doi.org/10.1103/physrevb.111.205143} {\bibfield  {journal} {\bibinfo
   {journal} {Physical Review B}\ }\textbf {\bibinfo {volume} {111}} (\bibinfo
  {year} {2025}),\ 10.1103/physrevb.111.205143}\BibitemShut {NoStop}%
\bibitem [{\citenamefont {van Leeuwen}\ and\ \citenamefont
  {Baerends}(1994{\natexlab{b}})}]{vanLeeuwen_1994}%
  \BibitemOpen
  \bibfield  {author} {\bibinfo {author} {\bibfnamefont {R.}~\bibnamefont {van
  Leeuwen}}\ and\ \bibinfo {author} {\bibfnamefont {E.~J.}\ \bibnamefont
  {Baerends}},\ }\bibfield  {title} {\enquote {\bibinfo {title}
  {Exchange-correlation potential with correct asymptotic behavior},}\ }\href
  {https://doi.org/10.1103/PhysRevA.49.2421} {\bibfield  {journal} {\bibinfo
  {journal} {Phys. Rev. A}\ }\textbf {\bibinfo {volume} {49}},\ \bibinfo
  {pages} {2421--2431} (\bibinfo {year} {1994}{\natexlab{b}})}\BibitemShut
  {NoStop}%
\bibitem [{\citenamefont {Rowlinson}(1989)}]{ROWLINSON_PA156_15}%
  \BibitemOpen
  \bibfield  {author} {\bibinfo {author} {\bibfnamefont {J.~S.}\ \bibnamefont
  {Rowlinson}},\ }\bibfield  {title} {\enquote {\bibinfo {title} {The {Y}ukawa
  potential},}\ }\href
  {https://doi.org/https://doi.org/10.1016/0378-4371(89)90108-8} {\bibfield
  {journal} {\bibinfo  {journal} {Physica A}\ }\textbf {\bibinfo {volume}
  {156}},\ \bibinfo {pages} {15--34} (\bibinfo {year} {1989})}\BibitemShut
  {NoStop}%
\bibitem [{\citenamefont {Penz}\ \emph {et~al.}(2019)\citenamefont {Penz},
  \citenamefont {Laestadius}, \citenamefont {Tellgren},\ and\ \citenamefont
  {Ruggenthaler}}]{KS_PRL_2019}%
  \BibitemOpen
  \bibfield  {author} {\bibinfo {author} {\bibfnamefont {M.}~\bibnamefont
  {Penz}}, \bibinfo {author} {\bibfnamefont {A.}~\bibnamefont {Laestadius}},
  \bibinfo {author} {\bibfnamefont {E.~I.}\ \bibnamefont {Tellgren}},\ and\
  \bibinfo {author} {\bibfnamefont {M.}~\bibnamefont {Ruggenthaler}},\
  }\bibfield  {title} {\enquote {\bibinfo {title} {{Guaranteed convergence of a
  regularized Kohn--Sham iteration in finite dimensions}},}\ }\href
  {https://doi.org/10.1103/PhysRevLett.123.037401} {\bibfield  {journal}
  {\bibinfo  {journal} {Phys. Rev. Lett.}\ }\textbf {\bibinfo {volume} {123}},\
  \bibinfo {pages} {037401} (\bibinfo {year} {2019})}\BibitemShut {NoStop}%
\bibitem [{\citenamefont {Pulay}(1980)}]{PULAY_CPL73_393}%
  \BibitemOpen
  \bibfield  {author} {\bibinfo {author} {\bibfnamefont {P.}~\bibnamefont
  {Pulay}},\ }\bibfield  {title} {\enquote {\bibinfo {title} {Convergence
  acceleration of iterative sequences---the case of {SCF} iteration},}\
  }\href@noop {} {\bibfield  {journal} {\bibinfo  {journal} {Chem. Phys.
  Lett.}\ }\textbf {\bibinfo {volume} {73}},\ \bibinfo {pages} {393--398}
  (\bibinfo {year} {1980})}\BibitemShut {NoStop}%
\bibitem [{\citenamefont {Pulay}(1982)}]{PULAY_JCC3_556}%
  \BibitemOpen
  \bibfield  {author} {\bibinfo {author} {\bibfnamefont {P.}~\bibnamefont
  {Pulay}},\ }\bibfield  {title} {\enquote {\bibinfo {title} {Improved {SCF}
  convergence acceleration},}\ }\href@noop {} {\bibfield  {journal} {\bibinfo
  {journal} {J. Comput. Chem.}\ }\textbf {\bibinfo {volume} {3}},\ \bibinfo
  {pages} {556--560} (\bibinfo {year} {1982})}\BibitemShut {NoStop}%
\bibitem [{\citenamefont {Helgaker}, \citenamefont {J{\o}rgensen},\ and\
  \citenamefont {Olsen}(2000)}]{HELGAKER00}%
  \BibitemOpen
  \bibfield  {author} {\bibinfo {author} {\bibfnamefont {T.}~\bibnamefont
  {Helgaker}}, \bibinfo {author} {\bibfnamefont {P.}~\bibnamefont
  {J{\o}rgensen}},\ and\ \bibinfo {author} {\bibfnamefont {J.}~\bibnamefont
  {Olsen}},\ }\href@noop {} {\emph {\bibinfo {title} {Molecular
  Electronic-Structure Theory}}}\ (\bibinfo  {publisher} {John Wiley \& Sons,
  Ltd},\ \bibinfo {year} {2000})\BibitemShut {NoStop}%
\bibitem [{\citenamefont {Helgaker}\ \emph {et~al.}(2000)\citenamefont
  {Helgaker}, \citenamefont {Larsen}, \citenamefont {Olsen},\ and\
  \citenamefont {J{\o}rgensen}}]{HELGAKER_CPL327_397}%
  \BibitemOpen
  \bibfield  {author} {\bibinfo {author} {\bibfnamefont {T.}~\bibnamefont
  {Helgaker}}, \bibinfo {author} {\bibfnamefont {H.}~\bibnamefont {Larsen}},
  \bibinfo {author} {\bibfnamefont {J.}~\bibnamefont {Olsen}},\ and\ \bibinfo
  {author} {\bibfnamefont {P.}~\bibnamefont {J{\o}rgensen}},\ }\bibfield
  {title} {\enquote {\bibinfo {title} {Direct optimization of the {AO} density
  matrix in {H}artree--{F}ock and {K}ohn--{S}ham theories},}\ }\href@noop {}
  {\bibfield  {journal} {\bibinfo  {journal} {Chem. Phys. Lett.}\ }\textbf
  {\bibinfo {volume} {327}},\ \bibinfo {pages} {397--403} (\bibinfo {year}
  {2000})}\BibitemShut {NoStop}%
\bibitem [{\citenamefont {Bohle}, \citenamefont {Lotfigolian},\ and\
  \citenamefont {Tellgren}(2025)}]{SableCode}%
  \BibitemOpen
  \bibfield  {author} {\bibinfo {author} {\bibfnamefont {O.}~\bibnamefont
  {Bohle}}, \bibinfo {author} {\bibfnamefont {M.}~\bibnamefont {Lotfigolian}},\
  and\ \bibinfo {author} {\bibfnamefont {E.~I.}\ \bibnamefont {Tellgren}},\
  }\href@noop {} {} (\bibinfo {year} {2025}),\ \bibinfo {note} {{S}able: a {1D}
  {H}artree--{F}ock code, \texttt{gitlab.com/et/sable-release}}\BibitemShut
  {NoStop}%
\end{thebibliography}
\end{document}